\documentclass[12pt]{article}
\usepackage[T1]{fontenc}
\usepackage{newtxtext,newtxmath}

\usepackage{graphicx}
\usepackage{subcaption}
\usepackage[bb=boondox,bbscaled=.95,cal=boondoxo]{mathalfa}
\usepackage[letterpaper,margin=1in]{geometry}

\renewenvironment{abstract}
	{\quotation}
	{\endquotation}

\date{}

\makeatletter
\renewcommand{\fnum@figure}{\textbf{Figure \thefigure}}
\renewcommand{\fnum@table}{\textbf{Table \thetable}}
\makeatother

\usepackage{scicite}

\usepackage{url}

\usepackage[most]{tcolorbox}
\usepackage{booktabs}
\usepackage[detect-all]{siunitx}
\usepackage{xspace}
\usepackage[pdftex,linkcolor=black,pdfborder={0 0 0}]{hyperref} 
\usepackage{amsmath}
\usepackage{setspace}
\usepackage{enumitem}
\usepackage{bm}
\usepackage[%
	sort&compress]
    {cleveref}
\Crefname{appendix}{Supplement}{Supplements}

\usepackage[left]{lineno}

\newcommand\ie{i.e.\xspace}
\newcommand\eg{e.g.\xspace}
\newcommand{\X}{$\mathbb{X}$\xspace}
\newcommand{\US}{{U.S.}\xspace}

\newcommand{\func}[1]{\mathit{#1}}

\def\scititle{ \singlespacing
	Location transparency reduces activity by accounts misrepresenting their location on X
}
\title{\bfseries \boldmath \scititle}

\author{
    Yuwei Chuai$^{1,2}$,
	Thomas Renault$^{3}$,
	David Rand$^{4,5}$,
	Mohsen Mosleh$^{2,4,\ast}$\and
	\footnotesize$^{1}$SnT, University of Luxembourg, Luxembourg, Luxembourg.\and
	\footnotesize$^{2}$Oxford Internet Institute, University of Oxford, Woodstock Road, Oxford, OX2 6GG.\and
	\footnotesize$^{3}$Université Paris-Saclay, Faculté Jean Monnet,
54 Boulevard Desgranges, 92330 Sceaux, France.\and
  \footnotesize$^{4}$Sloan School of Management, Massachusetts Institute of Technology, 100 Main Street Cambridge, MA 02142.\and
  \footnotesize$^{5}$Departments of Information Science,  Marketing, and Psychology, Cornell University, 107 Hoy Rd, Ithaca, NY 14850.\and
	\footnotesize$^\ast$Corresponding author. Email: mohsen.mosleh@oii.ox.ac.uk
}

\renewcommand{\thefigure}{\arabic{figure}}

\begin{document} 
\maketitle
\thispagestyle{empty}

\newpage
\begin{center} \bfseries \boldmath
    Abstract
\end{center}
\begin{abstract} 
Concerns about inauthentic accounts, including foreign actors posing as domestic voices, are central to debates about online discourse. Yet, little is known about accounts with inaccurate location claims and how they behave when discrepancies between their claimed and actual locations become publicly visible. In November 2025, \X introduced an ``About this account'' feature that discloses each account's platform-inferred location of operation. We leverage this intervention in a large-scale quasi-experimental study of 8,200 politically engaged accounts claiming a U.S. location, comparing accounts whose disclosed locations matched versus contradicted their claims across 1.3 million posts and 3.6 million replies over 21 weeks. Before disclosure, location-mismatched accounts posted more misleading, scam-related, and cryptocurrency-related content, but showed no distinctive partisan leaning. Difference-in-differences estimates show that disclosure reduced the posting activity of location-mismatched accounts by 13.1\% with the largest declines among accounts revealed to be in Africa (29.2\%) and Asia (24.4\%), and among accounts with VPN flags, username changes, or scam- and crypto-heavy content. Additionally, the decline in their replies was concentrated in interactions with U.S.-based recipients (10.3\%), whereas replies to non-U.S.-based recipients showed no statistically significant change. Conversely, there was no significant change in average audience engagement with their posts. Location transparency thus works primarily by inducing restraint among the disclosed accounts rather than by shifting audience behaviour, and the accounts it constrains look at least as much like cross-border fraud as foreign political influence.
\end{abstract}

\newpage
\section*{Introduction}
The integrity of online public discourse depends not only on what information users share, but also on the identities through which that information is produced and circulated~\cite{donath2002identity}. Social media platforms enable users to selectively construct and manage their identities in relation to different audiences, creating online presentations that may diverge from their offline attributes~\cite{zhao2008identity,marwick2011tweet}. Geographic location is a particularly consequential yet difficult-to-verify dimension of online identity: researchers generally lack direct information about where accounts are operated and have to instead rely on self-reported profile fields, language, time zones, posting patterns, network structure, classifier-based inferences or platform takedown data~\cite{jurgens2015geolocation,zheng2018survey}. In November 2025, \X (formerly Twitter) introduced an ``About this account'' feature that publicly displays the country or region from which an account is inferred to operate, based on aggregated IP addresses, alongside information such as username changes and potential VPN use~\cite{x2026change}. By rendering previously hidden discrepancies between users' self-presented and platform-inferred locations publicly observable, the roll-out provides a rare opportunity to study location mismatch at scale and examine the consequences of its disclosure.

Inauthentic location presentation matters because presenting oneself as locally embedded may increase credibility, facilitate participation in domestic political conversations, or obscure the cross-border character of commercial activity~\cite{zannettou2019disinformation,eady2023exposure,bail2020assessing,alizadeh2020content}. Coordinated influence operations, for instance, have used deceptive profiles to participate in domestic online discussions during the \US presidential elections, while commercially motivated networks use similar tactics to promote scams, spam, and other forms of fraud~\cite{eady2023exposure,appel2026deceptive,ferrara2022twitter}. Recent evidence also suggests that political and commercial motives can coexist within the same deceptive networks, with political content serving as a means of attracting attention for financial gain~\cite{appel2026deceptive}.

Here, we study the effects of \X revealing inauthentic locations. This transparency can alter online behaviour on both sides of an interaction, \ie, the activity of source accounts and the engagement decisions of their audiences. Research on disclosure and observability suggests that making information public can change the behaviour of those accounts being observed by increasing accountability and altering the reputational or strategic costs of actions~\cite{weil2006effectiveness,loewenstein2014disclosure,yoeli2013powering}. Thus, accounts whose self-presented and platform-inferred locations conflict may reduce or modify their activity when that discrepancy becomes publicly visible. It is also possible that audiences who notice the disclosure may update their beliefs about an account's identity or credibility and become more or less willing to engage with its content. However, these channels need not operate together: because \X displays location information on an account's profile page rather than alongside its individual posts, the disclosure may be salient to the account operator while remaining peripheral to -- and thus going unnoticed by -- audiences when they decide whether to engage.

Prior research makes both account- and audience-side responses plausible.
On the account side, evidence from mandatory IP-location disclosure on Chinese social media --  both on account profile pages and alongside users' contributions -- shows mixed results. On Weibo, disclosure reduces cross-province participation while increasing regionally discriminatory replies and location-based incivility~\cite{guo2025civilizing,yang2025user}. By contrast, experiments across three Chinese platforms find that displaying IP locations reduces cyberbullying by strengthening users' impression-management motives~\cite{meng2025combating}. Meanwhile, another study finds that disclosure dampens negative expression towards government accounts but amplifies it towards non-government accounts~\cite{li2026location}. On the audience side, prior work finds that identity cues can causally increase up-votes and reduce down-votes for accounts with high levels of contribution and reputation~\cite{taylor2023identity}, while platform labels identifying state-affiliated sources can reduce subsequent news sharing~\cite{liang2023effects}. Together, these mixed findings suggest that location transparency does not produce a uniform response.

Importantly, however, these studies examine the consequences of location visibility itself rather than the public exposure of an \textit{inconsistency} between an account's disclosed location and its self-presentation. Beyond this emerging literature, research on platform interventions aimed at improving information integrity, particularly in Western contexts, has largely focused on information attached to individual pieces of content through fact-checks, warning labels, accuracy prompts, or crowdsourced assessments~\cite{pennycook2019fighting,bak2022combining}. Community-based fact-checking, for example, can reduce the spread of annotated misleading posts~\cite{chuai2026community,slaughter2025community}, but has divergent downstream effects on the subsequent behaviour of corrected accounts~\cite{chuai2026corrections}. Unlike these content-level interventions, \X's location feature discloses information about an account's provenance rather than the accuracy of a particular claim. It therefore remains unclear whether revealing a geographic identity discrepancy changes the behaviour of the accounts whose discrepancy becomes visible and the engagement decisions of their audiences.

In this study, we focus on politically relevant accounts that claimed a \US location before the location transparency feature was introduced. This setting is substantively important because foreign actors presenting themselves as \US domestic participants in political discussions have received substantial public and scholarly attention, particularly in relation to coordinated manipulation, foreign influence operations, and attempts to exacerbate domestic political divisions~\cite{zannettou2019disinformation,bail2020assessing,eady2023exposure}. At the same time, misleading location presentation may also facilitate commercially motivated activities, including scams, spam, and cryptocurrency promotion~\cite{appel2026deceptive}. Consequently, we address two related questions. First, is location mismatch among politically relevant accounts associated primarily with indicators of political influence operations (\eg, posting divisive or misleading content), or does it predict more commercially-oriented activity like spam (or both)? Second, what are the effects of location mismatch becoming publicly visible? Do the affected accounts change their activity, and do audiences change their engagement? These questions allow us to examine what location mismatch signals and how disclosure affects behaviour (without having to assume that any individual mismatch constitutes deliberate deception).

To address our questions, we conduct a large-scale observational study using a quasi-experimental design based on \num{8200} location-matched and location-mismatched accounts, covering \num{1327286} original posts and \num{3631457} outbound replies observed over seven weeks before and fourteen weeks after the roll-out of location transparency on \X. Before disclosure, comparative analysis shows that location-mismatched accounts were more likely to have changed their usernames, to be flagged as potentially using a VPN, and to publish misleading, scam-related and cryptocurrency-related content, but did not exhibit stronger left- or right-leaning posting preferences. Our Difference-in-Differences (DiD) approach identifies that location transparency reduces the posting activity of location-mismatched accounts, while producing no statistically detectable change in the average audience engagement received per original post. Location-mismatched accounts' activity reductions are particularly pronounced among accounts disclosed as operating from Africa and Asia and among toxic replies from those accounts to \US-based recipients. Together, our findings show that location transparency primarily induces behavioural restraint among location-mismatched accounts, but does not produce a broad withdrawal of audience engagement. They further suggest that the location mismatch may characterise a broader mixture of cross-border inauthentic behaviours, such as scams, rather than being particularly driven by foreign political influence operations.

\clearpage
\section*{Results}

We rely on profile data collected prior to the introduction of \X's transparency feature and restrict the sample to users who claimed a \US location in their self-reported profile information and signalled engagement with \US politics in their bios (see \nameref{sec:methods} for details). This results in a balanced dataset of 8,200 accounts, comprising users whose self-reported locations matched their algorithmically revealed locations and users whose self-reported locations did not match. For each account, we collect transparency information (``About this account'' section of the user profile), including the algorithmically revealed location, the number of username changes, and whether the inferred location was flagged as potentially inaccurate, such as in cases of VPN use. Finally, we collect all original posts and replies from the selected accounts over a longitudinal period of seven weeks before and fourteen weeks after the transparency intervention.

\subsection*{Characteristics of location-mismatched accounts}

\begin{figure}[ht]
    \centering
    \captionsetup[subfloat]{font={bf, small}, skip=0pt, singlelinecheck=false, labelformat=simple, position=top}
    \subfloat{\includegraphics[width = \textwidth]{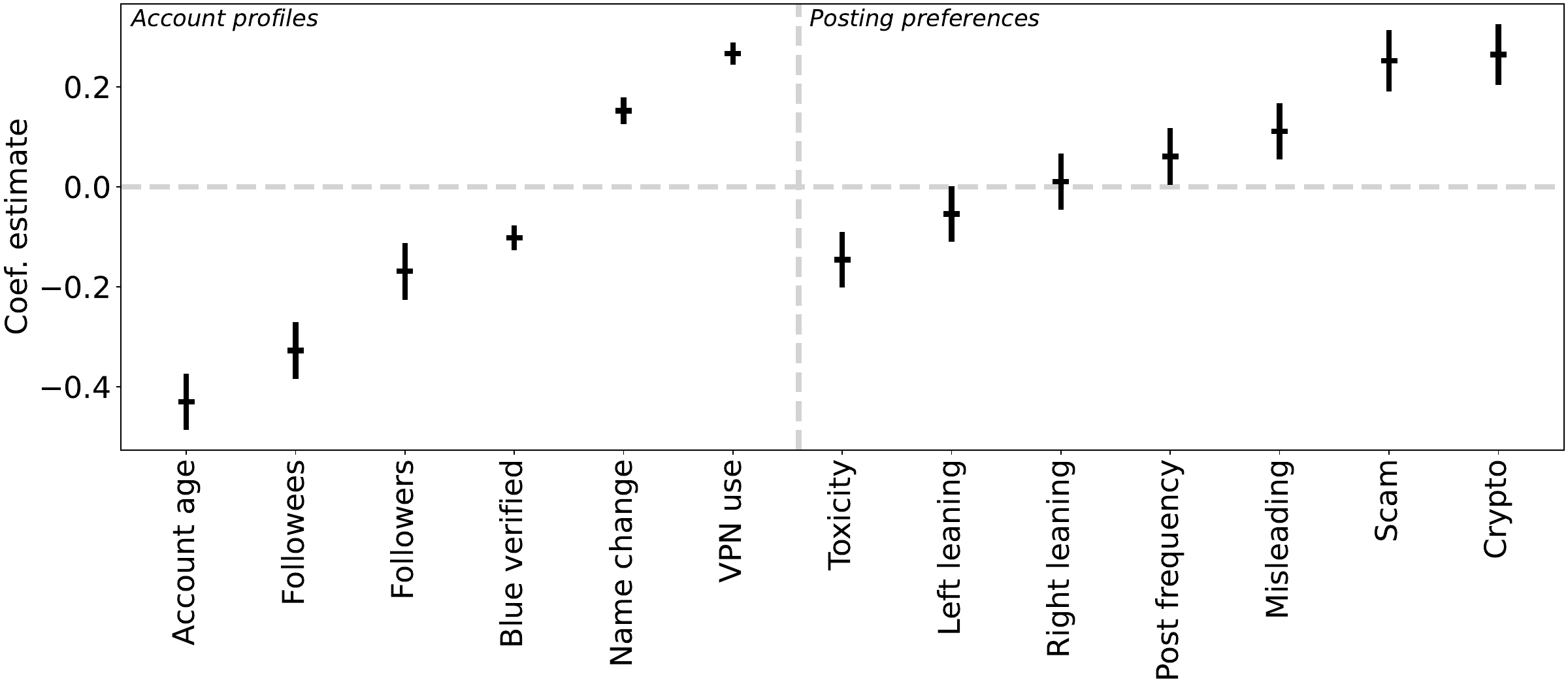}}
    \caption{\textbf{Comparison between location-mismatched and location-matched accounts.} Shown are coefficient estimates from univariate linear regression models across account profiles and posting preferences. The estimation is based on \num{4937} accounts without missing values across variables, and all continuous variables are log-transformed and $z$-standardised for better interpretability. The estimates are ordered in ascending order in the account-profile and posting-preference panels, respectively. The error bars represent 95\% Confidence Intervals (CIs).}
    \label{fig:account_linear}
\end{figure}

We begin by documenting descriptive differences between accounts that claim a \US location but are algorithmically revealed to be outside the \US, and accounts whose claimed and revealed locations are both \US domestic prior to the introduction of the ``About this account'' feature. For our analysis, we run a series of linear regression models across account characteristics and posting preferences from the content perspective (see \Cref{fig:account_linear}). 

We find that location-mismatched accounts were created more recently (Coef.~$=-0.430$, $z=-15.035$, $p<0.001$; 95\%~CI: $[-0.486, -0.374]$), have a lower number of followers (Coef.~$=-0.169$, $z=-5.823$, $p<0.001$; 95\%~CI: $[-0.226, -0.112]$) and followees (Coef.~$=-0.327$, $z=-11.230$, $p<0.001$; 95\%~CI: $[-0.384, -0.270]$), and are less likely to have a Blue Verified badge (Coef.~$=-0.102$, $z=-8.134$, $p<0.001$; 95\%~CI: $[-0.126, -0.077]$) than location-matched accounts. In contrast, location-mismatched accounts are more likely to have changed their usernames in the past (Coef.~$=0.152$, $z=11.091$, $p<0.001$; 95\%~CI: $[0.125, 0.179]$) and are more likely to use a VPN (Coef.~$=0.266$, $z=23.569$, $p<0.001$; 95\%~CI: $[0.244, 0.288]$), compared to location-matched accounts.

Additionally, we examine accounts' posting behaviour based on the original posts (without replies) published by these users in the seven weeks preceding the introduction of transparency information on \X. For each account, we use the weekly average of the number of original posts to denote its posting frequency and calculate the average shares of posts devoted to political leaning, toxicity, scams, cryptocurrency, and misleading content, respectively. We find that location-mismatched accounts publish less toxic content than location-matched accounts (Coef.~$=-0.146$, $z=-5.104$, $p<0.001$; 95\%~CI: $[-0.201, -0.090]$). Regarding political posts, we find no significant difference between the two groups in publishing either left-leaning (Coef.~$=-0.054$, $z=-1.921$, $p=0.055$; 95\%~CI: $[-0.110, 0.001]$) or right-leaning content (Coef.~$=0.011$, $z=0.375$, $p=0.707$; 95\%~CI: $[-0.046, 0.067]$). Finally, location-mismatched accounts post more misleading (Coef.~$=0.111$, $z=3.856$, $p<0.001$; 95\%~CI: $[0.055, 0.168]$), scam-related (Coef.~$=0.252$, $z=8.079$, $p<0.001$; 95\%~CI: $[0.191, 0.313]$), and cryptocurrency-related content (Coef.~$=0.264$, $z=8.605$, $p<0.001$; 95\%~CI: $[0.204, 0.325]$), compared to location-matched accounts.

In summary, these differences are present before disclosure of account location, suggesting that misrepresentation is not random but correlated with observable behavioural markers that may relate to strategic communication, amplification tactics, or attempts to appear domestically legitimate.

\clearpage

\clearpage
\subsection*{Impact of location transparency}

Next, we causally estimate the impact of account location disclosure on (i) account activity (\ie, original posts and outbound replies from the selected accounts) and (ii) inbound user engagement (\ie, reposts and replies received by the selected accounts) using a Difference-in-Differences (DiD) design. We define the treatment group as accounts whose self-reported location does not match their algorithmically revealed location, and the control group as accounts whose self-reported location matches their algorithmically revealed location. We leverage the introduction of the location-transparency feature on \X as an external shock that publicly disclosed this information, allowing us to compare how behaviour changed among treated accounts relative to comparable control accounts before and after the feature was introduced. The DiD design estimates the additional post-disclosure change in the treatment group, namely the Average Treatment Effect on the Treated (ATT), relative to the pre-disclosure period and compared to the control group. Specifically, we employ both leads-and-lags and pre-post DiD specifications to estimate weekly and aggregated effects of the location-disclosure feature over a period spanning seven weeks before and fourteen weeks after its implementation on \X (see model details in \nameref{sec:did_spec}).

\begin{figure}[ht]
    \centering
    \captionsetup[subfloat]{font={bf, small}, skip=0pt, singlelinecheck=false, labelformat=simple, position=top}
    \subfloat[]{\includegraphics[width = \textwidth]{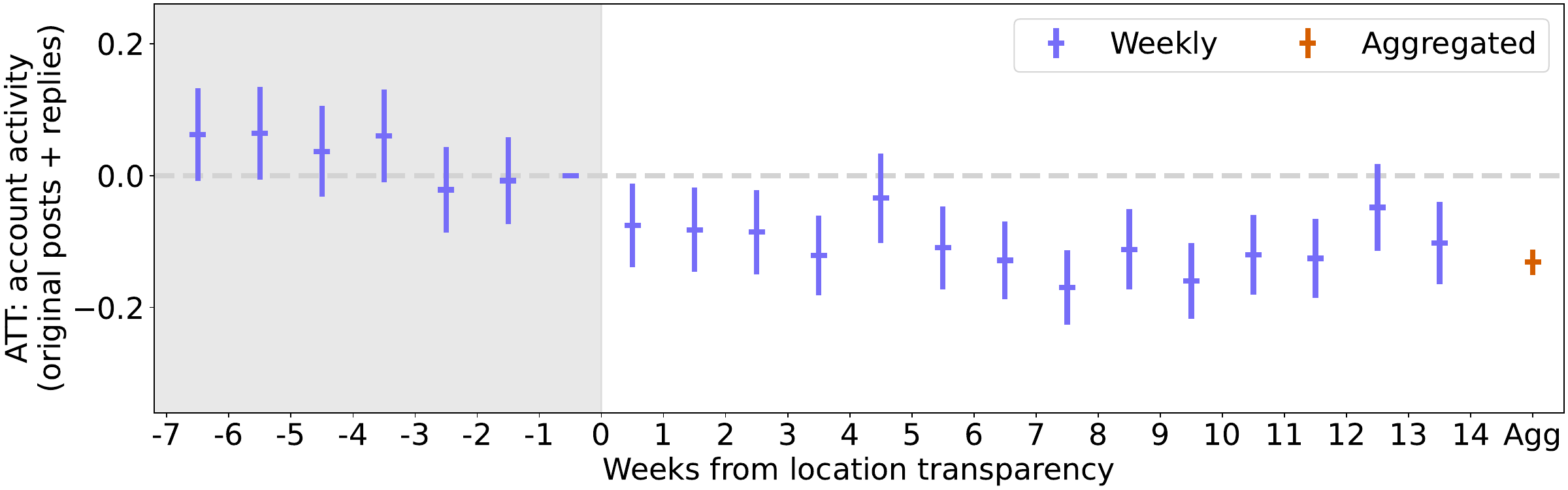}}\\
    \subfloat[]{\includegraphics[width = \textwidth]{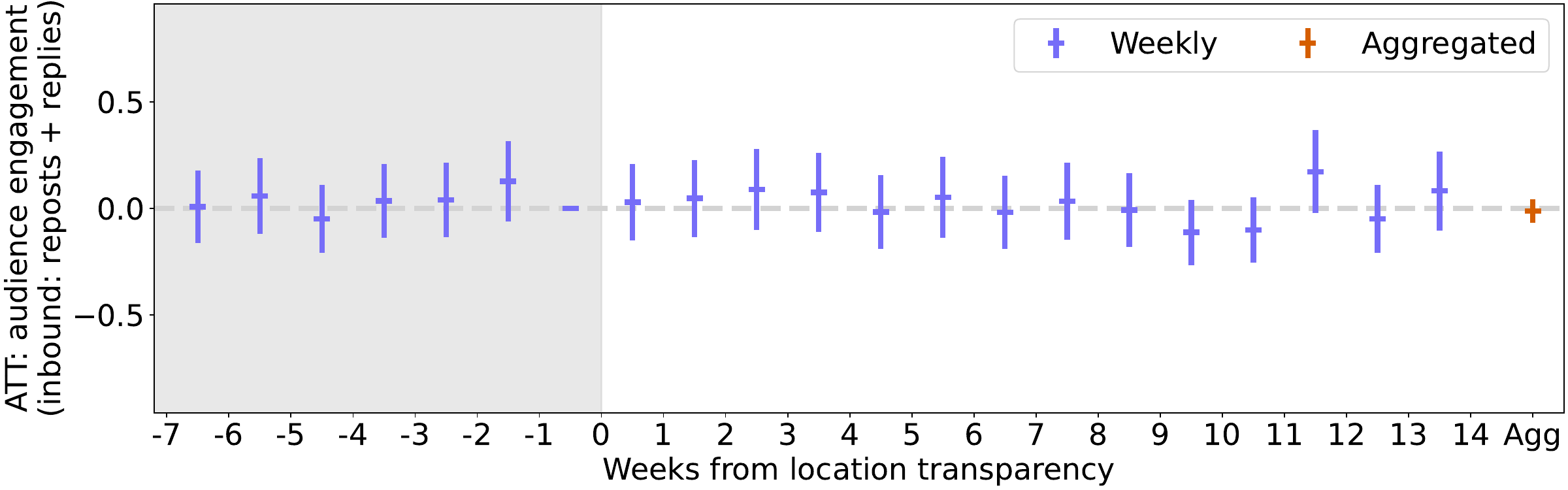}}
    \caption{\textbf{Location transparency reduces the posting activity of location-mismatched accounts but does not detectably affect the average audience engagement their posts receive.} \textbf{(a)}~Weekly and aggregated ATT estimates for the number of original posts and outbound replies from accounts in the treatment group (account activity). The estimation is based on \num{144543} count observations across \num{6883} accounts. \num{1317} accounts are omitted because they have no post observed during the observation window. \textbf{(b)}~Weekly and aggregated ATT estimates for the number of inbound reposts and replies in the treatment group (audience engagement). The estimation is based on \num{39512} count observations across \num{2929} accounts. \num{2196} accounts are dropped because of all zero outcomes, and \num{693} accounts are dropped due to only one observation. The error bars represent 95\% CIs. Full estimation results are reported in \Cref{tab:did_main_account,tab:did_main_engage}.}
    \label{fig:main}
\end{figure}

\Cref{fig:main}a shows the weekly and aggregated effects of location transparency on the posting activity of location-mismatched accounts. Before the disclosure, none of the weekly ATT estimations are significantly different from zero, consistent with parallel trends assumption between treatment and control groups before the treatment. However, after the disclosure, the weekly ATT estimations have a clear decreasing trend, showing that the treatment has a suppression effect on the posting activity of location-mismatched accounts (see weekly ATT estimates in \Cref{supp:did_estimations}). On average, the disclosure of account location reduces the posting activity of accounts in the treatment group by 13.1\% (ATT~$=-0.131$, $z=-12.534$, $p<0.001$; 95\%~CI: $[-0.150, -0.112]$), which is consistent across both original posts and outbound replies (see details in \Cref{supp:did_estimations}).

Additionally, we estimate the changes in audiences' average inbound engagement -- reposts and replies -- with the original posts published by the accounts in the treatment and control groups following the location disclosure (\Cref{fig:main}b). We find that the weekly ATT estimates (see detailed weekly ATT estimates in \Cref{supp:did_estimations}) and the aggregated ATT estimate (ATT~$=-0.012$, $z=-0.412$, $p=0.681$; 95\%~CI: $[-0.067, 0.046]$) are consistently and statistically indistinguishable from zero, irrespective of the introduction of location transparency. The ATT estimates remain statistically non-significant when we examine inbound reposts and replies separately (see details in \Cref{supp:did_estimations}). Therefore, we find no evidence that the disclosure of account location has a statistically significant effect on the average audience engagement with the original posts from the location-mismatched accounts in the treatment group. We also examine the total number of inbound reposts and replies received by each account per week, and find that the weekly audience engagement received by each account decreases by 14.2\% (ATT~$=-0.142$, $z=-7.293$, $p<0.001$; 95\%~CI: $[-0.176, -0.106]$; see details in \Cref{supp:did_estimations}), which is consistent with the overall reduction in posting activity of location-mismatched accounts following location transparency. Given that the average engagement per post remains stable over time, our findings suggest that the observed decline in total audience engagement is largely attributed to the concurrent reduction in posting activity, rather than a change in how audiences engage with each individual post.

We perform additional robustness checks to ensure the reliability of the causal estimates of the impact of location transparency on the posting activity of location-mismatched accounts: (i) We estimate the aggregated ATT at the daily level and using bootstrap inference (see \Cref{supp:long_daily}); (ii) We repeat our estimation using synthetic DiD~\cite{arkhangelsky2021synthetic} (see \Cref{supp:sdid}); (iii) We repeat our estimation after conducting propensity score matching based on account profiles and content characteristics (see \Cref{supp:psm}); (iv) We repeat our estimation using alternative pre-treatment periods (see \Cref{supp:alter_pre_treat}); (v) We repeat our estimation based on the data collected via the full-archive count endpoint of \X's API (see \Cref{supp:api_count}). All of these checks are robust and consistently support our main findings. Furthermore, we analyse the status of unavailable accounts during our data collection, and find that the treatment group includes approximately nine times as many suspended accounts as the control group: 125 out of \num{4100} treatment accounts (3.0\%) are suspended, compared with 14 out of \num{4100} control accounts (0.3\%). However, given that \X's API does not return the timestamps of the account sanctions, it is challenging to investigate whether the location transparency feature also increases the probability of being suspended by the platform for location-mismatched accounts. Nevertheless, overall, only 1.7\% of accounts in our dataset are suspended by the platform, which is unlikely to significantly affect our estimates of the effect of location transparency on users' posting activity (see \Cref{supp:account_status}).

\clearpage

\clearpage
\subsection*{Heterogeneity by platform-revealed non-\US locations}

Notably, some location mismatches may reflect genuine relocation rather than deliberate misrepresentation, particularly for accounts that have moved to geographically or culturally proximate regions such as other North American countries or Europe, where the probability of true relocation is higher than for more distant regions such as Asia or Africa. The heterogeneity in the plausibility of genuine relocation across regions provides a useful lens through which to interpret the differential suppressive effects of location transparency.

\begin{figure}[ht]
    \centering
    \captionsetup[subfloat]{font={bf, small}, skip=0pt, singlelinecheck=false, labelformat=simple, position=top}
    \subfloat[]{\includegraphics[width = .48\textwidth]{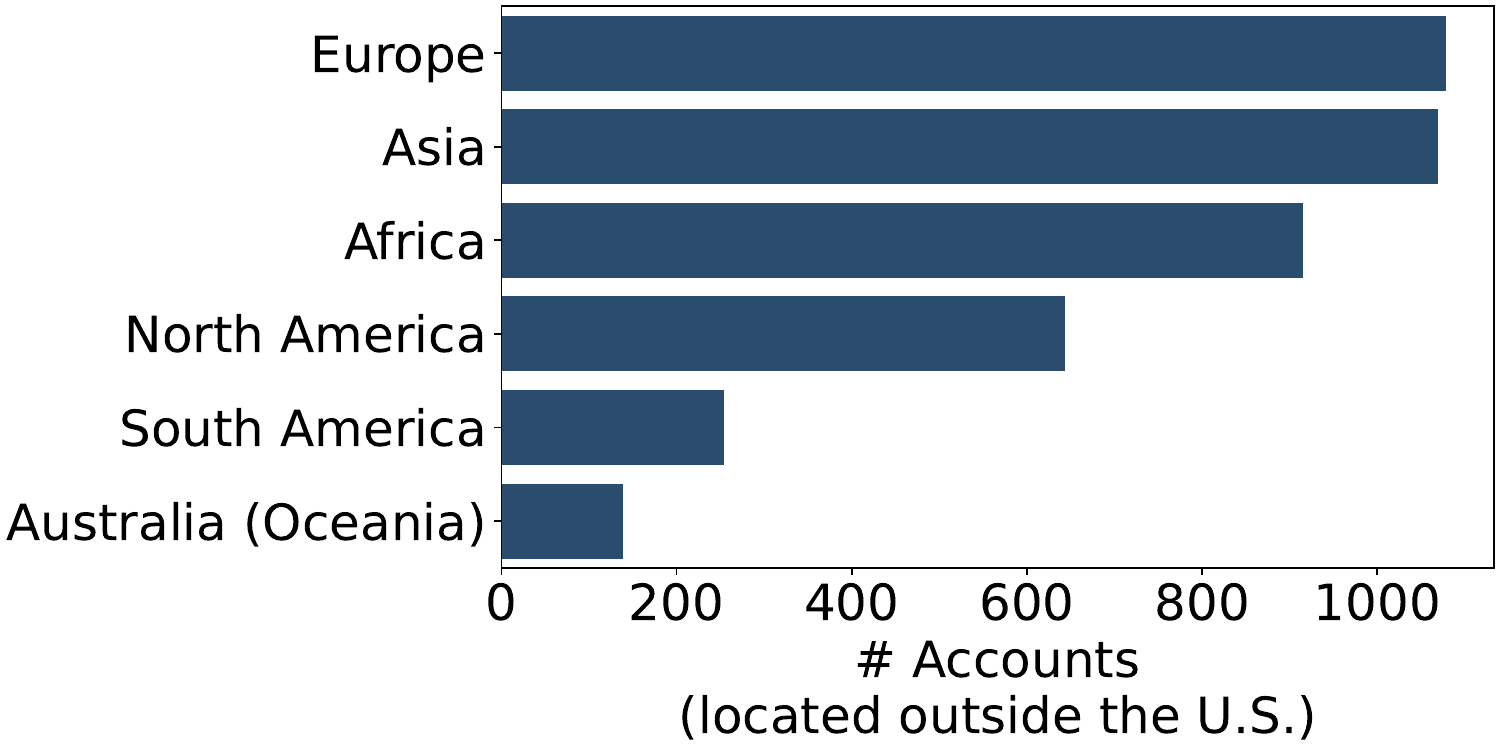}}
    \hfill
    \subfloat[]{\includegraphics[width = .48\textwidth]{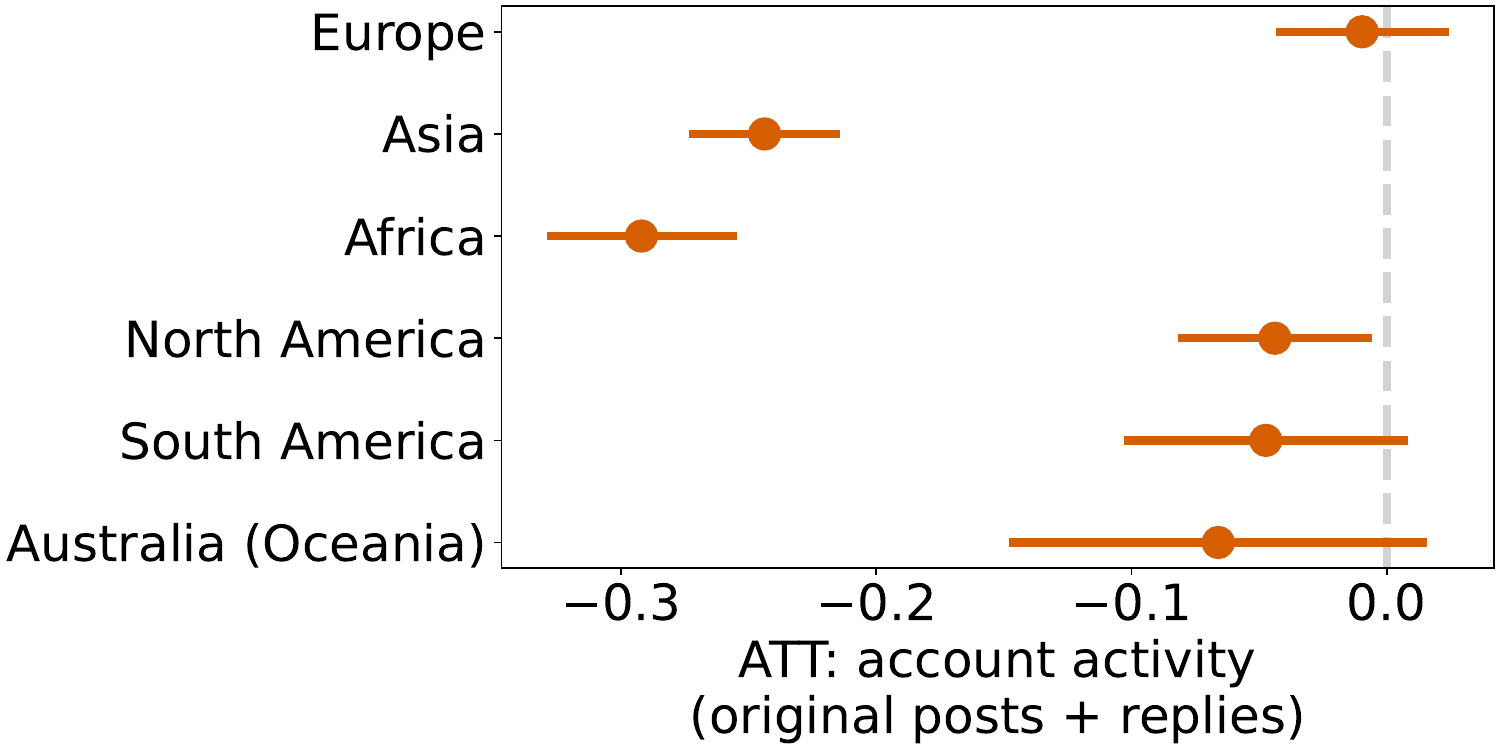}}
    \caption{\textbf{Heterogeneity effects of location transparency on account activity by revealed non-\US locations.} \textbf{(a)}~The distribution of the number of location-mismatched accounts across revealed non-\US locations at the continent level. \textbf{(b)}~Aggregated ATT estimates for the number of original posts and replies from location-mismatched accounts in the treatment group across revealed non-\US locations at the continent level. The error bars represent 95\% CIs.}
    \label{fig:he_location}
\end{figure}

Among the location-mismatched accounts in the treatment group, a total of 156 countries/regions outside of the \US are revealed by the platform, and the five most frequent ones are Nigeria, Canada, the United Kingdom, Africa, and India (see \Cref{fig:he_location_supp}a, \Cref{supp:he_location}). Notably, \X allows users to choose whether the ``About this account'' feature displays their inferred country or a broader geographic region~\cite{x2026change}. To facilitate our analysis, we harmonise the revealed locations into continent-level categories (\Cref{fig:he_location}a) and find that the location-mismatched accounts are mainly from Europe ($n=$ \num{1079}), Asia ($n=$ \num{1069}), and Africa ($n=$ 915), followed by non-\US North America ($n=$ 644), South America ($n=$ 254), and Australia/Oceania ($n=$ 139). 

We further examine the heterogeneity effects of location transparency on account activity across different locations, and the ATT estimates are shown in \Cref{fig:he_location}b. The suppression effect of location transparency is concentrated on location-mismatched accounts from Asia (ATT~$=-0.244$, $z=-14.079$, $p<0.001$; 95\%~CI: $[-0.273, -0.214]$) and Africa (ATT~$=-0.292$, $z=-12.872$, $p<0.001$; 95\%~CI: $[-0.328, -0.254]$). Specifically, the disclosure of account location reduces the posting activity of Asia- and Africa-based users by 24.4\% and 29.2\%, respectively. However, we find no significant or very small effect of location transparency in reducing the posting activity of accounts based on Europe (ATT~$=-0.010$, $z=-0.554$, $p=0.58$; 95\%~CI: $[-0.043, 0.025]$), North America (ATT~$=-0.044$, $z=-2.204$, $p=0.028$; 95\%~CI: $[-0.081, -0.005]$), South America (ATT~$=-0.047$, $z=-1.629$, $p=0.103$; 95\%~CI: $[-0.101, 0.010]$), and Australia/Oceania (ATT~$=-0.066$, $z=-1.527$, $p=0.127$; 95\%~CI: $[-0.144, 0.020]$).

We note that the platform's use of both country- and region-level labels introduces several classification nuances. Because users may choose to display a broader region rather than an inferred country, a \US-based account could, in principle, display ``North America'' and therefore appear location-mismatched under a label-based definition. However, ``North America'' does not appear among the platform-disclosed location labels in our dataset. Although the \US Virgin Islands and Puerto Rico are \US territories, \X reports them as separated location labels distinct from ``United States.'' Consequently, 41 accounts claiming a \US location but disclosed as operating from the \US Virgin Islands ($n=$ 3) or Puerto Rico ($n=$ 38) satisfy our label-based mismatch definition and are assigned to non-\US North America. Excluding these accounts leaves the regional estimate for non-\US North America substantively unchanged (ATT~$=-0.061$, $z=-2.975$, $p=0.003$; 95\%~CI: $[-0.099, -0.021]$). Additionally, under our regional harmonisation scheme, a single account labelled ``Russian Federation'' is assigned to Europe. Excluding this transcontinental case likewise leaves the estimated effect for Europe substantively unchanged (ATT~$=-0.009$, $z=-0.511$, $p=0.610$; 95\%~CI: $[-0.042, 0.026]$). Finally, VPN use may reduce the precision of platform-disclosed locations. We therefore re-estimate the region-specific effects after excluding accounts for which \X indicates that the disclosed location may be affected by VPN use. The estimated effects remain substantively similar across regions, except for non-\US North America, where the previously small negative estimate is no longer statistically distinguishable from zero (see \Cref{fig:he_location_supp}b, \Cref{supp:he_location}).

Taken together, our findings suggest that the suppressive effect of location transparency is strongest among accounts from regions where genuine relocation is less plausible, supporting the interpretation that the observed behavioural changes are driven primarily by non-\US accounts deliberately misrepresenting their locations.

\clearpage

\clearpage
\subsection*{Heterogeneity by account features, posting behaviours, and interaction targets}

In addition to variation across platform-revealed locations, we examine whether the impact of location transparency differs according to (i) the characteristics and pre-intervention posting behaviours of the selected accounts in the treatment and control groups and (ii) the external (recipient) accounts with which they interact. The first analysis identifies which types of accounts in the treatment group respond most strongly to location disclosure, while the second examines whether transparency changes whom location-mismatched accounts reply to, compared to location-matched accounts.

\begin{figure}[ht]
    \centering
    \includegraphics[width=\linewidth]{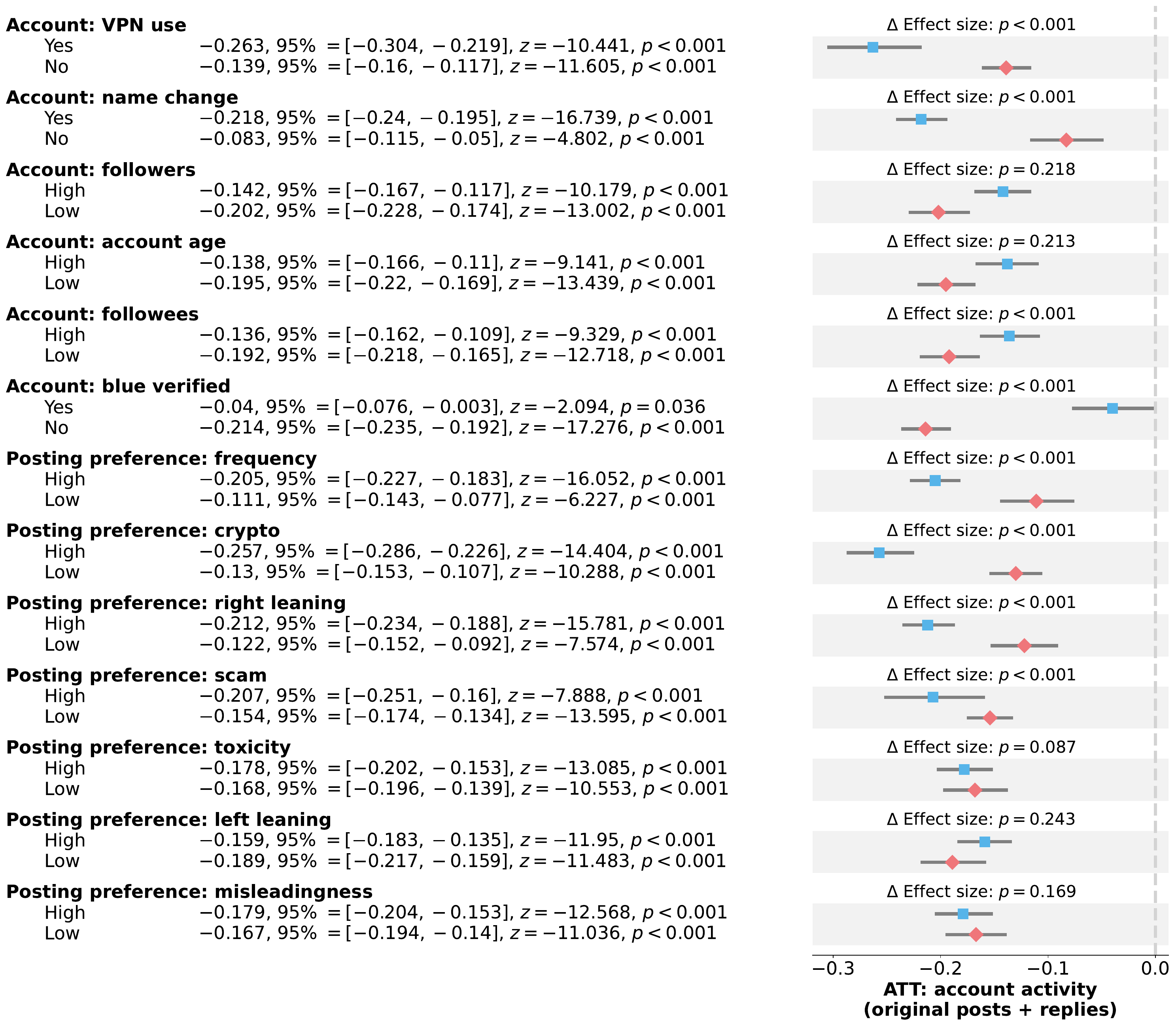}
    \caption{\textbf{Heterogeneity effects of location transparency on account activity by account subgroups.} Shown are aggregated ATT estimates across subgroups separated by account characteristics and posting preferences (see group sizes and estimation results for interaction analysis in \Cref{supp:he_effects}). The error bars represent 95\% CIs.}
    \label{fig:he_sensitivity_account}
\end{figure}

\vspace{1em}
\noindent \textbf{Account characteristics and posting behaviours.}\\
We start by investigating the impact of location transparency across subgroups separated by account characteristics (including followers, followees, Blue Verified status, account age, VPN use, and username change) and posting preferences (including post frequency, political leaning, toxicity, scams, cryptocurrency, and misleadingness). Posting preferences are measured using original posts published before the intervention, with related content-level variables averaged across these posts (for details on variable measurements, see \nameref{sec:methods}). We then divide continuous and count variables into High and Low groups based on their medians, while groups for account-level binary variables are defined by their indicator values. \Cref{fig:he_sensitivity_account} presents the subgroup-specific average treatment effects alongside interaction-based moderation tests. The subgroup-specific effects are estimated separately within each subgroup to quantify the magnitude of the treatment effect, whereas the interaction analyses use the full analytical sample to test whether the effect varies with each moderator (see the full results of interaction analysis in \Cref{supp:he_account}).

The subgroup analyses show that location transparency significantly reduces activity across all examined account types. The reduction is particularly pronounced among accounts flagged as potentially using a VPN: activity decreases by 26.3\% among flagged accounts, compared with 13.9\% among unflagged accounts, with the interaction analysis indicating significant moderation ($p<0.001$). Accounts that changed their usernames reduce their activity by 21.8\%, compared with 8.3\% among accounts without a username change ($p<0.001$ for the interaction). Similarly, activity decreases by 21.4\% among accounts without Blue Verified status but by only 4\% among Blue Verified accounts ($p<0.001$ for the interaction). Accounts with fewer followees also exhibit a larger subgroup-specific reduction than those with more followees (19.2\% vs. 13.6\%; $p<0.001$ for the interaction). By contrast, although the subgroup-specific reductions are larger among accounts with fewer followers (20.2\% vs. 14.2\%) and younger accounts (19.5\% vs. 13.8\%), the corresponding interaction tests are not statistically significant ($p=0.218$ and $p=0.213$, respectively).

Location transparency also produces greater reductions among accounts with higher pre-intervention posting frequency (20.5\% vs. 11.1\%), cryptocurrency-related content (25.7\% vs. 13\%), right-leaning content (21.2\% vs. 12.2\%), and scam-related content (20.7\% vs. 15.4\%); the interaction-based moderation tests are statistically significant for each of these dimensions (all $p<0.001$). In contrast, we find no statistically significant moderation by toxicity (17.8\% vs. 16.8\%; $p=0.087$), left-leaning content (15.9\% vs. 18.9\%; $p=0.243$), or misleadingness (17.9\% vs. 16.7\%; $p=0.169$). Taken together, these findings indicate that the suppressive effect of location transparency is broadly observed across account types and posting preferences but is especially pronounced among accounts exhibiting characteristics potentially associated with identity concealment or strategic platform use.

\clearpage

\clearpage
\noindent \textbf{Reply targets and interaction patterns.}\\
Beyond changing their overall level of activity, location-mismatched accounts may also change whom they interact with following disclosure. We therefore examine whether the effect of location transparency on outbound replies varies according to the disclosed location of the recipient account (\ie, \US vs. non-\US). Importantly, replies are particularly relevant for examining toxicity because, unlike original posts, they constitute directed interpersonal interactions in which hostile, insulting, or abusive language may emerge. Given this, we also assess whether location transparency changes the toxicity of outbound reply interactions. Specifically, we analyse replies from accounts in the treatment and control group over a period spanning four weeks before and four weeks after the introduction of location transparency, resulting in \num{815917} outbound replies to \num{152035} recipient accounts. We then extract the accounts to which these outbound replies are directed and identify whether their self-claimed locations match the platform-revealed locations. Using the same approach of LLM-based country extraction for account selection (see details in \nameref{sec:methods}), we successfully identify \num{45167} \US-based (location-matched) accounts and \num{2745} non-\US-based (location-mismatched) accounts among those receiving outbound replies from the treatment and control groups. 

\begin{figure}[ht]
    \centering
    \captionsetup[subfloat]{font={bf, small}, skip=0pt, singlelinecheck=false, labelformat=simple, position=top}
    \subfloat[]{\includegraphics[width = .32\textwidth]{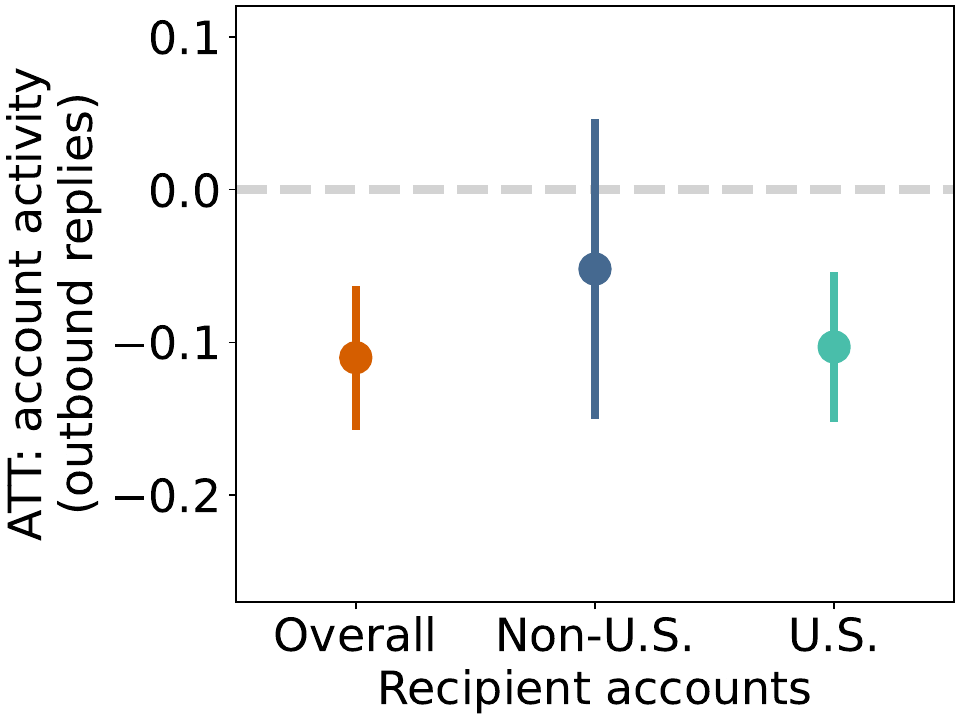}}
    \hfill
    \subfloat[]{\includegraphics[width = .32\textwidth]{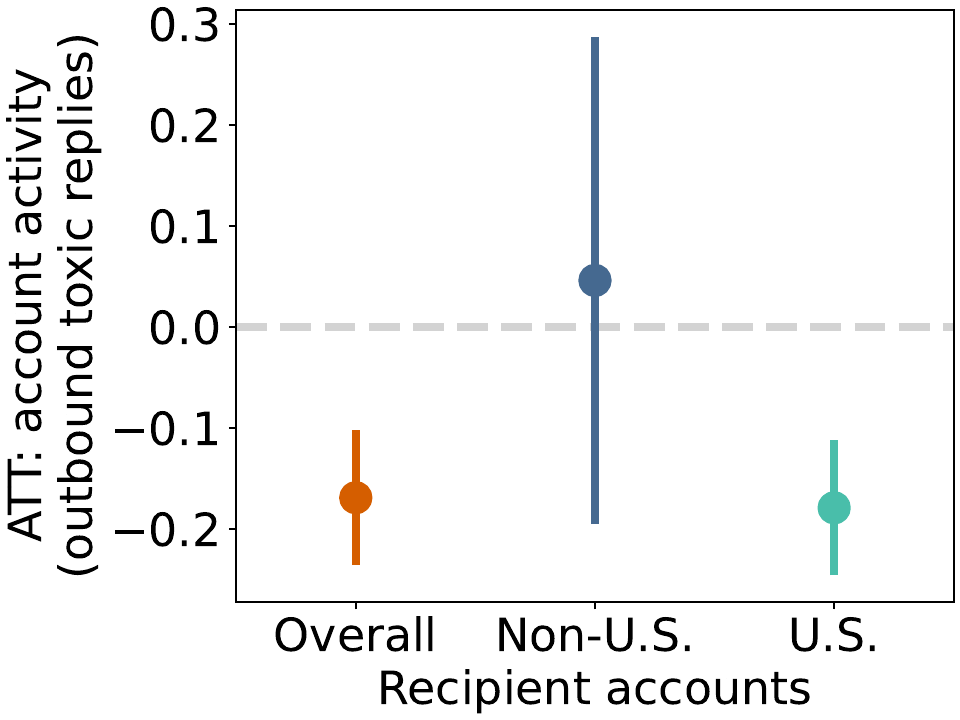}}
    \hfill
    \subfloat[]{\includegraphics[width = .32\textwidth]{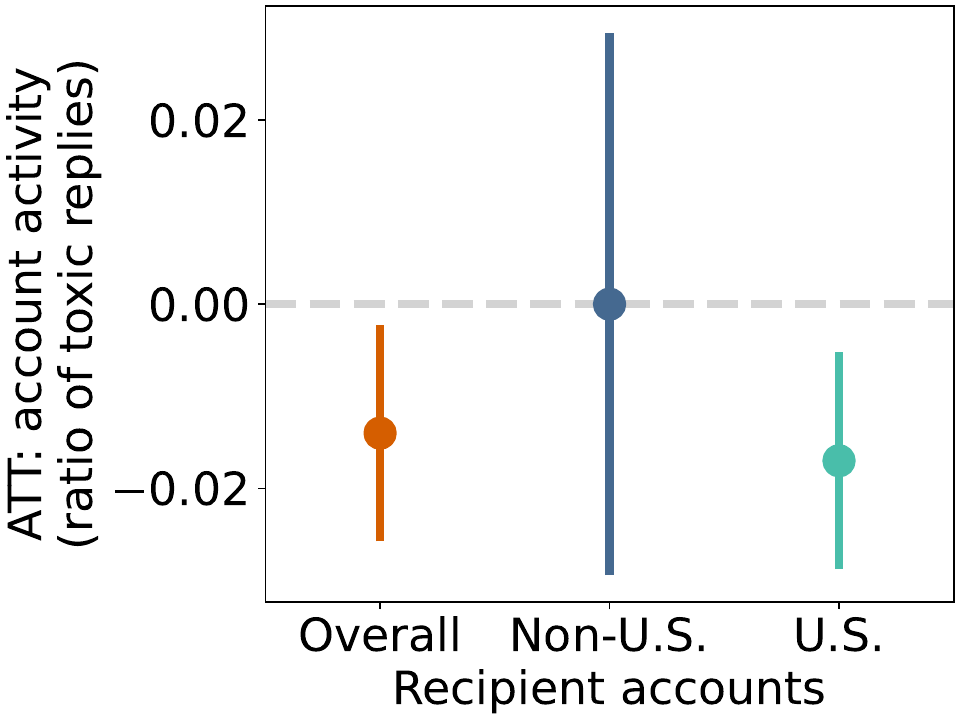}}
    \\
    \subfloat[]{\includegraphics[width = .48\textwidth]{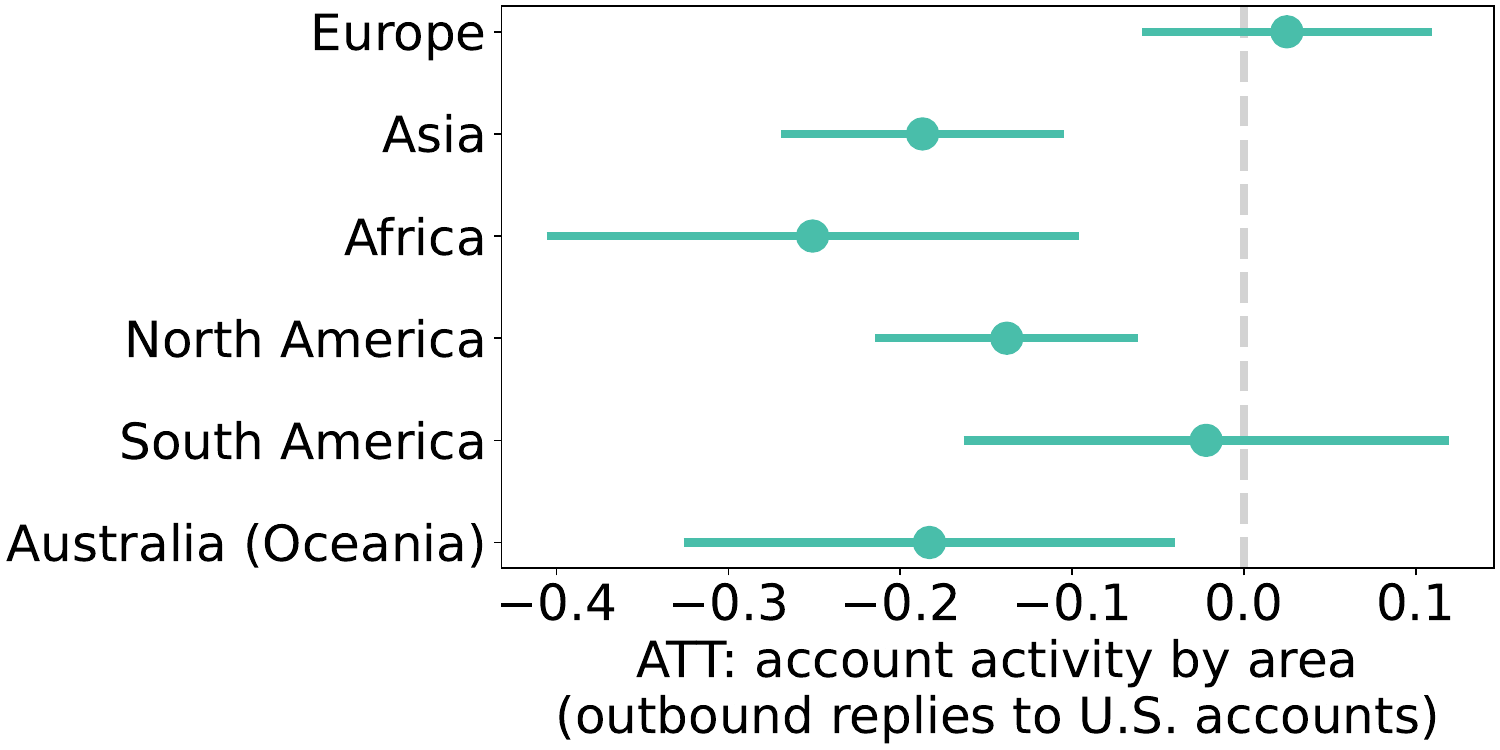}}
    \hfill
    \subfloat[]{\includegraphics[width = .48\textwidth]{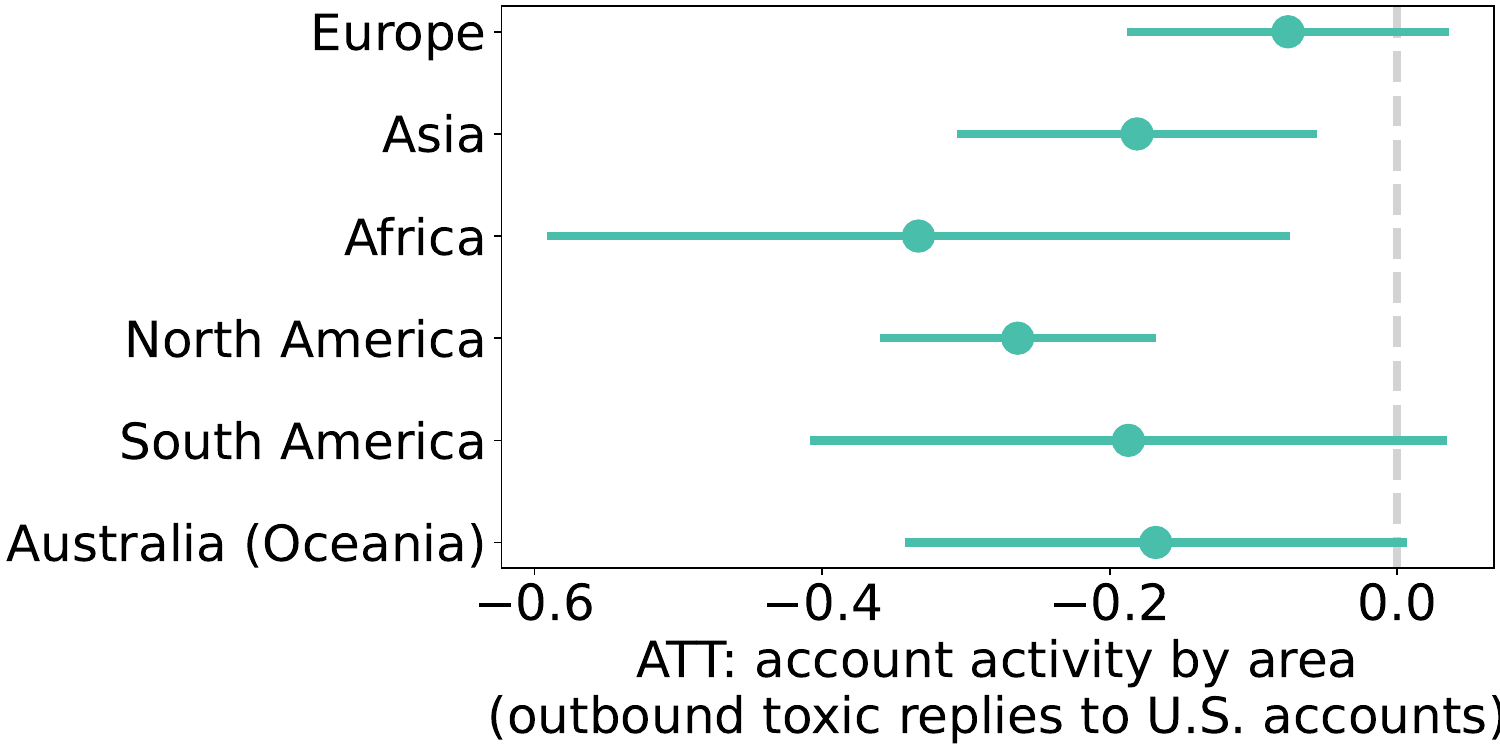}}
    \caption{\textbf{Location transparency reduces (toxic) replies from location-mismatched foreign accounts to \US domestic accounts.} \textbf{(a)}~Aggregated ATT estimates on outbound replies to different recipient accounts. \textbf{(b)}~Aggregated ATT estimates on outbound toxic replies to different recipient accounts. \textbf{(c)}~Aggregated ATT estimates on the ratio of toxicity in replies to different recipient accounts. \textbf{(d)}~Aggregated ATT estimates on outbound replies from non-\US accounts across continental areas to \US-based domestic accounts. \textbf{(e)}~Aggregated ATT estimates on outbound toxic replies from non-\US accounts across continental areas to \US-based domestic accounts. The error bars represent 95\% CIs, and the full estimation results are reported in \Cref{supp:he_replies}.}
    \label{fig:reply_interaction}
\end{figure}

Across all recipients with identifiable locations, location transparency reduces outbound replies from treated accounts by 11.0\% (ATT~$=-0.110$, $z=-4.293$, $p<0.001$; 95\%~CI: $[-0.155, -0.061]$; \Cref{fig:reply_interaction}a). Notably, this reduction in reply interactions is primarily driven by replies directed at \US-based accounts (ATT~$=-0.103$, $z=-3.933$, $p<0.001$; 95\%~CI: $[-0.15, -0.053]$), whereas the effect on replies to non-\US accounts is not statistically significant (ATT~$=-0.052$, $z=-1.016$, $p=0.310$; 95\%~CI: $[-0.145, 0.051]$). The reduction is more pronounced when we restrict the analysis to replies with toxicity scores above 0.5 (\Cref{fig:reply_interaction}b). The number of toxic replies declines by 16.9\% overall (ATT~$=-0.169$, $z=-4.598$, $p<0.001$; 95\%~CI: $[-0.233, -0.101]$) and 17.9\% (ATT~$=-0.179$, $z=-4.783$, $p<0.001$; 95\%~CI: $[-0.243, -0.110]$) for those directed at \US-based recipients. By contrast, we detect no corresponding change in toxic replies to non-\US recipients (ATT~$=0.046$, $z=0.385$, $p=0.700$; 95\%~CI: $[-0.168, 0.316]$). Furthermore, the ratio of toxic replies relative to total outbound replies also declines by 1.4 percentage points (ATT~$=-0.014$, $t=-2.524$, $p=0.012$; 95\%~CI: $[-0.026, -0.003]$; \Cref{fig:reply_interaction}c). For replies to \US-based recipients, this proportion declined by 1.7 percentage points (ATT~$=-0.017$, $t=-2.834$, $p=0.005$; 95\%~CI: $[-0.028, -0.005]$), equivalent to an 11.4\% reduction relative to its pre-intervention mean. By contrast, we detected no significant change in the ratio of toxic replies directed at non-\US-based recipients (ATT~$=0.000$, $t=-0.015$, $p=0.988$; 95\%~CI: $[-0.031, 0.030]$). Thus, the decline in toxic replies to \US-based accounts reflects both lower reply volume and a reduction in the proportion of replies containing toxic language (for a comparison between outbound replies and inbound replies, see \Cref{supp:he_replies}).

Stratifying treated accounts by their platform-disclosed regions (\Cref{fig:reply_interaction}d) further shows that reductions in replies to \US-based recipients are statistically significant among accounts located in Africa (ATT~$=-0.251$, $z=-2.737$, $p=0.006$; 95\%~CI: $[-0.39, -0.079]$), Asia (ATT~$=-0.187$, $z=-4.038$, $p<0.001$; 95\%~CI: $[-0.265, -0.101]$), non-\US North America (ATT~$=-0.138$, $z=-3.284$, $p=0.001$; 95\%~CI: $[-0.211, -0.058]$) and Australia/Oceania (ATT~$=-0.183$, $z=-2.245$, $p=0.025$; 95\%~CI: $[-0.315, -0.025]$). \Cref{fig:reply_interaction}e shows that reductions in toxic replies are clearest among accounts disclosed as being in Africa (ATT~$=-0.333$, $z=-2.048$, $p=0.041$; 95\%~CI: $[-0.548, -0.017]$), non-\US North America (ATT~$=-0.264$, $z=-4.617$, $p<0.001$; 95\%~CI: $[-0.354, -0.162]$), and Asia (ATT~$=-0.181$, $z=-2.546$, $p=0.011$; 95\%~CI: $[-0.298, -0.045]$). Together, these results show that location transparency is followed by fewer interactions -- and particularly fewer toxic interactions -- from location-mismatched accounts towards the \US-based recipients whose location they have claimed to share.

\clearpage

\clearpage
\section*{Discussion}

Inauthentic geographic identification is often invoked in debates about foreign influence, online fraud and coordinated manipulation, yet little is known about which accounts have inauthentic locations and how those accounts respond when discrepancies between their claimed and platform-inferred locations become publicly visible. Leveraging the location disclosure feature on \X, we perform a large-scale quasi-experimental study and find that location transparency reduces the posting activity of location-mismatched accounts by 13.1\%. By contrast, we detected no corresponding change in the average audience engagement received per original post. Disclosure is also followed by fewer and less toxic replies from location-mismatched accounts to \US-based recipients. Taken together, our findings indicate that the location transparency feature operates primarily by changing the behaviour of the disclosed accounts themselves rather than by shifting audience willingness to engage with their content. In the two-channel framework we outline at the outset -- increased accountability prompting behavioural restraint versus audience-driven belief updating -- our evidence favours the former. This is a notable departure from work on content-level interventions, where effects typically run through downstream audience sharing~\cite{pennycook2019fighting,bak2022combining,slaughter2025community,chuai2026community}.

Our subgroup and supplementary analyses show that the effect of location transparency on account activity is pronounced across the board and is consistent with three distinct pathways through which the broad ``accountability'' mechanism may have reduced activity. First, disclosure may impose social or reputational costs by making operators anticipate embarrassment, challenges from other users, or reputational damage associated with maintaining a visibly discrepant identity. Our findings show that replies explicitly calling attention to location discrepancies increased after disclosure, indicating that at least some accounts encountered direct social scrutiny; nevertheless, these replies remained rare, suggesting that visible social sanction was not widespread. 
Second, disclosure may encourage location-mismatched accounts to reduce their activity to avoid platform scrutiny, if they believe that publicly visible inconsistencies make their accounts more likely to be investigated, restricted, or suspended. Prior research shows that moderation sanctions and explanations can alter operators' subsequent participation and compliance~\cite{jhaver2019does,srinivasan2019content,horta2023automated,chuai2026corrections}, and that warnings about potential account suspension can reduce hate speech on Twitter (now \X)~\cite{yildirim2023short}. Consistent with this possibility, we find a higher suspension rate among location-mismatched accounts than among location-matched accounts when their status was subsequently queried. However, the absence of suspension timestamps prevents us from determining whether location disclosure increases suspension risk. Third, disclosure may reduce the economic or operational value of an account. If appearing locally embedded helps operators attract attention, establish trust, or convert audiences into customers or scam victims, exposing a mismatch may lower the expected return from continued posting~\cite{appel2026deceptive,cuevas2026chameleon}. The larger reductions among highly active accounts and those with stronger prior preferences for scam- and cryptocurrency-related content lend comparatively greater support to this economic explanation than to a uniform psychological response. Stronger responses among accounts flagged for potential VPN use, accounts with previous username changes, and accounts without Blue Verified status are also consistent with disclosure imposing greater operational costs on accounts whose activities depend on maintaining a credible identity.

The regional heterogeneity we observe strengthens the case that this suppression reflects deliberate misrepresentation rather than being an artefact of the disclosure itself. The effect is concentrated among accounts revealed to be operating from Africa (-29.2\%) and Asia (-24.4\%), regions for which claiming a \US location while operating from afar is least plausibly explained by ordinary relocation, compared to Europe, South America, and Australia/Oceania. If the suppression effect is mechanically produced by the mere act of disclosure -- for instance, through discomfort at any location mismatch being surfaced, regardless of its cause -- we would not expect this systematic gradient. Instead, the pattern is consistent with an account-level accountability mechanism that is selectively activated for accounts whose location claims were least defensible~\cite{loewenstein2014disclosure}. A further contribution of this study is evidence on whom location-mismatched accounts engage with less once disclosed. The reduction in outbound replies is concentrated on replies directed at \US-based recipients and is accompanied by an even larger decline in toxic replies to those recipients. Replies to non-\US recipients show no corresponding change on either dimension. This asymmetry suggests that transparency particularly curtails the interpersonal, and disproportionately hostile, interactions these accounts are directing at the very audience whose location they have falsely claimed to share. This is a behaviourally specific form of restraint: accounts appear less willing to continue directing themselves at \US domestic political and economic audiences once their claim to a shared geographic identity is publicly contradicted.

The temporal development of the suppression effect on account activity also weighs against interpreting the reduction as a short-lived response. The weekly treatment estimates became progressively more negative during the first four weeks following the intervention and then reached a relatively stable level. This pattern is consistent with gradual awareness of the feature, delayed behavioural adaptation, or cumulative attrition as accounts reduce or cease their activity during the initial stage. Notably, the account-level nature of these estimates places an important boundary on their interpretation. A 13.1\% decline among the observed location-mismatched accounts does not imply a corresponding decline in potentially inauthentic activity across \X as a whole. Related research on community bans shows that moderation can substantially reduce harmful participation on the focal platform~\cite{chandrasekharan2017you}. Cross-platform research nevertheless demonstrates that some affected users and communities migrate, rebuild social networks, or remain highly active on less regulated platforms~\cite{horta2021platform,monti2023online,mekacher2023systemic}. Although location disclosure is less restrictive than deplatforming, these findings illustrate why effects on identified accounts should be distinguished from effects on the broader information ecosystem.

Given these effects on the disclosed accounts' behaviour, why do we not find any corresponding effect on users' engagement with these accounts? The explanation for this null audience effect may lie in limited exposure to the disclosure because of where it is surfaced. \X displays location information on an account's profile rather than directly alongside each post. Users often encounter, reply to, and repost content without ever viewing the author's profile -- and thus never seeing the disclosure. In addition to this limited exposure to profile-level information, established follower relationships and pre-existing content preferences may help explain why disclosure produces no detectable change in average engagement. This interpretation remains provisional, as we cannot observe which users inspected the transparency information or how it changed their beliefs. Nevertheless, it underscores the principle of targeted transparency: information is unlikely to alter recipients' decisions when it is not incorporated into their normal decision environment~\cite{weil2013targeting,loewenstein2014disclosure}.

Our results speak to a broader theoretical question about how location-based inauthenticity online should be conceptualised. Much public and scholarly attention to foreign accounts posing as domestic \US users has centred on state-sponsored or politically motivated influence operations~\cite{ferrara2022twitter,eady2023exposure,appel2026deceptive,zannettou2019disinformation,bail2020assessing,alizadeh2020content}. Our findings suggest the coexistence of inauthentic behaviour targeting foreign political influence and commercially oriented activities~\cite{appel2026deceptive}. Location mismatch is not associated with a particular political lean at baseline (\Cref{fig:account_linear}), but political leaning does moderate the strength of the transparency suppression effect: accounts with a high pre-intervention share of right-leaning content show a substantially larger reduction in activity than those with a low share (21.2\% vs. 12.2\%), while left-leaning content shows no such difference (15.9\% vs. 18.9\%; \Cref{fig:he_sensitivity_account}). At the same time, we find robust evidence that location mismatch is associated with, and its behavioural response to disclosure is moderated by, commercially oriented content such as scams and cryptocurrency promotion. Together, these patterns point to a population in which politically motivated and commercially motivated misrepresentation coexist, which may blend partisan content with scam or cryptocurrency promotion to build an audience. This argues for treating location transparency not as a tool narrowly tailored to counter-influence-operations efforts, but as a general-purpose \textit{trust-and-safety intervention} relevant to both election-integrity concerns and platform-wide efforts against fraud, spam and scam networks.

Our findings also carry practical implications and position account-level transparency as a potentially useful, comparatively light-touch complement to content moderation and platform enforcement. Location disclosure can induce behavioural restraint without directly removing accounts or content. However, location mismatch should be treated as a contextual risk signal rather than an enforcement criterion, because travel, relocation, VPN use, and inference errors can produce benign discrepancies. The absence of broad audience disengagement further suggests that profile-level information may be insufficiently salient to shape routine engagement; platforms could therefore test more visible but proportionate designs while monitoring for call-outs, regional stereotyping, and harassment~\cite{weil2013targeting,loewenstein2014disclosure,yang2025user}. More fundamentally, location transparency cannot substitute for fact-checking, anti-fraud enforcement, or network-level investigation: although it reduces activity and some toxic interactions, addressing misleading content and coordinated manipulation requires complementary interventions~\cite{pennycook2019fighting,bak2022combining,slaughter2025community,chuai2026community}. Platforms should also communicate the uncertainty and granularity of inferred locations, distinguish countries from broader regions, and provide mechanisms for users to contest inaccurate disclosures, particularly given the risks of privacy loss and regional stigmatization~\cite{guo2025civilizing,yang2025user}.

Several limitations qualify these conclusions and motivate future research. First, although our design leverages a plausibly exogenous platform-wide roll-out and passes parallel-trends assumption and multiple robustness checks, it remains an observational quasi-experiment; we cannot rule out that unobserved, time-varying factors coinciding with the roll-out affected treatment and control accounts differentially, even though such confounds would need to track our specific regional and behavioural heterogeneity patterns to explain our results. Experimental variation in the timing, placement or salience of location disclosure could provide stronger causal evidence and directly compare author- and audience-side mechanisms. Second, ``location mismatch'' as we define it is not proof of deliberate deception: it is possible for self-reported and platform-inferred locations to diverge for benign reasons, including genuine relocation, shared or corporate accounts, or privacy-motivated VPN use unrelated to identity misrepresentation. We address this by examining regional heterogeneity under the assumption that genuine relocation is more plausible for culturally and geographically proximate destinations, but this remains an indirect test rather than ground truth on intent. Future research could combine behavioural data with validated location histories and measures of perceived accountability. Third, platform enforcement may have contributed to the decline in activity: suspended accounts are more common in the treatment group, but the absence of suspension timestamps prevents us from investigating whether the location transparency feature also increases suspension risk, which finer-grained enforcement records would help distinguish. Fourth, our balanced sample of politically relevant, \US-claiming accounts facilitates group comparison but does not allow us to estimate the population prevalence of location mismatch, and may not generalise to other users or platforms. Larger samples and cross-platform comparisons are beneficial to assess the generalizability of these effects.

In conclusion, our findings show that a comparatively unobtrusive transparency intervention induces substantial behavioural restraint among accounts whose self-presented and platform-inferred locations diverge. This response is concentrated among accounts bearing independent markers of platform-policy-relevant misconduct and directed disproportionately at reducing hostile engagement with the very audience whose geographical identity they have claimed to share. While location-based inauthenticity is often framed in the public conversation as a problem of foreign political influence, our evidence suggests it is at least as strongly entangled with commercially motivated deception. Platform governance and future research on inauthentic behaviour may therefore benefit from treating geographic identity transparency as a general trust-and-safety tool against cross-border platform abuse, rather than one narrowly tailored to counter-influence-operations efforts alone.

\clearpage
\section*{Methods}
\label{sec:methods}

\subsection*{Data collection}

\textbf{Account selection.}\\
To examine the impact of location transparency on accounts misrepresenting their locations on \X, we extract an initial list of 
\num{2903966} users and their profile biographies from three independent datasets constructed prior to November 2025: (i) a large-scale public dataset including 22 million \X posts related to 2024 \US Presidential Election~\cite{balasubramanian2024public}; (ii) another large-scale dataset including over 1.6 million \X posts for which LLMs (\eg, Grok) have been asked to fact-check~\cite{renault2025grok}; and (iii) the third dataset collected using keywords related to climate/vaccine misinformation and conspiracy theories. Relying on these pre-existing datasets allows us to observe accounts' self-declared locations before the introduction of the ``About this account'' transparency feature on \X and to avoid attenuation bias that could arise if users strategically change their declared location after treatment. This step is necessary because only the current declared location is observable on the platform, not the history of location changes. Subsequently, we use two consecutive LLM classifiers based on the GPT-5~series models to identify accounts' claimed location and whether they share content related to \US politics:
\begin{itemize}[leftmargin=*,noitemsep,topsep=0pt]
    \item \emph{Identification of self-reported country information.} We use the first LLM classifier to infer each account's claimed country of operation from the self-reported location field in the profile biography (see details in \Cref{supp:country_extraction}).
    \item \emph{Identification of accounts related to \US politics.} For accounts whose self-reported location indicates the \US, we use the second LLM classifier to further identify accounts that clearly signalled engagement with U.S. politics in their profile biography (see details in \Cref{supp:country_extraction}).
\end{itemize}
As a result, this procedure yields \num{261250} potential candidate accounts. For these politically relevant \US-claiming accounts, we then collect profile-transparency information from their profiles on \X (\ie, ``About this account''), including the platform's algorithmically inferred country/region of operation, the number of prior username changes, and whether the platform indicated that the inferred location may be inaccurate due to VPN use. Finally, we classify accounts whose self-reported and platform-disclosed locations differ as the treatment group, resulting in \num{4100} eligible location-mismatched accounts. From the remaining eligible accounts whose two location signals match, we randomly select \num{4100} accounts for the control group. This procedure produces a balanced analytical sample of \num{8200} accounts, for which we subsequently collect original posts and replies.

\vspace{1em}
\noindent \textbf{Post collection.}\\
Given that the roll-out date of the location transparency feature is November 22, 2025~\cite{bier2025exp,bier2025roll}, we collect all \textit{original posts} (\ie, those posts starting the conversation thread) and \textit{reply posts} from these accounts (\ie, outbound replies to other accounts) over a longitudinal period spanning seven weeks preceding and fourteen weeks following the date of implementation. The collection results in \num{1327286} original posts and \num{3631457} reply posts (see an overview of the dataset in \Cref{tab:data_overview}, \Cref{supp:data_overview}).

\subsection*{Account characteristics and content features}
Based on the collected accounts and their posts, we extract a wide range of factors across account-level characteristics and post-level content features for our analysis.
\begin{itemize}[leftmargin=*,noitemsep,topsep=0pt]
    \item \textit{Followers.} The built-in account characteristic indicates the number of followers.
    \item \textit{Followees.} The built-in account characteristic indicates the number of followees.
    \item \textit{Blue verified.} The built-in account characteristic indicates whether the account is blue verified ($=$1, \ie, active subscription to \X Premium) or not ($=$0).
    \item \textit{Account ages.} The age (in weeks) from the creation date of the account to the implementation date of the location transparency feature.
    \item \textit{VPN.} The account-level variable indicates whether the account uses a VPN ($=$1) or not ($=$0). This is included in the ``About this account'' feature.
    \item \textit{Name change.} The account-level variable indicates whether the account has changed its username by the date of data collection ($=$1) or not ($=$0). It is included in the ``About this account'' feature.
    \item \textit{Toxicity.} We use a state-of-the-art Detoxify model to predict toxicity scores for the collected posts. The model has a high AUC score of 0.921~\cite{hanu2020detoxify}.
    \item \textit{Politics.} We use a topic model fine-tuned based on the pre-trained TwHIN-BERT (large) model to classify whether the original posts are related to politics, with a high accuracy of $0.844$ and an $F_{1}$ score of $0.816$~\cite{chuai2026community}.
    \item \textit{Political leaning.} For the politics-related posts, we use a DeBERTaV3-based classifier to identify their political leanings (see details in \Cref{supp:political_align})~\cite{volf2025political,lee2026llms}.
    \item \textit{Scam.} We use GPT-5.4 mini to identify scam content in the original posts and validate its reliability based on its full-sized model and a broader  DistilBERT-based spam detection model (see details in \Cref{supp:content_features}).
    \item \textit{Crypto.} We utilise a list of cryptocurrency-related keywords to identify crypto-related content in the original posts (see details in \Cref{supp:content_features}).
    \item \textit{Misleadingness.} Previous work shows that LLMs can assess content misleadingness at scale, achieving performance comparable to expert-verified fact-checks (accuracy $=0.801$; $F_1=0.857$)~\cite{chuai2026request,chuai2026corrections}. Following the same framework, we use GPT-5.4 mini to assign each original post a continuous misleadingness score ranging from 0 (not misleading) to 1 (extremely misleading).
\end{itemize}

\subsection*{Linear regression specification}
To examine differences between the treatment and control groups, we estimate a series of univariate linear regression models, each using one account characteristic or posting preference as the dependent variable and the treatment indicator as the independent variable. Specifically, for each characteristic ($Y_{i,j}$), we use the following specification:
\begin{equation}
Y_{i,j} = \beta_0 + \beta_1 \text{Treated}_i + \varepsilon_i,
\end{equation}
where $Y_{i,j}$ denotes the characteristic $j$ (\eg, followers, followees, verified status, account age, VPN usage, username changes, political leaning, toxicity, political content, scam-related content, cryptocurrency-related content, or content misleadingness) for account $i$, and $\text{Treated}_{i}$ is a dummy variable indicating whether account $i$ belongs to the treatment group ($=1$) or control group ($=0$). The intercept ($\beta_0$) represents the mean value of the characteristic in the control group, while the coefficient ($\beta_1$) measures the difference in the mean value of the characteristic between the treatment and control groups. $\varepsilon_i$ is the error term. Because the variance of each characteristic may differ between the treatment and control groups, all models are estimated using OLS with HC3 heteroskedasticity-robust standard errors. All continuous variables are standardised with log1p and then $z$-score for better interpretability.

\subsection*{Difference-in-differences specification}
\label{sec:did_spec}
We adopt a DiD design to examine the impact of location transparency and specify both weekly leads-and-lags and aggregated pre-post models. The leads-and-lags model estimates the temporal weekly dynamics of account activity and engagement they receive before and after the location transparency, allowing us to assess the parallel trends assumption and track the temporal evolution of treatment effect after the feature implementation. The aggregated pre-post model captures the overall effect of location transparency during the post-treatment period. 

The leads-and-lags DiD model is specified using a negative binomial regression to account for the count nature of the dependent variable:
\begin{equation}
\begin{aligned}
    \func{log}(E[Y_{i,t}|\bm{x}_{i,t}]) = & \, \beta_{0} + \beta_{1}\text{Treated}_{i} + \textbf{b}_{1}^{\intercal}\textbf{Before}_{t} + \textbf{b}_{2}^{\intercal}\textbf{After}_{t} + \textbf{b}_{3}^{\intercal}(\text{Treated}_{i} \times \textbf{Before}_{t}) \\
    &+ \textbf{b}_{4}^{\intercal}(\text{Treated}_{i} \times \textbf{After}_{t}) + \alpha_{i},
\end{aligned}
\end{equation}
where $\text{Treated}_{i}$ is the same as in the linear regression above. The vectors $\textbf{Before}_{t}$ and $\textbf{After}_{t}$ denote week dummies relative to the implementation of location transparency. Specifically, $\textbf{Before}_{t}$ covers Weeks $-7$ to $-2$, with Week $-1$ serving as the reference period, while $\textbf{After}_{t}$ spans Weeks 1 to 14. Their corresponding coefficient estimates, $\textbf{b}_{1}$ and $\textbf{b}_{2}$, represent weekly fixed effects. The DiD interaction coefficient estimates ($\textbf{b}_{3}$) capture pre-treatment (lead) effects, used to assess the parallel trends assumption, whereas estimates in $\textbf{b}_{4}$ capture post-treatment (lag) effects, measuring the weekly impact of location transparency on the dependent variable. Finally, $\alpha_{i}$ denotes account-specific fixed effects. In addition to the leads-and-lags DiD model, we estimate the overall effect of location transparency over the whole period after its implementation and specify the following pre-post DiD model with two periods, \ie, pre- and post-treatment:
\begin{equation}
\begin{aligned}
    \func{log}(E[Y_{i,t}|\bm{x}_{i,t}]) = \, \beta_{0} + \beta_{1}\text{Treated}_{i} + \beta_{2}\text{After}_{t} + \beta_{3}\text{Treated}_{i} \times \text{After}_{t} + \alpha_{i} + \gamma_{t},
\end{aligned}
\end{equation}
where $\text{After}_{t}$ is a post-treatment indicator that equals 1 after the implementation of location transparency and 0 before. The DiD interaction coefficient estimate ($\beta_{3}$) captures the aggregated effect of location transparency. $\gamma_{t}$ represents weekly fixed effects in the pre-post model. Additionally, to better understand the effect size of location transparency, we exponentially transform the coefficient estimates of the DiD terms in the models:
\begin{equation}
\begin{aligned}
    \text{ATT} = e^{\beta} - 1,
\end{aligned}
\end{equation}
where $\beta$ is the coefficient estimate for the specific DiD interaction term. ATT indicates the ratio of additional change in the count dependent variable with the treatment of location transparency relative to the counts that are expected to receive without the treatment.


\clearpage 

%
\bibliography{references} 
\bibliographystyle{sciencemag}

\clearpage
\section*{Acknowledgements}

\paragraph*{Ethics statement:} This study was approved by Oxford Internet Institute DREC (ref.~3111054). All analyses are based on publicly available data.

\paragraph*{Author contributions:} Y.C., T.R., D.R., and M.M designed research; Y.C., T.R., D.R., and M.M. performed research; Y.C., T.R., and M.M. analysed data; Y.C., T.R., D.R., and M.M wrote the manuscript.

\paragraph*{Data and code availability:} Upon publication of this work, all pseudonymised data, materials, and analysis scripts required to reproduce our study will be made available publicly.

\paragraph*{Competing interests:} D.R. was an unpaid consultant for Twitter in 2021 and 2022. M.M. and D.R. have previously received funding from Meta and Google to conduct research related to reducing the spread of misinformation online and identifying inauthentic accounts.





\newpage
\appendix
\crefalias{section}{appendix}
\crefalias{subsection}{appendix}


\renewcommand{\thefigure}{S\arabic{figure}}
\renewcommand{\thetable}{S\arabic{table}}
\renewcommand{\theequation}{S\arabic{equation}}
\renewcommand{\thepage}{S\arabic{page}}

\setcounter{table}{0}
\setcounter{equation}{0}
\setcounter{page}{1} 


\begin{center}
\section*{Supplementary Materials for\\ \scititle}

Yuwei Chuai,
Thomas Renault,
David Rand,
Mohsen Mosleh$^{\ast}$\\ 
\small$^\ast$Corresponding author. Email: mohsen.mosleh@oii.ox.ac.uk\\
\end{center}




\newpage
\tableofcontents

\clearpage
\setcounter{figure}{0}
\renewcommand{\thefigure}{S\arabic{figure}}
\setcounter{section}{0}
\renewcommand{\thesection}{S\arabic{section}}
\renewcommand{\thesubsection}{S\arabic{section}.\arabic{subsection}}

\section{``About this account'' feature on X (formerly Twitter)}
\label{supp:about}

In November 2025, \X introduced a new feature, ``About this account,'' to improve transparency and reduce inauthentic engagement on its platform. The location is an inferred geographic location for users based on the aggregated IP addresses at the country or region level~\cite{x2026change}.

This feature offers several advantages over the disclosed location field. First, it provides coverage for a much larger fraction of users, as many users do not declare a location or provide ambiguous entries in their profile. Second, because the location is inferred from observed behaviour rather than self-reported information, it is less susceptible to strategic misreporting or outdated entries. Third, the feature is designed to capture a user's predominant location over time, rather than a transient or symbolic affiliation, thereby improving the accuracy of geographic classification.

\begin{figure}[ht]
    \centering
    \includegraphics[width=.7\linewidth]{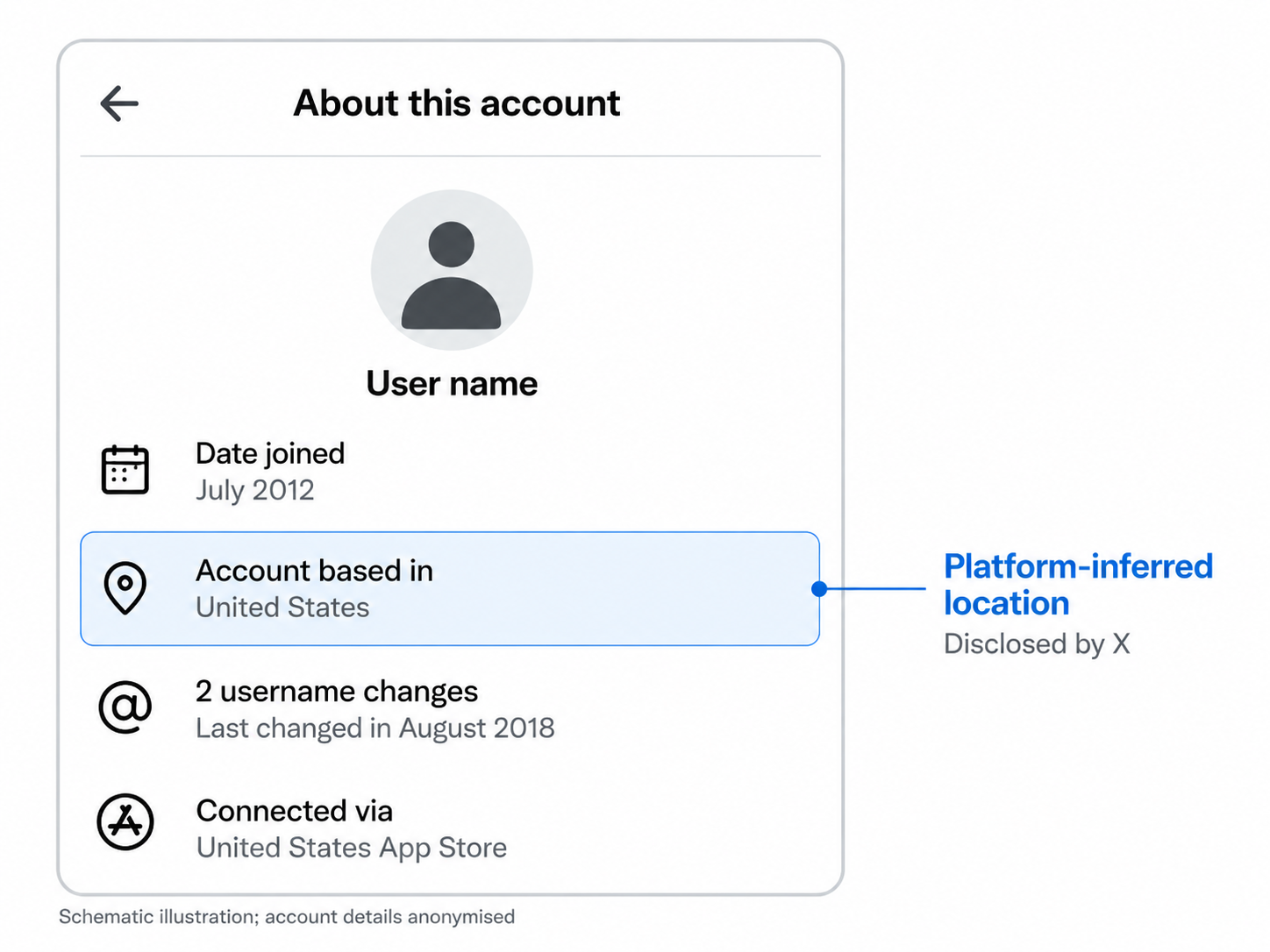}
    \caption{\textbf{The illustration of ``About this account'' feature on \X.} This picture is sourced from \X~\cite{bier2025roll}. The user's avatar and name are anonymised to protect privacy.}
    \label{fig:placeholder}
\end{figure}

\clearpage
\section{Data overview}
\label{supp:data_overview}
To examine the impact of location transparency on location-mismatched accounts' posting activity and audience engagement on \X, we collect all original posts and replies from the selected \num{8200} accounts in the treatment and control groups over a period of seven weeks before and fourteen weeks after the roll-out of the transparency feature. 
As a result, we successfully collect \num{1327286} original posts and \num{3631457} replies. The summary statistics of our dataset are reported in \Cref{tab:data_overview}. Notably, not all selected accounts have posts within the observation window. For instance, the collected original posts come from \num{5818} unique accounts, while the remaining accounts have no original posts recorded during this period. Given that only a small number of accounts are suspended or protected (see details in \Cref{supp:account_status}), accounts with no original posts were likely either inactive on the platform or had deleted their posts prior to data collection. In this study, we focus on the behaviour changes of active accounts in response to location transparency; the exclusion of these inactive accounts is unlikely to introduce systematic bias into our estimates. Additionally, we conduct several robustness checks (\eg, analysis through post count endpoint) to ensure the robustness of our findings (see \Cref{supp:robustness})

\begin{table}[ht]
    \centering
    \setlength{\tabcolsep}{18pt}
    \renewcommand{\arraystretch}{.6}
    \caption{\textbf{Data overview.} Reported are count numbers (numbers of account, original posts, outbound replies, inbound reposts, and inbound replies), means with standard deviations in parentheses (followers, followees, account ages, post frequency, political leanings in content, toxicity, scam, crypto, and misleadingness), or percentages (blue verified status, name change, and VPN use).}
    \begin{tabular}{l*3{S}}
    \toprule
    &{Total}&{Treatment}&{Control}\\
    \midrule
    \# Accounts&{8,200}&{4,100}&{4,100}\\
    \\
    \multicolumn{4}{l}{\underline{Posting activity of the selected accounts}}\\
    \quad \# Original posts&{1,327,286}&{662,320}&{664,966}\\
    \quad \# Outbound replies&{3,631,457}&{1,473,588}&{2,157,869}\\
    \\
    \multicolumn{4}{l}{\underline{Inbound engagement with original posts from selected accounts}}\\
    \quad \# Inbound reposts&{39,433,759}&{30,734,595}&{8,699,164}\\
    \quad \# Inbound replies&{21,689,432}&{18,903,506}&{2,785,926}\\
    \\
    \multicolumn{4}{l}{\underline{Account characteristics}}\\
    \quad \# Followers&{9,160.896}&{7,541.493}&{10,483.013}\\
    &{(84,904.314)}&{(40,001.455)}&{(108,562.371)}\\
    \quad \# Followees&{2,718.405}&{2,529.522}&{2,872.613}\\
    &{(6,924.579)}&{(6,933.012)}&{(6,914.943)}\\
    \quad Blue verified&{25.4\%}&{19.8\%}&{30.0\%}\\
    \quad Account ages (in weeks)&{415.560}&{355.702}&{464.429}\\
    &{(287.007)}&{(277.371)}&{(285.524)}\\
    \quad VPN&{21.8\%}&{36.0\%}&{7.6\%}\\
    \quad Name change&{63.0\%}&{71.9\%}&{54.1\%}\\
    \\
    \multicolumn{4}{l}{\underline{Posting preference based on original posts before treatment}}\\
    \quad Post frequency (weekly)&{13.799}&{15.591}&{12.385}\\
    \quad &{(56.695)}&{(78.033)}&{(30.687)}\\
    \quad Left-leaning&{0.095}&{0.089}&{0.100}\\
    \quad &{(0.173)}&{(0.160)}&{(0.183)}\\
    \quad Right-leaning&{0.170}&{0.172}&{0.169}\\
    \quad &{(0.223)}&{(0.224)}&{(0.221)}\\
    \quad Toxicity&{0.105}&{0.093}&{0.114}\\
    \quad &{(0.160)}&{(0.158)}&{(0.162)}\\
    \quad Scam&{0.009}&{0.017}&{0.002}\\
    \quad &{(0.062)}&{(0.085)}&{(0.031)}\\
    \quad Crypto&{0.023}&{0.038}&{0.010}\\
    \quad &{(0.109)}&{(0.146)}&{(0.064)}\\
    \quad Misleadingness&{0.447}&{0.463}&{0.434}\\
    \quad &{(0.229)}&{(0.237)}&{(0.222)}\\
    \bottomrule
    \end{tabular}
    \label{tab:data_overview}
\end{table}

\clearpage
\section{Methods}
\label{supp:methods}

\subsection{Identification of country information and \US politics}
\label{supp:country_extraction}

Prompt~1 and Prompt~2 report the full instructions used to identify accounts' country information and relevance of \US politics, respectively. Given the large volume of our dataset, we use the Batch API with GPT-5 mini model to perform the tasks. Additionally, to ensure the reliability of the model outputs, we use GPT-5.5 model to check 200 random samples with the same instructions. Based on the outputs from GPT-5.5, the precisions of the two tasks, \ie, country extraction and relevance of \US politics, for GPT-5 mini reach 0.835 and 0.84, respectively. 

\begin{tcolorbox}[colback=blue!5, colframe=blue!40, title=\textbf{Prompt~1. Instruction for country extraction}]
\setlength{\parskip}{0.8em}   
\setlength{\parindent}{0pt}   
\linespread{1}\selectfont     

You are an AI system tasked with identifying the country or countries
referenced in the Twitter/X profile.

You will be given a list of locations. For each location:\\
- Output a SINGLE valid JSON object.\\
- Each key MUST be exactly the original location string.\\
- Each value MUST be the ISO 3166-1 alpha-2 country codes.\\
- If no country is clearly mentioned or you are unsure, answer with an empty list.

Do NOT output anything except the JSON object.
\end{tcolorbox}

\begin{tcolorbox}[colback=blue!5, colframe=blue!40, title=\textbf{Prompt~2. Instruction for identification of \US politics in account profile}]
\setlength{\parskip}{0.8em}   
\setlength{\parindent}{0pt}   
\linespread{1}\selectfont     

You are a strict classifier.

Decide whether this X/Twitter bio indicates the account shares content about US politics.

Labels:\\
- 2: clear US politics focus.\\
- 1: hints of politics/ideology but US link is weak or ambiguous.\\
- 0: no evidence of US politics focus.\\

Rules:\\
- "Politics" without US anchor => 1 (unless clearly non-US).\\
- Any explicit US anchor (US parties, politicians, elections, Congress/SCOTUS, MAGA, etc.) => 2.\\
- If uncertain => 0.

Return only the label.
\end{tcolorbox}

\newpage
\subsection{Identification of political leaning}
\label{supp:political_align}

In the main text, we focus on the political leaning expressed in political posts and use a text political leaning classifier based on DeBERTaV3~\cite{volf2025political}. This classifier has an average accuracy of 0.87 and $F_1$ score of 0.872 across ten diverse datasets. For validation at the account-level, we average these post-level scores within each account and compare the resulting account-level classifications with those independently inferred by an LLM from pre-intervention original posts and replies. Specifically, the DeBERTaV3-based classifier processes \num{382023} political posts from \num{4439} unique accounts. We average the leaning scores for posts from the same account (left$=-1$, center$=0$, and right$=1$), and classify the account as left-leaning ($<0$), right-leaning $>0$ or centre ($=0$). This results in \num{1180} left-leaning accounts, \num{2980} right-leaning accounts, and \num{279} centred accounts.

We then identify each account's political alignment by using the approach proposed and validated by a previous study~\cite{lee2026llms}. They used GPT-4o to infer text-level political alignment and aggregated multiple text-level inferences into a user-level prediction. The inference approach achieves an $F_1$ score of 0.799 for general texts using the maximum-confidence method (\ie, the majority vote to the subset of texts with the highest confidence). Here, we adopt OpenAI's API with GPT-5.4~mini model to infer accounts' political alignment, and the full prompt is reported in Prompt~3. We use \textit{tiktoken} tokenizer to process the original posts and replies from accounts in the treatment and control groups and retain those posts/replies with tokens between 10 and \num{1000} ($10<\text{tokens}<1000$), resulting in \num{1081920} original posts and \num{3007817} replies from \num{6783} unique accounts. To isolate the potential contamination of the treatment on the leaning change, we use the original posts and replies created before the location transparency to infer accounts' political alignments. Specifically, we randomly select 10 posts (original posts or replies) per week for each account, and aggregate each account's posts within the same week into a single text. This yields \num{32706} texts for the GPT model to infer. We then use the maximum-confidence method to aggregate text-level inferences and predict account-level political alignment. Finally, we successfully identify the political alignment for \num{6123} accounts in total, with \num{4186} on the right, \num{1729} on the left, and \num{208} in the center.

We consider left- and right-leaning accounts predicted by the two models and do the cross-check with \num{3871} in common. The two models have a substantial inter-annotator agreement with Cohen's $\kappa=0.613$. We conduct a robustness check for the impact of location transparency based on the LLM-inferred political alignment; the results remain robust and consistent with our main findings (see \Cref{supp:he_account})

\begin{tcolorbox}[colback=blue!5, colframe=blue!40, title=\textbf{Prompt~3. Instruction for the identification of political alignment}]
\setlength{\parskip}{0.8em}   
\setlength{\parindent}{0pt}   
\linespread{1}\selectfont     

You are a X/Twitter expert specialized in identifying a user's political alignment in the U.S.

\#\#\# Input\\
You will be provided with a list of posts from the user.

\#\#\# Task\\
Based on the given posts, infer whether the user's opinions align more with the Republican Party or the Democratic Party.

\#\#\# Guidelines\\
- Choose "Republican" or "Democratic" based on the dominant political leaning expressed in the posts.\\
- Provide your confidence level for the classification on a scale from 1 (very low confidence) to 5 (very high confidence). This must be an integer.

\#\#\# Output\\
Return ONLY a JSON format with "party" and "confidence" as keys. For example: \{"party" : "Democratic", "confidence" : 5\}\\
Do not include any additional text or explanation.
\end{tcolorbox}

\newpage
\subsection{Extraction of content features}
\label{supp:content_features}

\noindent \textbf{Identification of scam posts.}\\
We use OpenAI's GPT-5.4 mini model to identify scam-related content across all the original posts from the selected accounts. The model is instructed using Prompt~4 and successfully assigns scam labels to \num{1297958} posts. The results show that \num{259715} posts are potentially related to scams, \num{8530} posts clearly contain scams or scam attempts, and the majority of posts ($=$ \num{1029713}) are not scam-related. We note that scam identification is more challenging than country extraction and politics identification. To ensure the consistency and reliability of the model outputs, we repeat the scam identification task with GPT-5.4 mini and find that the outputs across the two rounds are highly consistent (Cohen's $\kappa=1$). Additionally, we randomly select 300 posts (100 per label) and use the full-sized GPT-5.4 model to identify scams. The mini model achieves comparable performance to the full-sized model across the three labels (Cohen's $\kappa=0.65$, macro $F_1=0.76$). Moreover, if we consider only those posts clearly identified as scams and treat all others as not scam-related, thereby mitigating the uncertainty introduced by potentially scam-related posts across different models, the agreement between the mini model and full-sized model significantly improves (Cohen's $\kappa=0.717$, macro $F_1=0.86$). Therefore, to ensure the reliability of our results, we consider only those posts assigned label 2 by the GPT-5.4 mini model as scam posts in the main analysis. \Cref{tab:scam_examples} shows five random scam post examples identified by GPT-5.4 mini. Nevertheless, we expand our scope for scam posts and repeat our analysis with another DistilBERT-based model. The results remain consistent and robust (see \Cref{supp:he_account}).

\vspace{1em}
\noindent \textbf{Identification of crypto posts.}\\
To identify if the post is about cryptocurrency (Crypto), we check if any of the following terms appeared in the text:
\begin{itemize}[leftmargin=*,noitemsep,topsep=0pt]
    \item CRYPTO TERMS: "crypto", "cryptocurrency", "blockchain", "web3", "defi", "cefi",
    "token", "tokens", "coin", "coins", "altcoin", "altcoins", "stablecoin", "stablecoins"
    \item CRYPTO COINS: "bitcoin", "btc", "ethereum", "eth", "solana", "sol", "bnb", "binance", "ripple", "xrp", "dogecoin", "doge", "cardano", "ada", "polkadot", "dot", "litecoin", "ltc", "tron", "trx"
    \item TRADING SLANG: "hodl", "moon", "mooning", "dump", "pump", "pumping", "bullish", "bearish", "dip", "buythedip", "ath", "atl", "rekt", "bagholder", "whale", "whales", "liquidation", "fomo", "fud"
    \item CRYPTO ECOSYSTEM: "nft", "nfts", "mint", "minting", "airdrop", "staking", "yield", "apy", "liquidity", "liquid", "pool", "dao", "governance", "smartcontract", "smartcontracts", "layer2", "l2", "rollup", "zk", "zkp"
\end{itemize}
As a result, we find that \num{23266} posts (1.8\%) out of total original posts are related to cryptocurrency.

\begin{table}[ht]
    \centering
    \caption{\textbf{Examples of scam posts identified by GPT-5.4 mini model.} Exact URLs are omitted and replaced with a placeholder in the examples.}
    \begin{tabular}{p{.05\linewidth}p{.85\linewidth}}
    \toprule
    1&{3 odds banker available for pay after send a dm if you want knack am with me... Let's win together 5k stakes above https://t.co/} \\
    2&{Cut 2... Any winners?? Drop slips https://t.co/}\\
    3&{Retweet and drop aza for small staking power!!!!}\\
    4&{Donald Trump advised everyone to withdraw out from the bank and enroll to the \#QFS to enable you to purchase the ISO20022 coins \$XRP and \$XLM that will be used for all transactions with the new system \#QFS QUANTUM FINANCIAL SYSTEM is the future, and the future is now!!! appropr}\\
    5&{One Stop to Apply for ALL Gov’t Programs to Pay Your Bills! Free Download of 100 Free Money Programs in Your City. https://t.co/ Join our Loving Community https://t.co/ https://t.co/}\\
    \bottomrule
    \end{tabular}
    \label{tab:scam_examples}
\end{table}

\begin{tcolorbox}[colback=blue!5, colframe=blue!40, title=\textbf{Prompt~4. Instruction for the identification of scam posts}]
\setlength{\parskip}{0.8em}   
\setlength{\parindent}{0pt}   
\linespread{1}\selectfont     

You classify social media posts about scams.

Return ONLY valid JSON:
\{"label": <0|1|2>, "types": [list of strings]\}

Label definitions:\\
0 = Not about scams.\\
1 = Maybe scam / scam-adjacent / discussing scams.\\
2 = Certain scam or clear scam attempt.\\

Scam means ANY attempt or solicitation involving deception for money, credentials, crypto, goods, or personal data.

Label = 2 when the post:\\
• asks for money, crypto, gift cards, fees, deposits\\
• promises guaranteed profits / fast money / passive income\\
• asks users to DM for investment/job/offer\\
• contains giveaway/prize claims requiring action\\
• impersonates brands, celebrities, support accounts\\
• asks for login codes, passwords, wallet keys\\
• promotes "recovery services" for lost crypto/money\\
• pushes suspicious links with urgency\\
• instructs off-platform contact (Telegram, WhatsApp, DM)\\
• advertises clearly fraudulent job or marketplace offers

Label = 1 when the post:\\
• discusses scams, warns about scams, or reports being scammed\\
• asks if something is legit\\
• mentions scams/news/research/education\\
• is ambiguous or lacks enough evidence

If unsure between 1 and 2 → choose 1.

types must be chosen from:
["crypto","investment","forex","get\_rich\_quick","impersonation",
"giveaway","phishing","romance","job","marketplace","recovery","other"]

If label = 0 → types must be [].

Output JSON only. No extra text.
\end{tcolorbox}

\newpage
\section{Estimation results for the impact of location transparency}
\label{supp:did_estimations}

\subsection{Estimation results for main analysis}
The estimation results of both leads-and-lags and pre-post DiD in the main analysis are reported in \Cref{tab:did_main_account,tab:did_main_engage}.

\begin{table*}[ht]
    \centering
    \setlength{\tabcolsep}{10pt}
    \caption{\textbf{DiD estimates for the number of original posts and outbound replies from the accounts in the treatment group.} The estimation is based on \num{144543} count observations across \num{6883} accounts. Weekly and account-level fixed effects are included.}
    \begin{tabular}{l*{4}{c}}
    \toprule
    &   DiD estimate   &   $z$  &   $p$   &   95\% CI\\
    \midrule
    Week $-7$& $0.062$& $z=1.78$& $p=0.075$& $[-0.006, 0.135]$\\
    Week $-6$& $0.064$& $z=1.842$& $p=0.065$& $[-0.004, 0.137]$\\
    Week $-5$& $0.037$& $z=1.064$& $p=0.287$& $[-0.03, 0.108]$\\
    Week $-4$& $0.06$& $z=1.717$& $p=0.086$& $[-0.008, 0.133]$\\
    Week $-3$& $-0.022$& $z=-0.642$& $p=0.521$& $[-0.084, 0.046]$\\
    Week $-2$ & $-0.008$ & $z=-0.225$ & $p=0.822$ & $[-0.072, 0.061]$\\
    Week $1$  & $-0.076$ & $z=-2.258$ & $p=0.024$ & $[-0.137, -0.01]$\\
    Week $2$  & $-0.082$ & $z=-2.421$ & $p=0.015$ & $[-0.144, -0.016]$\\
    Week $3$  & $-0.086$ & $z=-2.519$ & $p=0.012$ & $[-0.147, -0.02]$\\
    Week $4$  & $-0.121$ & $z=-3.691$ & $p<0.001$ & $[-0.18, -0.059]$\\
    Week $5$  & $-0.034$ & $z=-0.968$ & $p=0.333$ & $[-0.1, 0.036]$\\
    Week $6$  & $-0.11$  & $z=-3.21$  & $p=0.001$ & $[-0.17, -0.044]$\\
    Week $7$  & $-0.129$ & $z=-4.002$ & $p<0.001$ & $[-0.185, -0.068]$\\
    Week $8$  & $-0.17$  & $z=-5.335$ & $p<0.001$ & $[-0.225, -0.111]$\\
    Week $9$  & $-0.112$ & $z=-3.404$ & $p<0.001$ & $[-0.171, -0.049]$\\
    Week $10$ & $-0.16$  & $z=-5.018$ & $p<0.001$ & $[-0.215, -0.101]$\\
    Week $11$ & $-0.12$  & $z=-3.651$ & $p<0.001$ & $[-0.179, -0.058]$\\
    Week $12$ & $-0.126$ & $z=-3.841$ & $p<0.001$ & $[-0.184, -0.064]$\\
    Week $13$ & $-0.048$ & $z=-1.398$ & $p=0.162$ & $[-0.112, 0.02]$\\
    Week $14$ & $-0.102$ & $z=-3.038$ & $p=0.002$ & $[-0.162, -0.037]$\\
    Aggregated& $-0.131$ & $z=-12.534$& $p<0.001$ & $[-0.15, -0.112]$\\
    \bottomrule
    \end{tabular}
    \label{tab:did_main_account}
\end{table*}

\begin{table*}[ht]
    \centering
    \setlength{\tabcolsep}{10pt}
    \caption{\textbf{DiD estimates for the number of inbound replies and reposts from the audiences to the accounts in the treatment group.} The estimation is based on \num{39512} count observations across \num{2929} accounts. Weekly and account-level fixed effects are included.}
    \begin{tabular}{l*{4}{c}}
    \toprule
    &   DiD estimate   &   $z$  &   $p$   &   95\% CI\\
    \midrule
    Week $-7$ & $0.008$  & $z=0.095$  & $p=0.925$ & $[-0.148, 0.193]$\\
    Week $-6$ & $0.06$   & $z=0.672$  & $p=0.501$ & $[-0.105, 0.254]$\\
    Week $-5$ & $-0.048$ & $z=-0.578$ & $p=0.563$ & $[-0.195, 0.125]$\\
    Week $-4$ & $0.036$  & $z=0.417$  & $p=0.677$ & $[-0.123, 0.225]$\\
    Week $-3$ & $0.04$   & $z=0.459$  & $p=0.646$ & $[-0.121, 0.23]$\\
    Week $-2$ & $0.128$  & $z=1.402$  & $p=0.161$ & $[-0.047, 0.335]$\\
    Week $1$  & $0.03$   & $z=0.335$  & $p=0.737$ & $[-0.134, 0.225]$\\
    Week $2$  & $0.048$  & $z=0.53$   & $p=0.596$ & $[-0.118, 0.245]$\\
    Week $3$  & $0.089$  & $z=0.955$  & $p=0.34$  & $[-0.086, 0.296]$\\
    Week $4$  & $0.075$  & $z=0.827$  & $p=0.408$ & $[-0.095, 0.278]$\\
    Week $5$  & $-0.017$ & $z=-0.186$ & $p=0.852$ & $[-0.176, 0.173]$\\
    Week $6$  & $0.053$  & $z=0.559$  & $p=0.576$ & $[-0.121, 0.261]$\\
    Week $7$  & $-0.017$ & $z=-0.194$ & $p=0.846$ & $[-0.175, 0.171]$\\
    Week $8$  & $0.034$  & $z=0.374$  & $p=0.708$ & $[-0.132, 0.232]$\\
    Week $9$  & $-0.007$ & $z=-0.085$ & $p=0.933$ & $[-0.166, 0.181]$\\
    Week $10$ & $-0.112$ & $z=-1.344$ & $p=0.179$ & $[-0.253, 0.056]$\\
    Week $11$ & $-0.1$   & $z=-1.22$  & $p=0.222$ & $[-0.241, 0.066]$\\
    Week $12$ & $0.173$  & $z=1.883$  & $p=0.06$  & $[-0.006, 0.385]$\\
    Week $13$ & $-0.049$ & $z=-0.582$ & $p=0.561$ & $[-0.197, 0.126]$\\
    Week $14$ & $0.083$  & $z=0.913$  & $p=0.361$ & $[-0.087, 0.286]$\\
    Aggregated& $-0.012$ & $z=-0.412$ & $p=0.681$ & $[-0.067, 0.046]$\\
    \bottomrule
    \end{tabular}
    \label{tab:did_main_engage}
\end{table*}

\clearpage
\subsection{Estimation results for separated original posts and outbound replies}
\begin{figure}[ht]
    \centering
    \captionsetup[subfloat]{font={bf, small}, skip=0pt, singlelinecheck=false, labelformat=simple, position=top}
    \subfloat[]{\includegraphics[width = \textwidth]{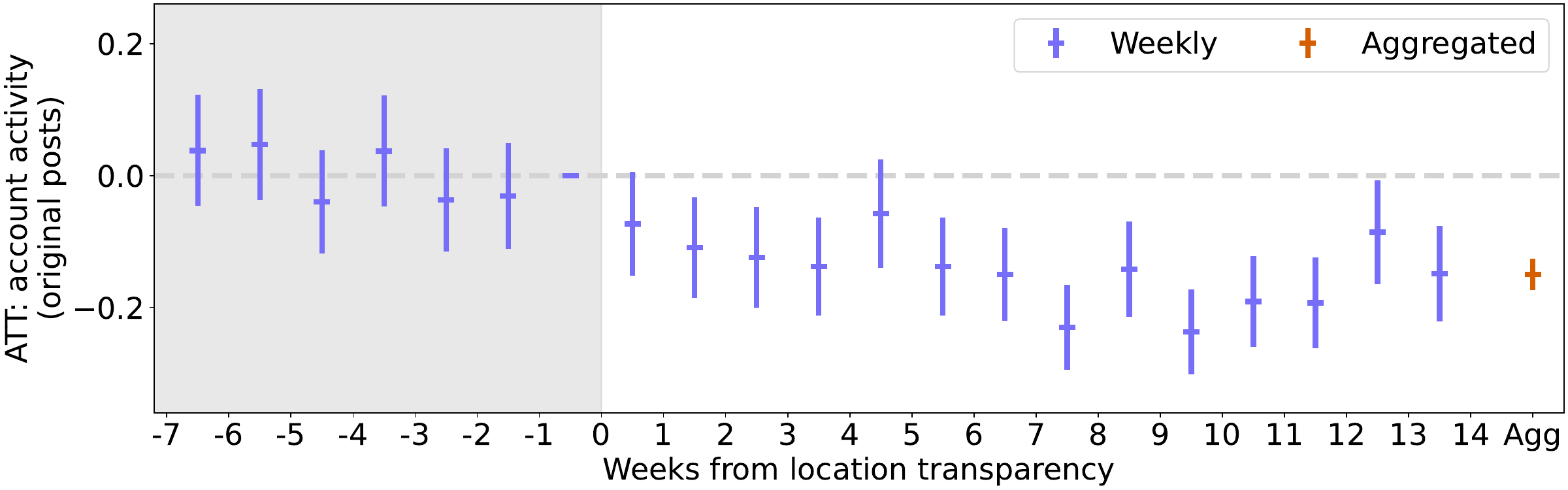}}

    \subfloat[]{\includegraphics[width = \textwidth]{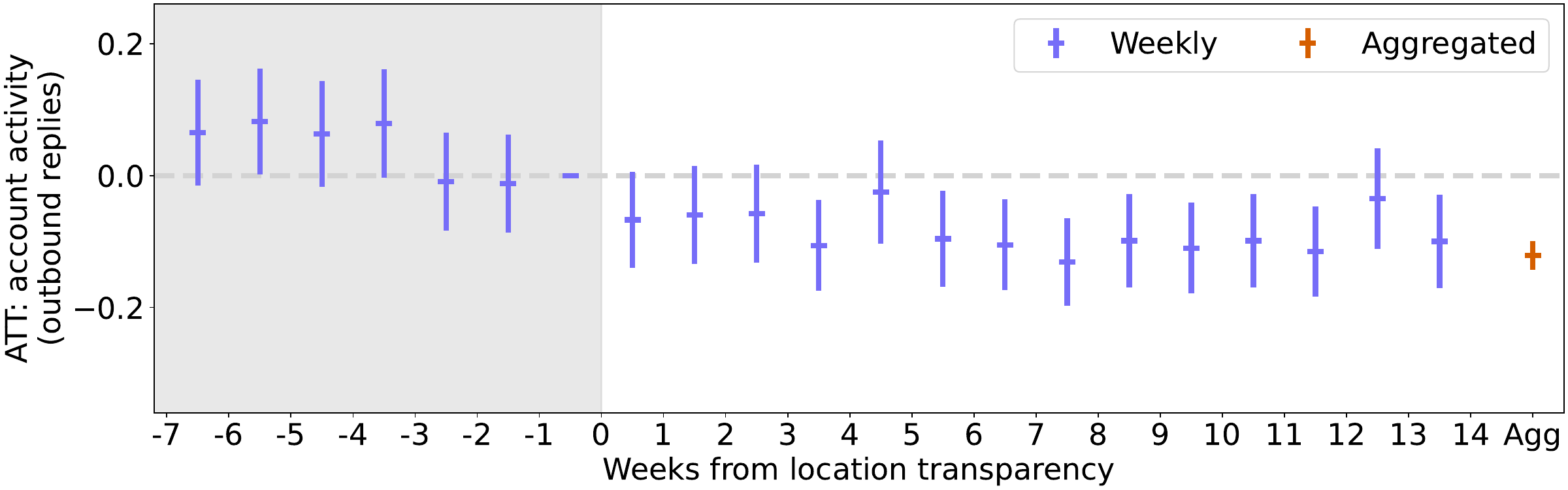}}

    \caption{\textbf{Separated effects of location transparency on account posting activity by post type.} \textbf{(a)}~Weekly and aggregated ATT estimates for the number of original posts from the accounts in the treatment group. The estimation is based on 122,178 count observations across 5,818 accounts \textbf{(b)}~Weekly and aggregated ATT estimates for the number of outbound replies from the accounts in the treatment group. The estimation is based on 131,397 count observations across 6,257 accounts. The error bars represent 95\% CI. The estimation results are reported in \Cref{tab:did_account_posts,tab:did_account_replies}.}
    \label{fig:account_post_replies}
\end{figure}

We examine the impact of location transparency on overall account activity by combining original posts and outbound replies in the main analysis. To validate whether the reduction applies to each post type separately, we conduct additional analyses on original posts and outbound replies independently. As shown in \Cref{fig:account_post_replies}, the introduction of location transparency has consistent reduction effects on both original posts (ATT~$=-0.15$, $z=-11.858$, $p<0.001$; 95\%~CI: $[-0.173, -0.127]$) and outbound replies (ATT~$=-0.121$, $z=-10.115$, $p<0.001$; 95\%~CI: $[-0.143, -0.099]$). Notably, \Cref{fig:account_post_replies}b suggests a mild violation of parallel trends assumption for outbound replies. To ensure that this pre-trend does not confound the estimated effect, we apply Synthetic 
DiD as a robustness check (see details in \Cref{supp:sdid}), and our findings remain robust.

\begin{table*}[ht]
    \centering
    \setlength{\tabcolsep}{10pt}
    \caption{\textbf{DiD estimates for the number of original posts from the accounts in the treatment group.} The estimation is based on \num{122178} count observations across \num{5818} accounts. Weekly and account-level fixed effects are included.}
    \begin{tabular}{l*{4}{c}}
    \toprule
    &   DiD estimate   &   $z$  &   $p$   &   95\% CI\\
    \midrule
    Week $-7$ & $0.038$  & $z=0.895$  & $p=0.371$ & $[-0.043, 0.125]$\\
    Week $-6$ & $0.047$  & $z=1.097$  & $p=0.273$ & $[-0.035, 0.135]$\\
    Week $-5$ & $-0.04$  & $z=-0.988$ & $p=0.323$ & $[-0.116, 0.041]$\\
    Week $-4$ & $0.037$  & $z=0.88$   & $p=0.379$ & $[-0.044, 0.126]$\\
    Week $-3$ & $-0.037$ & $z=-0.895$ & $p=0.371$ & $[-0.112, 0.045]$\\
    Week $-2$ & $-0.031$ & $z=-0.736$ & $p=0.462$ & $[-0.107, 0.053]$\\
    Week $1$  & $-0.073$ & $z=-1.741$ & $p=0.082$ & $[-0.149, 0.01]$\\
    Week $2$  & $-0.109$ & $z=-2.639$ & $p=0.008$ & $[-0.183, -0.029]$\\
    Week $3$  & $-0.124$ & $z=-3.018$ & $p=0.003$ & $[-0.197, -0.045]$\\
    Week $4$  & $-0.138$ & $z=-3.406$ & $p<0.001$ & $[-0.209, -0.061]$\\
    Week $5$  & $-0.058$ & $z=-1.347$ & $p=0.178$ & $[-0.137, 0.028]$\\
    Week $6$  & $-0.138$ & $z=-3.33$  & $p<0.001$ & $[-0.21, -0.059]$\\
    Week $7$  & $-0.15$  & $z=-3.878$ & $p<0.001$ & $[-0.216, -0.077]$\\
    Week $8$  & $-0.23$  & $z=-6.126$ & $p<0.001$ & $[-0.291, -0.163]$\\
    Week $9$  & $-0.142$ & $z=-3.583$ & $p<0.001$ & $[-0.211, -0.067]$\\
    Week $10$ & $-0.237$ & $z=-6.334$ & $p<0.001$ & $[-0.298, -0.17]$\\
    Week $11$ & $-0.191$ & $z=-4.917$ & $p<0.001$ & $[-0.257, -0.12]$\\
    Week $12$ & $-0.193$ & $z=-5.007$ & $p<0.001$ & $[-0.258, -0.122]$\\
    Week $13$ & $-0.086$ & $z=-2.062$ & $p=0.039$ & $[-0.16, -0.004]$\\
    Week $14$ & $-0.149$ & $z=-3.718$ & $p<0.001$ & $[-0.218, -0.073]$\\
    Aggregated& $-0.15$ & $z=-11.858$ & $p<0.001$ & $[-0.173, -0.127]$\\
    \bottomrule
    \end{tabular}
    \label{tab:did_account_posts}
\end{table*}

\begin{table*}[ht]
    \centering
    \setlength{\tabcolsep}{10pt}
    \caption{\textbf{DiD estimates for the number of outbound replies from the accounts in the treatment group.} The estimation is based on \num{131397} count observations across \num{6257} accounts. Weekly and account-level fixed effects are included.}
    \begin{tabular}{l*{4}{c}}
    \toprule
    &   DiD estimate   &   $z$  &   $p$   &   95\% CI\\
    \midrule
    Week $-7$ & $0.065$  & $z=1.632$  & $p=0.103$ & $[-0.012, 0.148]$\\
    Week $-6$ & $0.082$  & $z=2.069$  & $p=0.038$ & $[0.004, 0.167]$\\
    Week $-5$ & $0.063$  & $z=1.59$   & $p=0.112$ & $[-0.014, 0.145]$\\
    Week $-4$ & $0.079$  & $z=1.978$  & $p=0.048$ & $[0.001, 0.164]$\\
    Week $-3$ & $-0.009$ & $z=-0.237$ & $p=0.813$ & $[-0.081, 0.069]$\\
    Week $-2$ & $-0.012$ & $z=-0.308$ & $p=0.758$ & $[-0.084, 0.066]$\\
    Week $1$  & $-0.067$ & $z=-1.745$ & $p=0.081$ & $[-0.136, 0.009]$\\
    Week $2$  & $-0.06$  & $z=-1.543$ & $p=0.123$ & $[-0.131, 0.017]$\\
    Week $3$  & $-0.058$ & $z=-1.474$ & $p=0.141$ & $[-0.129, 0.02]$\\
    Week $4$  & $-0.106$ & $z=-2.831$ & $p=0.005$ & $[-0.173, -0.034]$\\
    Week $5$  & $-0.025$ & $z=-0.613$ & $p=0.54$  & $[-0.099, 0.056]$\\
    Week $6$  & $-0.096$ & $z=-2.474$ & $p=0.013$ & $[-0.165, -0.021]$\\
    Week $7$  & $-0.105$ & $z=-2.835$ & $p=0.005$ & $[-0.17, -0.034]$\\
    Week $8$  & $-0.131$ & $z=-3.56$  & $p<0.001$ & $[-0.196, -0.061]$\\
    Week $9$  & $-0.099$ & $z=-2.64$  & $p=0.008$ & $[-0.167, -0.027]$\\
    Week $10$ & $-0.11$  & $z=-2.966$ & $p=0.003$ & $[-0.176, -0.039]$\\
    Week $11$ & $-0.099$ & $z=-2.632$ & $p=0.008$ & $[-0.167, -0.026]$\\
    Week $12$ & $-0.115$ & $z=-3.085$ & $p=0.002$ & $[-0.181, -0.044]$\\
    Week $13$ & $-0.035$ & $z=-0.894$ & $p=0.371$ & $[-0.109, 0.044]$\\
    Week $14$ & $-0.1$   & $z=-2.603$ & $p=0.009$ & $[-0.168, -0.026]$\\
    Aggregated& $-0.121$ & $z=-10.115$& $p<0.001$ & $[-0.143, -0.099]$\\
    \bottomrule
    \end{tabular}
    \label{tab:did_account_replies}
\end{table*}

\clearpage
\subsection{Estimation results for separated inbound reposts and replies}

\begin{figure}[ht]
    \centering
    \captionsetup[subfloat]{font={bf, small}, skip=0pt, singlelinecheck=false, labelformat=simple, position=top}
    \subfloat[]{\includegraphics[width = \textwidth]{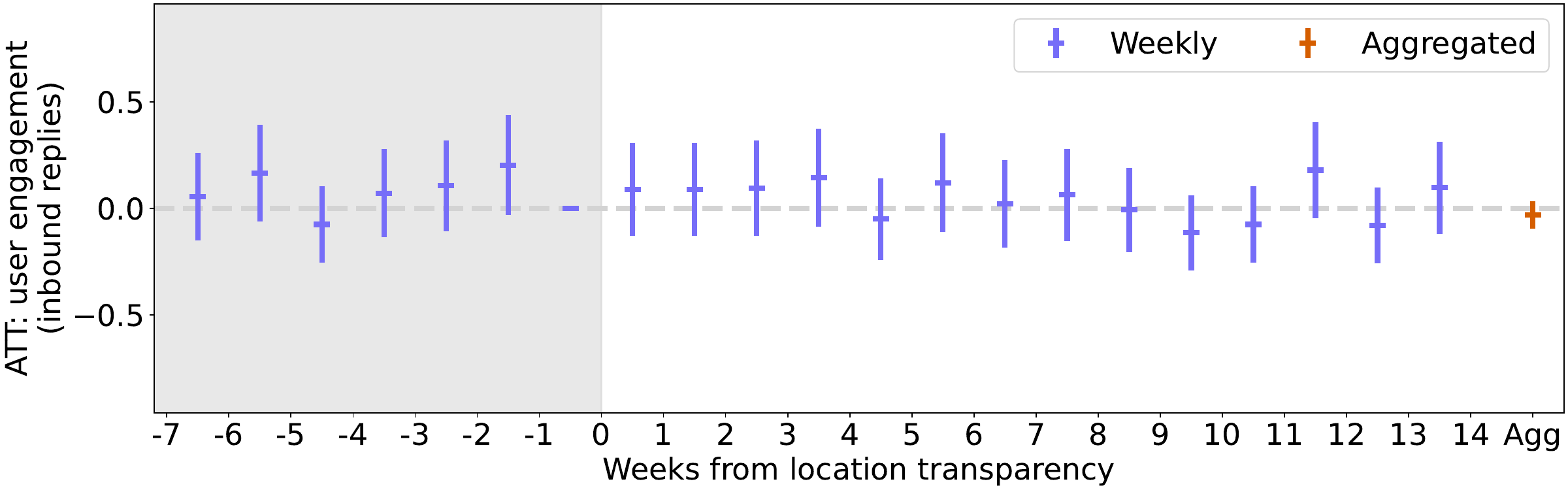}}

    \subfloat[]{\includegraphics[width = \textwidth]{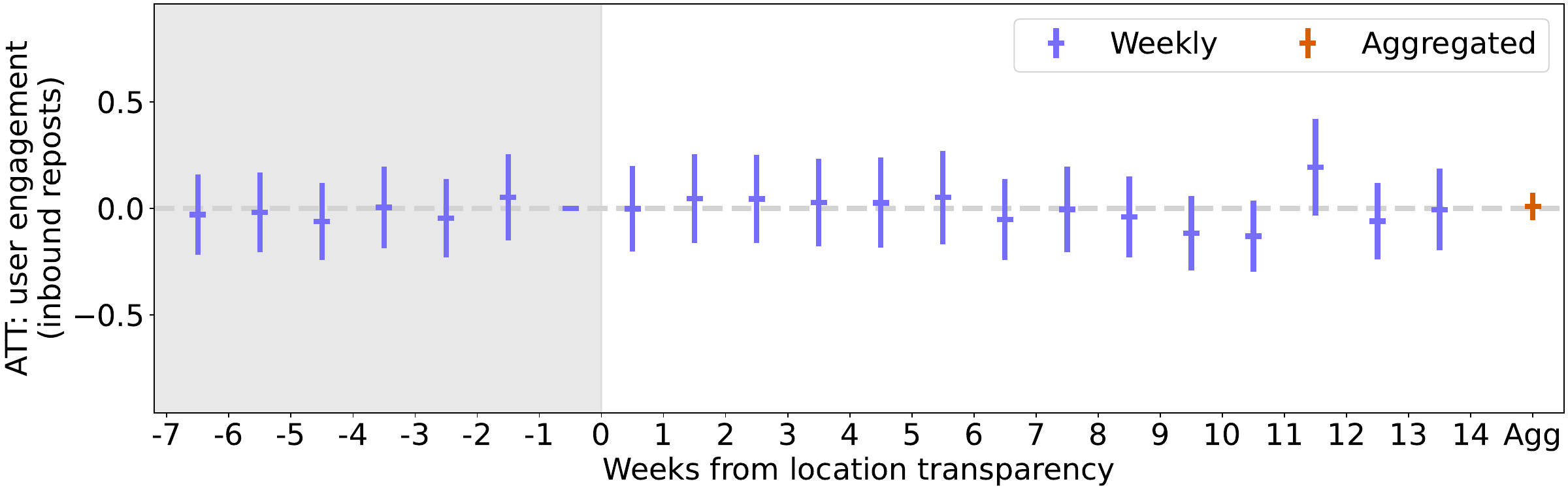}}

    \subfloat[]{\includegraphics[width = \textwidth]{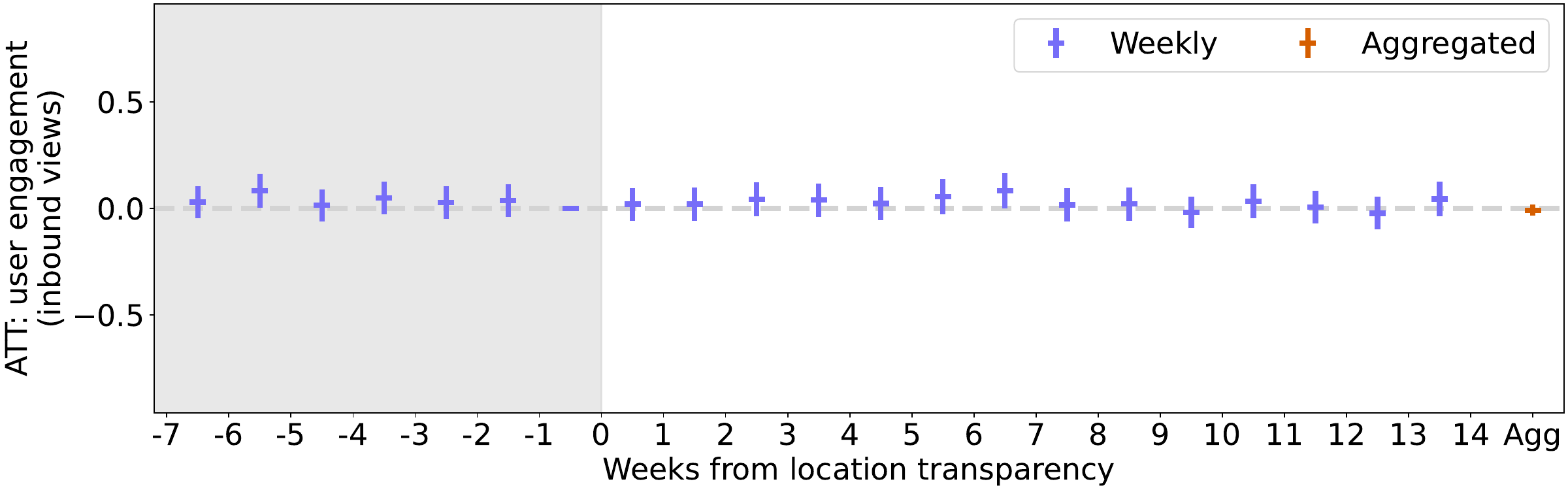}}
    \caption{\textbf{Separated effects of location transparency on audiences' engagement to the accounts in the treatment group by engagement types.} \textbf{(a)}~Weekly and aggregated ATT estimates for the number of inbound replies from the audiences. \textbf{(b)}~Weekly and aggregated ATT estimates for the number of inbound reposts from the audiences. \textbf{(c)}~Weekly and aggregated ATT estimates for the number of inbound views received from the audiences. The error bars represent 95\% CIs. The estimation results are reported in \Cref{tab:did_audience_replies,tab:did_audience_reposts,tab:did_audience_views}.}
    \label{fig:audience_engagement}
\end{figure}

\textbf{Changes in average inbound reposts and replies per post.}\\
To assess whether the null effect on aggregate audience engagement masks different responses across engagement types, we separately estimate the effects on inbound replies and reposts. For both outcomes, the weekly ATT estimates are consistently statistically indistinguishable from zero. The aggregated estimates are likewise non-significant for inbound replies (ATT~$=-0.03$, $z=-0.918$, $p=0.359$; 95\% CI: $[-0.092, 0.036]$) and inbound reposts (ATT~$=0.01$, $z=0.291$, $p=0.771$; 95\% CI: $[-0.054, 0.077]$). Thus, the null effect on overall audience engagement does not appear to conceal offsetting changes in replies and reposts (see \Cref{fig:audience_engagement}).

\begin{figure}[ht]
    \centering
    \captionsetup[subfloat]{font={bf, small}, skip=0pt, singlelinecheck=false, labelformat=simple, position=top}
    {\includegraphics[width = \textwidth]{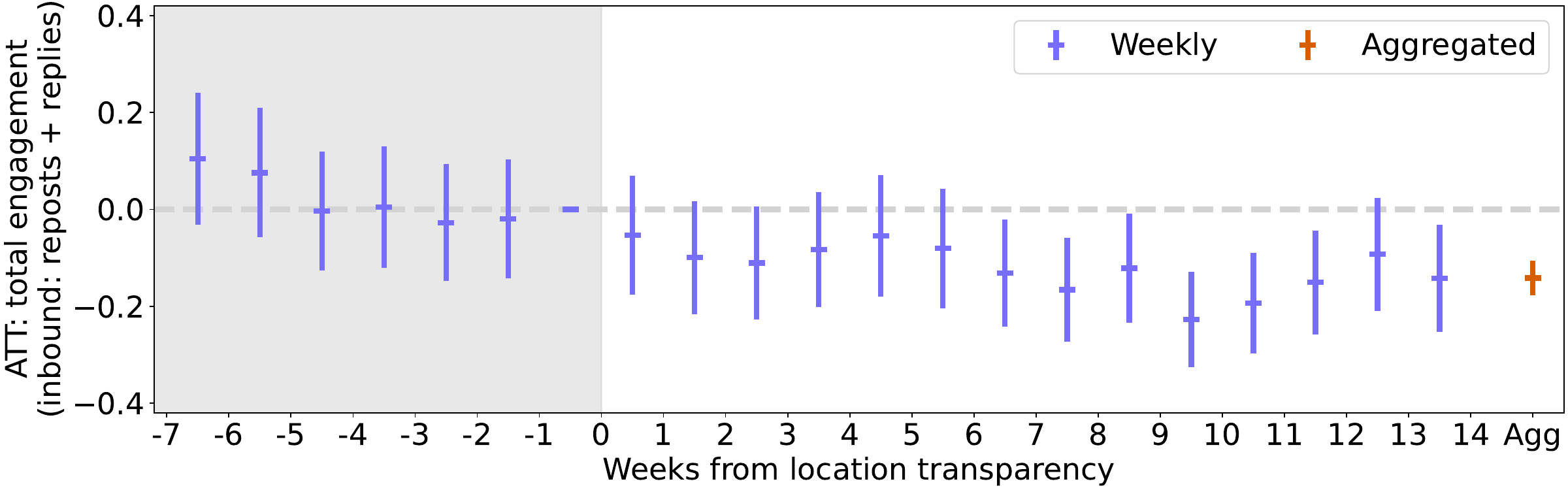}}
    \caption{\textbf{Effects of location transparency on audiences' total engagement per week to the accounts in the treatment group.} The error bars represent 95\% CIs.}
    \label{fig:audience_engagement_total}
\end{figure}

\noindent \textbf{Changes in total inbound reposts and replies per user-week.}\\
We further estimate the effects of location transparency on total audience engagement accounts receive over time (\Cref{fig:audience_engagement_total}). Following the decrease in posting activity of affected accounts, the total engagement they receive also decreases accordingly. On average, the weekly audience engagement received by each account reduces by 14.2\% (ATT~$=-0.142$, $z=-7.293$, $p<0.001$; 95\%~CI: $[-0.176, -0.106]$).

\begin{table*}[ht]
    \centering
    \setlength{\tabcolsep}{10pt}
    \caption{\textbf{DiD estimates for the number of inbound replies from the audiences to the accounts in the treatment group.} The estimation is based on \num{31799} count observations across \num{2334} accounts. Weekly and account-level fixed effects are included.}
    \begin{tabular}{l*{4}{c}}
    \toprule
    &   DiD estimate   &   $z$  &   $p$   &   95\% CI\\
    \midrule
    Week $-7$ & $0.055$  & $z=0.54$   & $p=0.59$  & $[-0.131, 0.281]$\\
    Week $-6$ & $0.166$  & $z=1.549$  & $p=0.121$ & $[-0.04, 0.417]$\\
    Week $-5$ & $-0.075$ & $z=-0.792$ & $p=0.428$ & $[-0.236, 0.121]$\\
    Week $-4$ & $0.072$  & $z=0.708$  & $p=0.479$ & $[-0.116, 0.302]$\\
    Week $-3$ & $0.107$  & $z=1.025$  & $p=0.306$ & $[-0.088, 0.343]$\\
    Week $-2$ & $0.204$  & $z=1.862$  & $p=0.063$ & $[-0.01, 0.464]$\\
    Week $1$  & $0.09$   & $z=0.85$   & $p=0.395$ & $[-0.107, 0.33]$\\
    Week $2$  & $0.089$  & $z=0.835$  & $p=0.404$ & $[-0.108, 0.33]$\\
    Week $3$  & $0.096$  & $z=0.889$  & $p=0.374$ & $[-0.105, 0.343]$\\
    Week $4$  & $0.144$  & $z=1.325$  & $p=0.185$ & $[-0.063, 0.397]$\\
    Week $5$  & $-0.049$ & $z=-0.487$ & $p=0.627$ & $[-0.223, 0.165]$\\
    Week $6$  & $0.121$  & $z=1.087$  & $p=0.277$ & $[-0.088, 0.378]$\\
    Week $7$  & $0.022$  & $z=0.215$  & $p=0.83$  & $[-0.164, 0.25]$\\
    Week $8$  & $0.064$  & $z=0.598$  & $p=0.55$  & $[-0.131, 0.303]$\\
    Week $9$  & $-0.006$ & $z=-0.056$ & $p=0.955$ & $[-0.186, 0.214]$\\
    Week $10$ & $-0.113$ & $z=-1.177$ & $p=0.239$ & $[-0.273, 0.083]$\\
    Week $11$ & $-0.075$ & $z=-0.782$ & $p=0.434$ & $[-0.239, 0.124]$\\
    Week $12$ & $0.18$   & $z=1.71$   & $p=0.087$ & $[-0.024, 0.428]$\\
    Week $13$ & $-0.08$  & $z=-0.841$ & $p=0.4$   & $[-0.242, 0.117]$\\
    Week $14$ & $0.098$  & $z=0.933$  & $p=0.351$ & $[-0.098, 0.336]$\\
    Aggregated&$-0.03$   & $z=-0.918$ & $p=0.359$ & $[-0.092, 0.036]$\\
    \bottomrule
    \end{tabular}
    \label{tab:did_audience_replies}
\end{table*}

\begin{table*}[ht]
    \centering
    \setlength{\tabcolsep}{10pt}
    \caption{\textbf{DiD estimates for the number of inbound reposts from the audiences to the accounts in the treatment group.} The estimation is based on 30,112 count observations across \num{2112} accounts. Weekly and account-level fixed effects are included.}
    \begin{tabular}{l*{4}{c}}
    \toprule
    &   DiD estimate   &   $z$  &   $p$   &   95\% CI\\
    \midrule
    Week $-7$ & $-0.029$ & $z=-0.299$ & $p=0.765$ & $[-0.2, 0.178]$\\
    Week $-6$ & $-0.018$ & $z=-0.187$ & $p=0.852$ & $[-0.19, 0.19]$\\
    Week $-5$ & $-0.061$ & $z=-0.638$ & $p=0.523$ & $[-0.225, 0.138]$\\
    Week $-4$ & $0.005$  & $z=0.053$  & $p=0.958$ & $[-0.17, 0.217]$\\
    Week $-3$ & $-0.045$ & $z=-0.472$ & $p=0.637$ & $[-0.213, 0.157]$\\
    Week $-2$ & $0.053$  & $z=0.53$   & $p=0.596$ & $[-0.13, 0.275]$\\
    Week $1$  & $-0.001$ & $z=-0.013$ & $p=0.989$ & $[-0.183, 0.22]$\\
    Week $2$  & $0.047$  & $z=0.453$  & $p=0.651$ & $[-0.142, 0.277]$\\
    Week $3$  & $0.045$  & $z=0.437$  & $p=0.662$ & $[-0.143, 0.274]$\\
    Week $4$  & $0.029$  & $z=0.281$  & $p=0.778$ & $[-0.157, 0.256]$\\
    Week $5$  & $0.027$  & $z=0.256$  & $p=0.798$ & $[-0.164, 0.262]$\\
    Week $6$  & $0.052$  & $z=0.477$  & $p=0.634$ & $[-0.146, 0.296]$\\
    Week $7$  & $-0.052$ & $z=-0.523$ & $p=0.601$ & $[-0.225, 0.159]$\\
    Week $8$  & $-0.004$ & $z=-0.037$ & $p=0.971$ & $[-0.185, 0.218]$\\
    Week $9$  & $-0.039$ & $z=-0.396$ & $p=0.692$ & $[-0.212, 0.171]$\\
    Week $10$ & $-0.117$ & $z=-1.24$  & $p=0.215$ & $[-0.275, 0.075]$\\
    Week $11$ & $-0.13$  & $z=-1.426$ & $p=0.154$ & $[-0.282, 0.054]$\\
    Week $12$ & $0.194$  & $z=1.834$  & $p=0.067$ & $[-0.012, 0.443]$\\
    Week $13$ & $-0.059$ & $z=-0.628$ & $p=0.53$  & $[-0.223, 0.139]$\\
    Week $14$ & $-0.005$ & $z=-0.05$  & $p=0.96$  & $[-0.18, 0.208]$\\
    Aggregated& $0.01$ & $z=0.291$ & $p=0.771$ & $[-0.054, 0.077]$\\
    \bottomrule
    \end{tabular}
    \label{tab:did_audience_reposts}
\end{table*}

\begin{table*}[ht]
    \centering
    \setlength{\tabcolsep}{10pt}
    \caption{\textbf{DiD estimates for the number of inbound views from the audiences to the accounts in the treatment group.} The estimation is based on 63,401 count observations across \num{5125} accounts. Weekly and account-level fixed effects are included.}
    \begin{tabular}{l*{4}{c}}
    \toprule
    &   DiD estimate   &   $z$  &   $p$   &   95\% CI\\
    \midrule
    Week $-7$ & $0.03$   & $z=0.791$  & $p=0.429$ & $[-0.043, 0.108]$\\
    Week $-6$ & $0.084$  & $z=2.141$  & $p=0.032$ & $[0.007, 0.166]$\\
    Week $-5$ & $0.015$  & $z=0.401$  & $p=0.689$ & $[-0.057, 0.093]$\\
    Week $-4$ & $0.05$   & $z=1.308$  & $p=0.191$ & $[-0.024, 0.13]$\\
    Week $-3$ & $0.028$  & $z=0.742$  & $p=0.458$ & $[-0.045, 0.107]$\\
    Week $-2$ & $0.038$  & $z=0.989$  & $p=0.323$ & $[-0.036, 0.118]$\\
    Week $1$  & $0.02$   & $z=0.511$  & $p=0.609$ & $[-0.054, 0.1]$\\
    Week $2$  & $0.02$   & $z=0.516$  & $p=0.606$ & $[-0.055, 0.102]$\\
    Week $3$  & $0.044$  & $z=1.103$  & $p=0.27$  & $[-0.033, 0.128]$\\
    Week $4$  & $0.04$   & $z=1.015$  & $p=0.31$  & $[-0.036, 0.123]$\\
    Week $5$  & $0.024$  & $z=0.588$  & $p=0.557$ & $[-0.053, 0.106]$\\
    Week $6$  & $0.056$  & $z=1.359$  & $p=0.174$ & $[-0.024, 0.141]$\\
    Week $7$  & $0.083$  & $z=2.065$  & $p=0.039$ & $[0.004, 0.168]$\\
    Week $8$  & $0.017$  & $z=0.437$  & $p=0.662$ & $[-0.058, 0.098]$\\
    Week $9$  & $0.021$  & $z=0.537$  & $p=0.591$ & $[-0.054, 0.103]$\\
    Week $10$ & $-0.018$ & $z=-0.47$  & $p=0.638$ & $[-0.091, 0.06]$\\
    Week $11$ & $0.034$  & $z=0.856$  & $p=0.392$ & $[-0.042, 0.117]$\\
    Week $12$ & $0.007$  & $z=0.175$  & $p=0.861$ & $[-0.067, 0.086]$\\
    Week $13$ & $-0.022$ & $z=-0.556$ & $p=0.579$ & $[-0.095, 0.057]$\\
    Week $14$ & $0.045$  & $z=1.111$  & $p=0.267$ & $[-0.033, 0.131]$\\
    Aggregated & $-0.008$ & $z=-0.641$ & $p=0.521$ & $[-0.033, 0.017]$\\
    \bottomrule
    \end{tabular}
    \label{tab:did_audience_views}
\end{table*}

\clearpage

\clearpage
\section{Robustness checks for the effect on account activity}
\label{supp:robustness}

\subsection{Overall effect at daily level and using bootstrap inference}
\label{supp:long_daily}
In the main analysis, we estimate the effect of location transparency using weekly-level count observations to absorb excessive zero counts and potential day-of-week periodic patterns, while also reducing computational cost. Here, to ensure the robustness of our findings, we repeat our estimation of the aggregated effect using daily count observations. The estimation is based on $1,011,801$ count observations across $6,883$ accounts and over 147 days, and the aggregated ATT estimate at the daily level is $-10.9\%$ (ATT~$=-0.109$, $z=-19.195$, $p<0.001$; 95\%~CI: $[-0.12, -0.099]$), which is consistent with the aggregated ATT estimate at the weekly level (ATT~$=-0.131$, $z=-12.534$, $p<0.001$; 95\%~CI: $[-0.15, -0.112]$).

We additionally assess the robustness of statistical inference to within-account dependence. Because the conditional fixed-effects negative binomial estimator used in our analysis does not support cluster-robust standard errors, we re-estimate the standard errors for the pre-post weekly model using a panel-level bootstrap with \num{1000} replications. The aggregated ATT estimate remains robust and consistent with our main analysis (ATT~$=-0.131$, $z=-6.018$, $p<0.001$; 95\%~CI: $[-0.170, -0.091]$).

\subsection{Robustness checks with synthetic DiD}
\label{supp:sdid}

\begin{figure}[h]
    \centering
    \captionsetup[subfloat]{font={bf, small}, skip=0pt, singlelinecheck=false, labelformat=simple, position=top}
    \subfloat[]{\includegraphics[width = .48\textwidth]{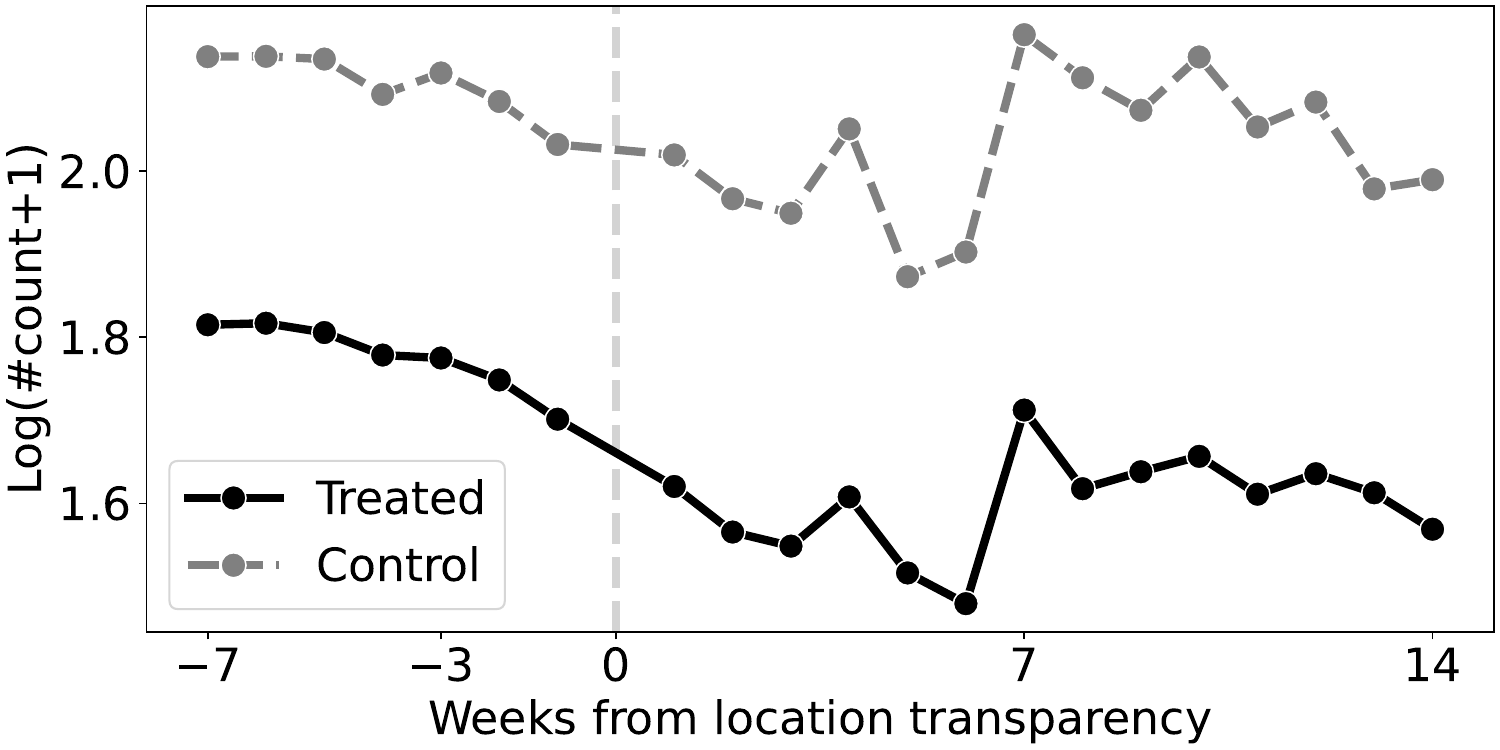}}
    \hfill
    \subfloat[]{\includegraphics[width = .48\textwidth]{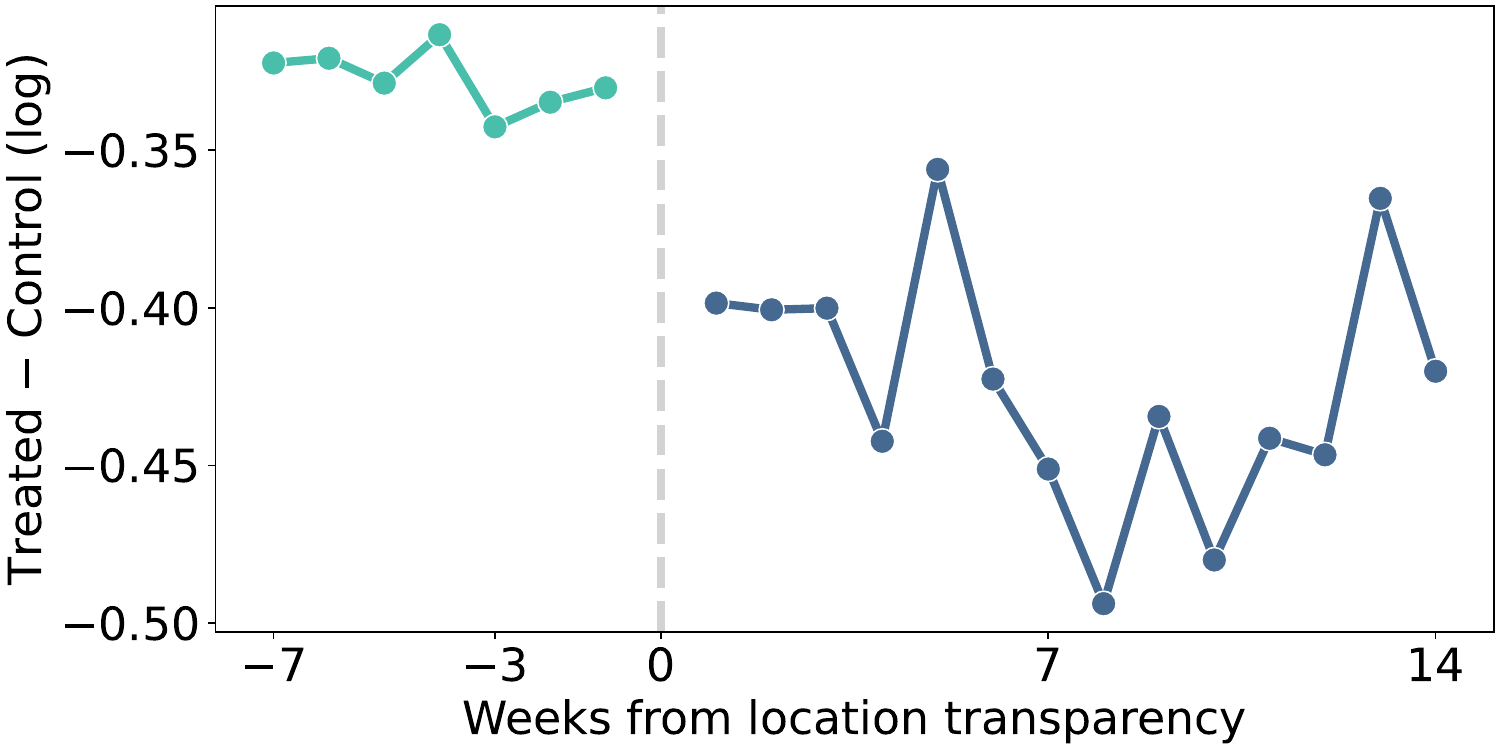}}

    \caption{\textbf{The estimation results of synthetic DID.} \textbf{(a)}~The SDiD-estimated changes in the number of original posts and outbound replies between treatment and synthetic control groups over time. \textbf{(b)}~The SDiD-estimated differences in the number of original posts and outbound replies between treatment and synthetic control groups. The estimation is based on 144,543 count observations across 6,883 accounts with 100 bootstraps.}
    \label{fig:sdid_account_weekly}
\end{figure}

While the weekly ATT estimates from the leads-and-lags DiD approach statistically support the parallel trends assumption (see \Cref{tab:did_main_account}), a slight pre-disclosure downward trend is visible in \Cref{fig:main}a, and a mild violation of the assumption is observed in \Cref{fig:account_post_replies}b. We therefore conduct two robustness checks using log-transformed count variable (\ie, the number of original posts and outbound replies per week): log-transformed pre-post DiD estimation and Synthetic DiD (SDiD) estimation that relaxes the requirement of parallel trends. First, we estimate the effect of location transparency via a log-transformed pre-post DiD, yielding an estimate of $-0.103$ ($t=-11.086$, $p<0.001$; 95\%~CI: $[-0.121, -0.085]$). The log-transformed DiD estimate can provide a comparable baseline for SDiD. Second, we apply SDiD to the log-transformed variable, obtaining a consistent estimate of $-0.096$ ($t=-4.77$, $p<0.001$; 95\%~CI: $[-0.136, -0.057]$). The convergence of these estimates indicates the robustness of our main findings. Additionally, \Cref{fig:sdid_account_weekly}a shows the estimated changes in the treatment and synthetic control groups over time, and \Cref{fig:sdid_account_weekly}b shows the additional changes in the treatment group relative to the synthetic control group. The two panels reveal a sharp decline in the number of original posts and replies following the introduction of location transparency, further supporting our results. 

Meanwhile, we use SDiD to estimate the separated effects of location transparency by post types. For original posts, the log-transformed pre-post DiD estimate is $-0.087$ ($t=-11.136$, $p<0.001$; 95\%~CI: $[-0.102, -0.071]$), and the SDiD estimate is $-0.086$ ($t=-5.51$, $p<0.001$; 95\%~CI: $[-0.117, -0.056]$). For outbound replies, the log-transformed pre-post DiD estimate is $-0.086$ ($t=-8.89$, $p<0.001$; 95\%~CI: $[-0.105, -0.067]$), and the SDiD estimate is $-0.075$ ($t=-3.82$, $p<0.001$; 95\%~CI: $[-0.114, -0.037]$). Taken together, the introduction of location transparency has consistent and robust reduction effects on both original posts and outbound replies.

\clearpage
\subsection{Robustness check with propensity score matching}
\label{supp:psm}

\begin{figure}[ht]
    \centering
    \captionsetup[subfloat]{font={bf, small}, skip=0pt, singlelinecheck=false, labelformat=simple, position=top}
    \subfloat[]{\includegraphics[width = .43\textwidth]{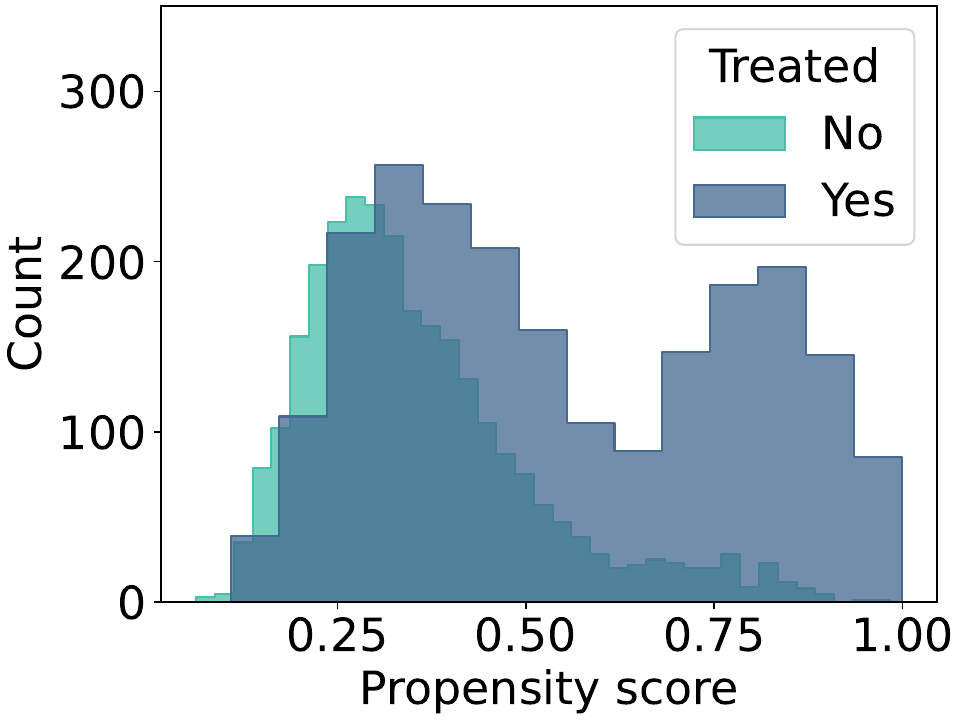}}
    \hspace{1cm}
    \subfloat[]{\includegraphics[width = .43\textwidth]{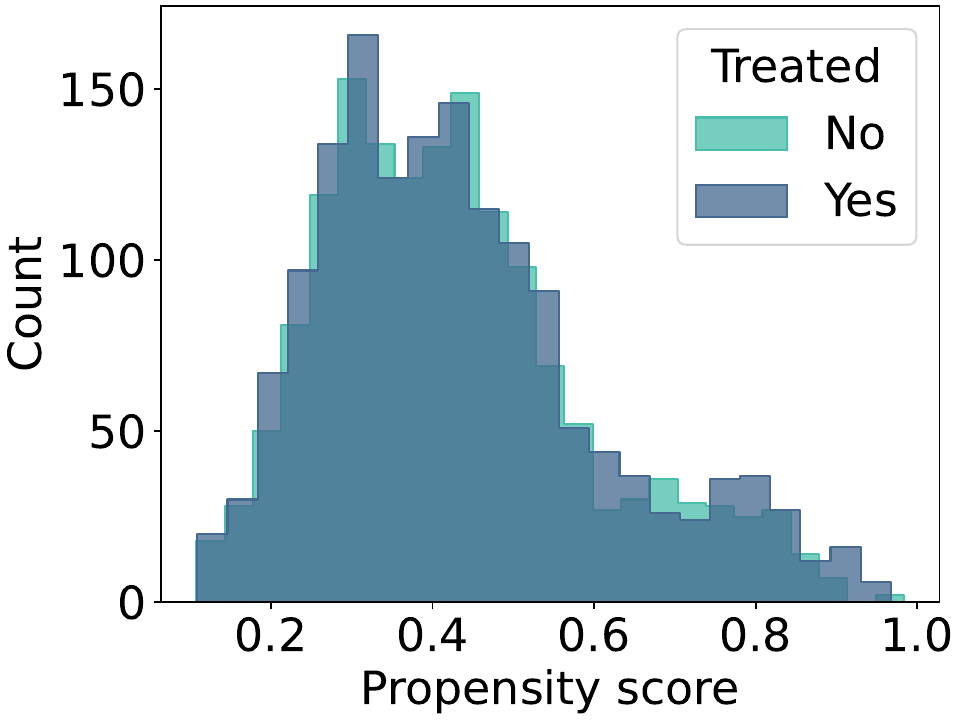}}
    \caption{\textbf{The distributions of propensity scores.} \textbf{(a)}~The distributions of propensity scores in the treatment and control groups before matching. \textbf{(b)}~The distributions of propensity scores in the treatment and control groups after matching.}
    \label{fig:ps_matching}
\end{figure}

We use a series of extracted variables across author characteristics and their posting preferences to conduct propensity score matching, further mitigating potential confounding between the treatment and control groups. Before the matching, the propensity scores in the treatment and control groups have significantly different distributions ($t(4935)=-33.85$, $p<0.001$; \Cref{fig:ps_matching}a). After the matching, the distributions of propensity scores in the treatment and control groups are well balanced and have no significant difference ($t(3092)=1.297$, $p=0.195$; \Cref{fig:ps_matching}b). This results in \num{3094} accounts (\num{1547} treated-control pairs) .

\begin{figure}[ht]
    \centering
    \captionsetup[subfloat]{font={bf, small}, skip=0pt, singlelinecheck=false, labelformat=simple, position=top}
    \subfloat{\includegraphics[width = \textwidth]{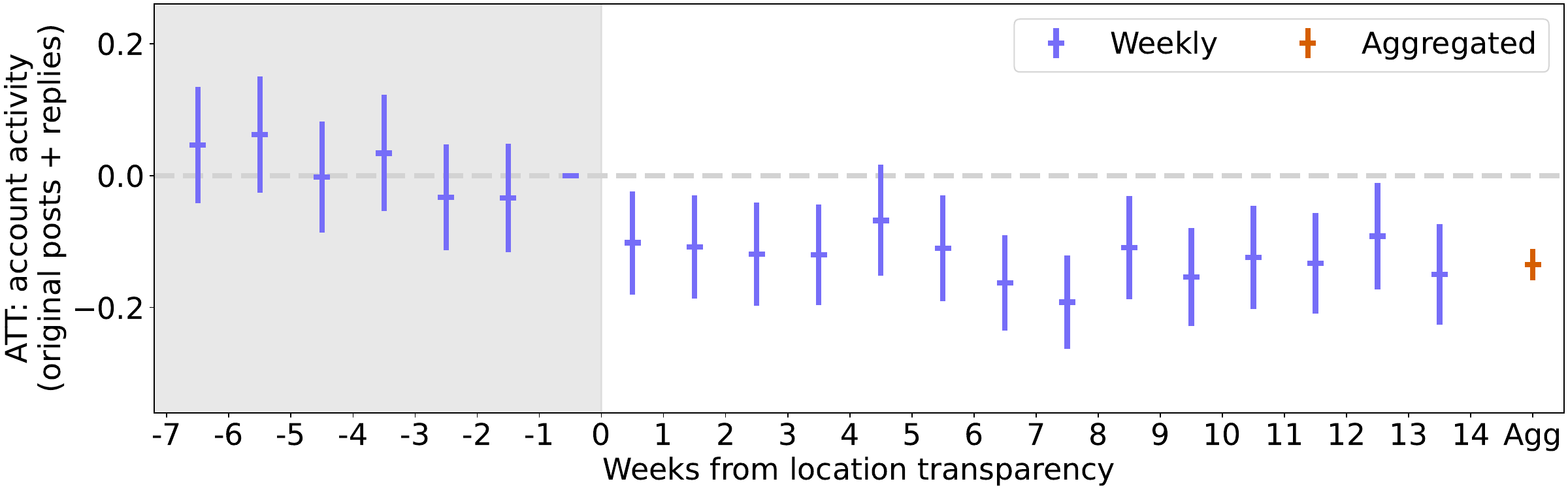}}
    
    \caption{\textbf{ATT estimates for the impact of location transparency on account activity after propensity score matching.} Weekly and aggregated ATT estimates for the number of original posts and replies from the selected accounts. The error bars represent 95\% CIs. The full estimation results are reported in \Cref{tab:did_total_posts_psm}.}
    \label{fig:did_psm}
\end{figure}

As a robustness check, we conduct our DiD estimation after propensity score matching. \Cref{fig:did_psm} shows the weekly and aggregated ATT estimates for the impact of location transparency on account activity. Overall, the treatment effect is $-13.5\%$ (ATT~$=-0.135$, $z=-10.144$, $p<0.001$; 95\%~CI: $[-0.159, -0.11]$), which is consistent with the estimated effect in our main analysis.
 
\begin{table*}[ht]
    \centering
    \setlength{\tabcolsep}{10pt}
    \caption{\textbf{DiD estimates for the number of original posts and replies from accounts in the treatment group after propensity score matching.} The estimation is based on 64,974 count observations across 3,094 accounts. Weekly and account-level fixed effects are included.}
    \begin{tabular}{l*{4}{c}}
    \toprule
    &   DiD estimate   &   $z$  &   $p$   &   95\% CI\\
    \midrule
    Week $-7$ & $0.046$  & $z=1.047$  & $p=0.295$ & $[-0.038, 0.137]$\\
    Week $-6$ & $0.062$  & $z=1.405$  & $p=0.16$  & $[-0.023, 0.154]$\\
    Week $-5$ & $-0.002$ & $z=-0.043$ & $p=0.966$ & $[-0.082, 0.085]$\\
    Week $-4$ & $0.034$  & $z=0.785$  & $p=0.432$ & $[-0.049, 0.126]$\\
    Week $-3$ & $-0.033$ & $z=-0.787$ & $p=0.431$ & $[-0.111, 0.052]$\\
    Week $-2$ & $-0.034$ & $z=-0.788$ & $p=0.43$  & $[-0.112, 0.052]$\\
    Week $1$  & $-0.102$ & $z=-2.422$ & $p=0.015$ & $[-0.178, -0.02]$\\
    Week $2$  & $-0.108$ & $z=-2.532$ & $p=0.011$ & $[-0.184, -0.026]$\\
    Week $3$  & $-0.119$ & $z=-2.789$ & $p=0.005$ & $[-0.194, -0.037]$\\
    Week $4$  & $-0.12$  & $z=-2.873$ & $p=0.004$ & $[-0.194, -0.04]$\\
    Week $5$  & $-0.068$ & $z=-1.521$ & $p=0.128$ & $[-0.148, 0.02]$\\
    Week $6$  & $-0.11$  & $z=-2.496$ & $p=0.013$ & $[-0.187, -0.025]$\\
    Week $7$  & $-0.163$ & $z=-4.029$ & $p<0.001$ & $[-0.232, -0.087]$\\
    Week $8$  & $-0.192$ & $z=-4.758$ & $p<0.001$ & $[-0.26, -0.118]$\\
    Week $9$  & $-0.109$ & $z=-2.578$ & $p=0.01$  & $[-0.184, -0.027]$\\
    Week $10$ & $-0.154$ & $z=-3.752$ & $p<0.001$ & $[-0.225, -0.077]$\\
    Week $11$ & $-0.124$ & $z=-2.914$ & $p=0.004$ & $[-0.198, -0.042]$\\
    Week $12$ & $-0.133$ & $z=-3.156$ & $p=0.002$ & $[-0.206, -0.053]$\\
    Week $13$ & $-0.092$ & $z=-2.115$ & $p=0.034$ & $[-0.17, -0.007]$\\
    Week $14$ & $-0.15$  & $z=-3.551$ & $p<0.001$ & $[-0.223, -0.07]$\\
    Aggregated & $-0.135$ & $z=-10.144$ & $p<0.001$ & $[-0.159, -0.11]$\\
    \bottomrule
    \end{tabular}
    \label{tab:did_total_posts_psm}
\end{table*}

\clearpage
\subsection{Alternative pre-treatment periods}
\label{supp:alter_pre_treat}

\begin{figure}[ht]
    \centering
    \captionsetup[subfloat]{font={bf, small}, skip=0pt, singlelinecheck=false, labelformat=simple, position=top}
    \subfloat[]{\includegraphics[width = .43\textwidth]{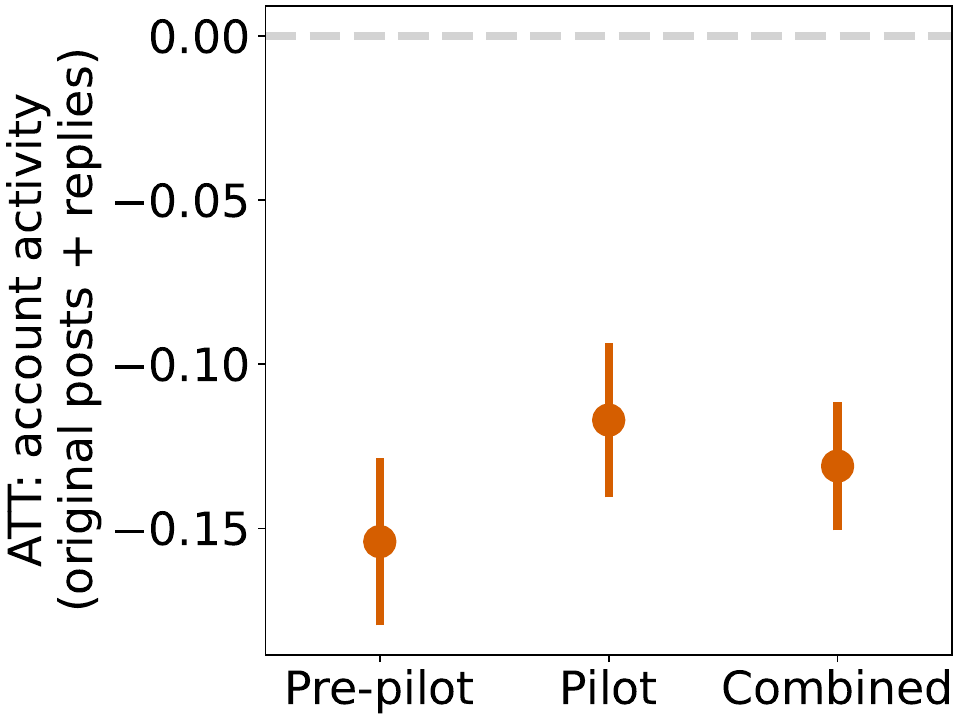}}
    \hspace{1cm}
    \subfloat[]{\includegraphics[width = .43\textwidth]{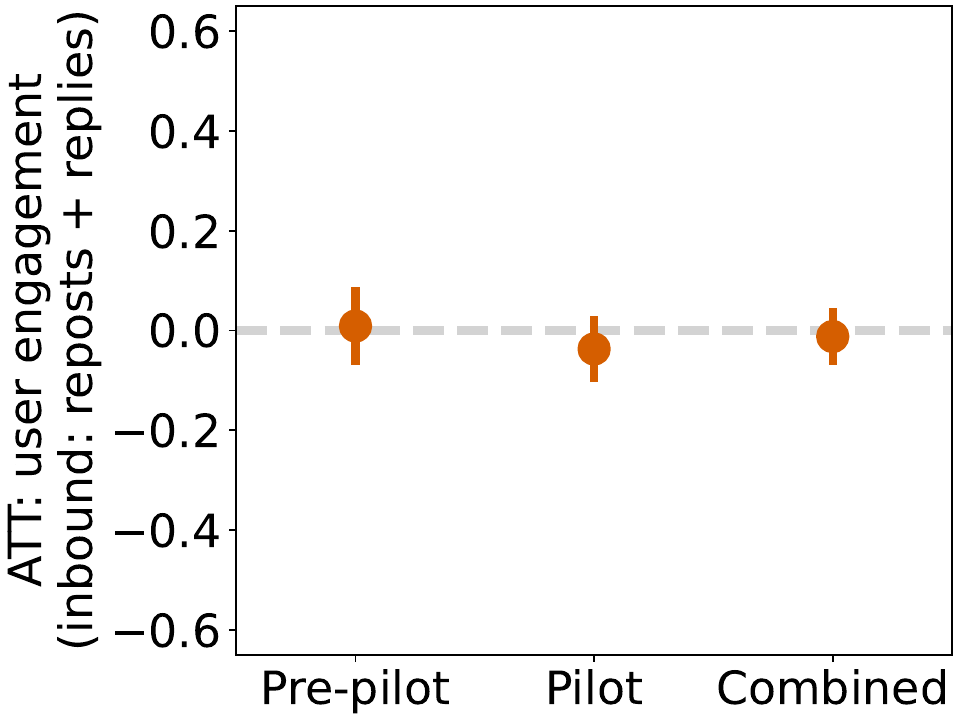}}
    \caption{\textbf{Aggregated ATT estimates across different pre-treatment periods.} \textbf{(a)}~Aggregated ATT estimates for the number of original posts and replies from the accounts in the treatment group. \textbf{(b)}~Aggregated ATT estimates for the number of inbound reposts and replies from the audiences to the accounts in the treatment group. The error bars represent 95\% CIs.}
    \label{fig:alter_pre_treat}
\end{figure}

We note that \X began piloting the account transparency feature on some \X's team members in late October 2025~\cite{bier2025exp}, nearly one month before the official roll-out of the feature. This pilot should not have affected accounts in the treatment group; however, the mild pre-trend detected in the pre-disclosure period may be related to it. Given this, we split the seven-week pre-intervention period into two sub-periods, pre-pilot (Weeks $-7$ to $-5$) and pilot (Weeks $-4$ to $-1$), and check the robustness of the treatment effect across different pre-treatment periods. \Cref{fig:alter_pre_treat}a shows that the ATT estimates for the changes in account activity are consistent and robust across pre-pilot (ATT~$=-0.154$, $z=-11.15$, $p<0.001$; 95\%~CI: $[-0.179, -0.129]$), pilot (ATT~$=-0.117$, $z=-9.238$, $p<0.001$; 95\%~CI: $[-0.14, -0.093]$), and combined periods (ATT~$=-0.131$, $z=-12.534$, $p<0.001$; 95\%~CI: $[-0.15, -0.112]$). Similarly, the ATT estimates for the audiences' engagement are consistent and robust across pre-pilot (ATT~$=0.009$, $z=0.222$, $p=0.824$; 95\%~CI: $[-0.067, 0.09]$), pilot (ATT~$=-0.037$, $z=-1.084$, $p=0.278$; 95\%~CI: $[-0.101, 0.031]$), and combined periods (ATT~$=-0.012$, $z=-0.412$, $p=0.681$; 95\%~CI: $[-0.067, 0.046]$).

\clearpage
\subsection{Robustness check with API count endpoints}
\label{supp:api_count}

In our main analysis, we count all collected original posts and replies to examine the impact of location transparency, allowing us to understand not only overall activity levels but also the underlying mechanisms in terms of author profiles and content characteristics. However, we recognise that collecting specific posts may omit deleted or suspended content. To ensure the robustness of our findings, we query \X's count endpoint with full archive access via its API, collecting time series post counts for the accounts in the treatment and control groups spanning four weeks before and four weeks after the transparency date. Here, we collect the time series counts for the combination of original posts and replies (corresponding to our main analysis), as well as for reposts alone to expand our findings to location-mismatched users' outbound reposting behaviour.

\begin{figure}[ht]
    \centering
    \captionsetup[subfloat]{font={bf, small}, skip=0pt, singlelinecheck=false, labelformat=simple, position=top}
    \subfloat[]{\includegraphics[width = .48\textwidth]{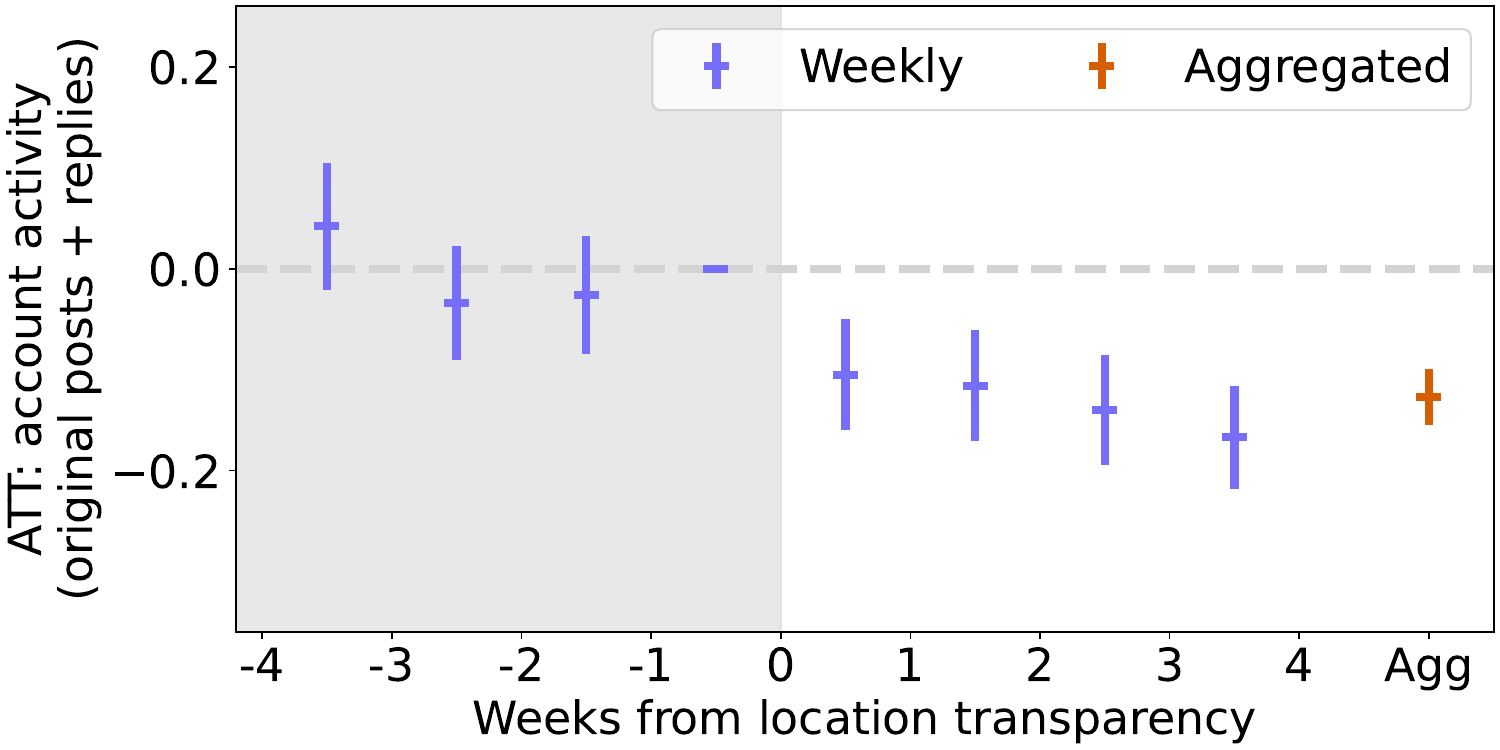}}
    \hfill
    \subfloat[]{\includegraphics[width = .48\textwidth]{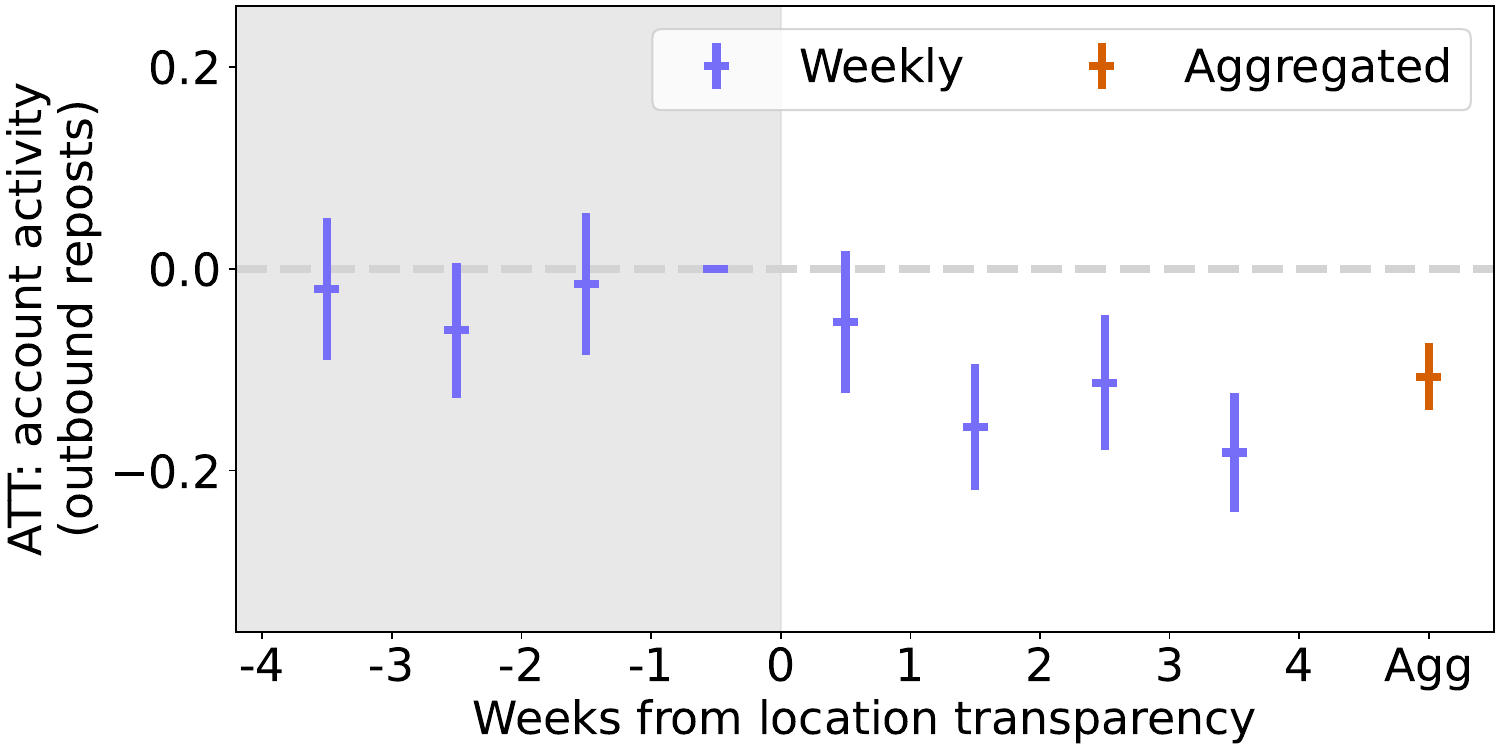}}

    \subfloat[]{\includegraphics[width = .48\textwidth]{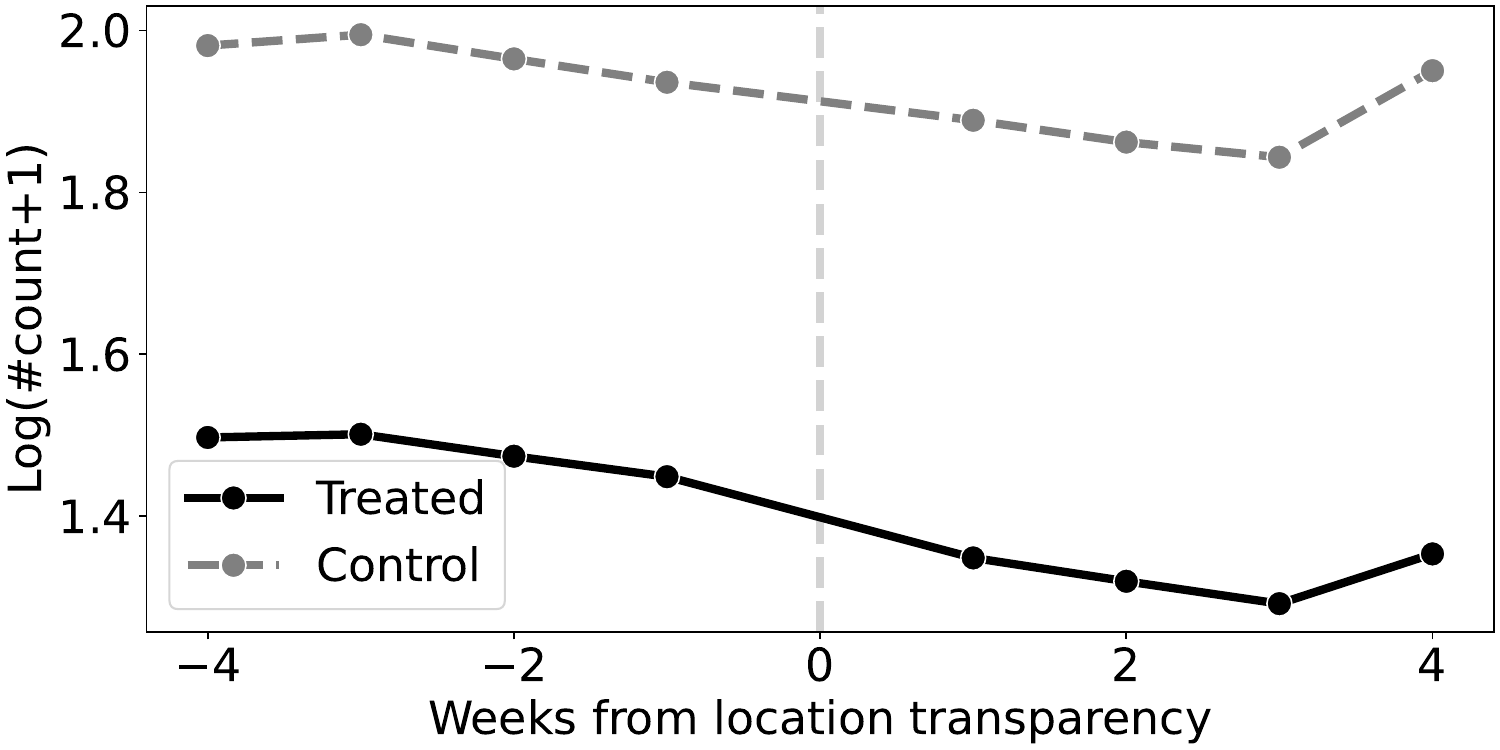}}
    \hfill
    \subfloat[]{\includegraphics[width = .48\textwidth]{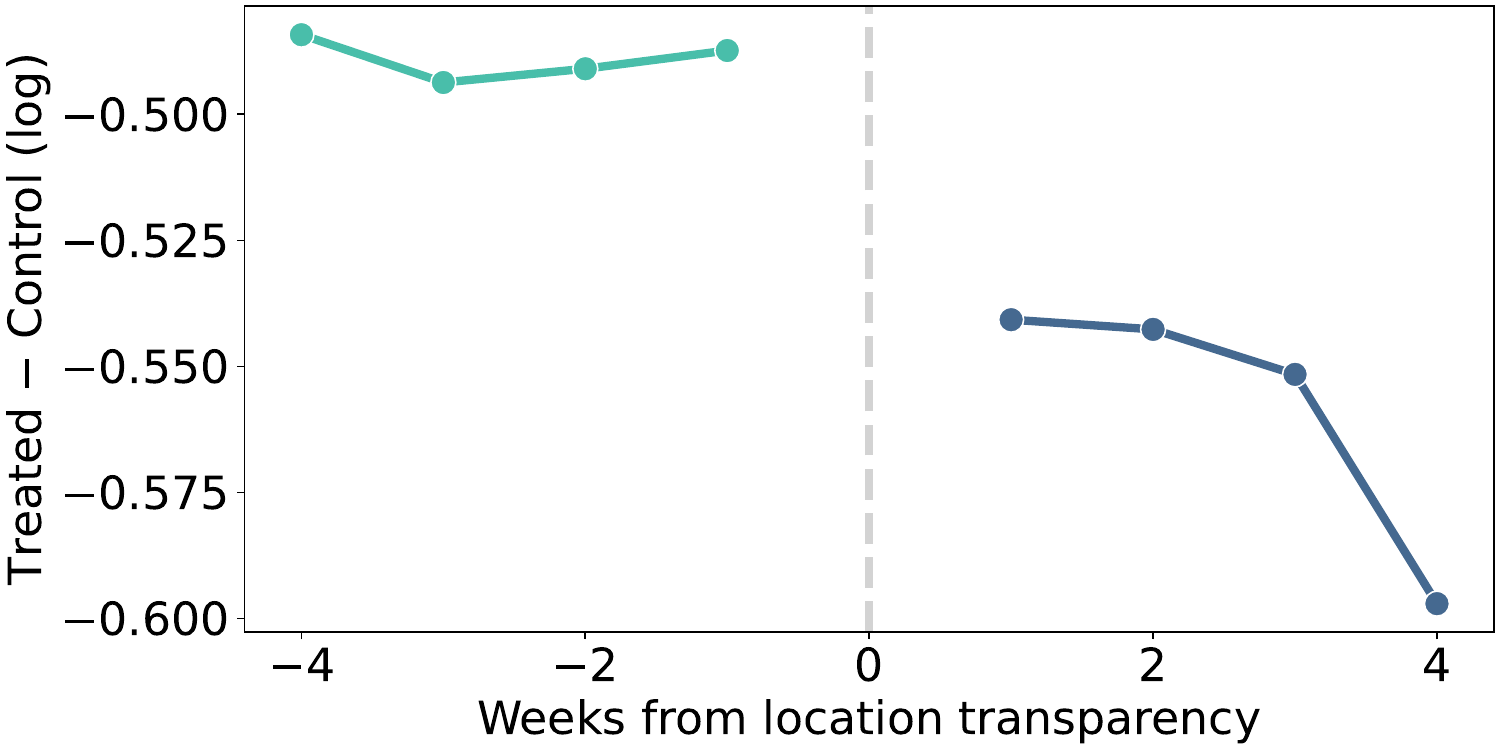}}

    \caption{\textbf{Robustness check based on \X's API-returned counts.} \textbf{(a)}~The weekly and aggregated ATT estimates for the number of original posts and outbound replies from the accounts in the treatment group (corresponding to the main analysis). The weekly and aggregated ATT estimations are based on 50,616 count observations across 6,327 accounts. \textbf{(b)}~The weekly and aggregated ATT estimates for the number of outbound reposts from the accounts in the treatment group (post type additional to the main analysis). The weekly and aggregated ATT estimations are based on 41,072 count observations across 5,134 accounts. \textbf{(c)}~The SDiD-estimated changes in the number of original posts and outbound replies between treatment and synthetic control groups over time. The estimation is based on \num{63672} log-transformed count observations across \num{7959} accounts. \textbf{(d)}~The SDiD-estimated differences in the number of original posts and outbound replies between treatment and synthetic control groups. In (a) and (b), the error bars represent 95\% CIs.}
    \label{fig:api_count}
\end{figure}

\vspace{1em}
\noindent \textbf{API-based count estimation.}\\
We use the same DiD approach (see \nameref{sec:did_spec}) to estimate the impact of location transparency based on the API-returned counts. We first examine the changes in the number of original posts and outbound replies over time. \Cref{fig:api_count}a shows that the weekly ATT estimates before the location transparency are consistently and statistically insignificant: Week~$-4$: $0.042$, $z=1.336$, $p=0.182$, 95\%~CI: $[-0.019, 0.106]$; Week~$-3$: $-0.034$, $z=-1.157$, $p=0.247$, 95\%~CI: $[-0.09, 0.025]$; and Week~$-2$: $-0.026$, $z=-0.877$, $p=0.38$, 95\%~CI: $[-0.083, 0.034]$. After the location transparency, the weekly ATT estimates are statistically significant and show a clear decreasing trend: Week~1: $-0.105$, $z=-3.485$, $p<0.001$, 95\%~CI: $[-0.159, -0.047]$; Week~2: $-0.116$, $z=-3.835$, $p<0.001$, 95\%~CI: $[-0.17, -0.058]$; Week~3: $-0.14$, $z=-4.656$, $p<0.001$, 95\%~CI: $[-0.193, -0.084]$; and Week~4: $-0.167$, $z=-5.774$, $p<0.001$, 95\%~CI: $[-0.217, -0.114]$. Overall, the implementation of location transparency reduces the number of original posts and outbound replies in the treatment group by 12.7\% (ATT~$=-0.127$, $z=-8.619$, $p<0.001$; 95\%~CI: $[-0.154, -0.1]$)

Additionally, we examine whether location transparency affects the reposting behaviour of location-mismatched accounts (see \Cref{fig:api_count}b), finding a similar effect to that observed for original posts and replies. Before the location transparency, the weekly ATT estimates are consistently and statistically indistinguishable from zero: Week~$-4$: $-0.02$, $z=-0.562$, $p=0.574$, 95\%~CI: $[-0.088, 0.052]$; Week~$-3$: $-0.061$, $z=-1.741$, $p=0.082$, 95\%~CI: $[-0.126, 0.008]$; and Week~$-2$: $-0.015$, $z=-0.406$, $p=0.684$, 95\%~CI: $[-0.083, 0.059]$. Following the location transparency, the weekly ATT estimates show a decreasing trend: Week~1: $-0.053$, $z=-1.432$, $p=0.152$, 95\%~CI: $[-0.121, 0.02]$; Week~2: $-0.157$, $z=-4.509$, $p<0.001$, 95\%~CI: $[-0.217, -0.092]$; Week~3: $-0.113$, $z=-3.17$, $p=0.002$, 95\%~CI: $[-0.177, -0.045]$; and Week~4: $-0.182$, $z=-5.389$, $p<0.001$, 95\%~CI: $[-0.239, -0.12]$. Overall, the treatment effect on the outbound reposts from the location-mismatched accounts is, on average, -10.7\% (ATT~$=-0.107$, $z=-6.091$, $p<0.001$; 95\%~CI: $[-0.139, -0.074]$).

Taken together, our findings show that location transparency reduces activity of location-mismatched accounts on \X across original posts, replies, and reposts.

\vspace{1em}
\noindent \textbf{Synthetic DiD estimation.}\\
We log-transform the number of original posts and replies returned by the count API and repeat our estimation using Synthetic DiD (SDiD) with 100 bootstraps. The estimate for the overall effect is significantly negative (ATT(log) $=-0.070$, $t=-5.11$, $p<0.001$; 95\%~CI: $[-0.096, -0.043]$). \Cref{fig:api_count}c shows the estimated changes in the treatment and synthetic control groups over time, and \Cref{fig:api_count}d shows the additional changes in the treatment group relative to the synthetic control group. We observe a sharp decrease in the number of original posts and replies following the location transparency, further supporting our main findings.

\clearpage
\subsection{Analysis of account status}
\label{supp:account_status}

\begin{figure}[ht]
    \centering
    \captionsetup[subfloat]{font={bf, small}, skip=0pt, singlelinecheck=false, labelformat=simple, position=top}
    \subfloat[]{\includegraphics[width = .48\textwidth]{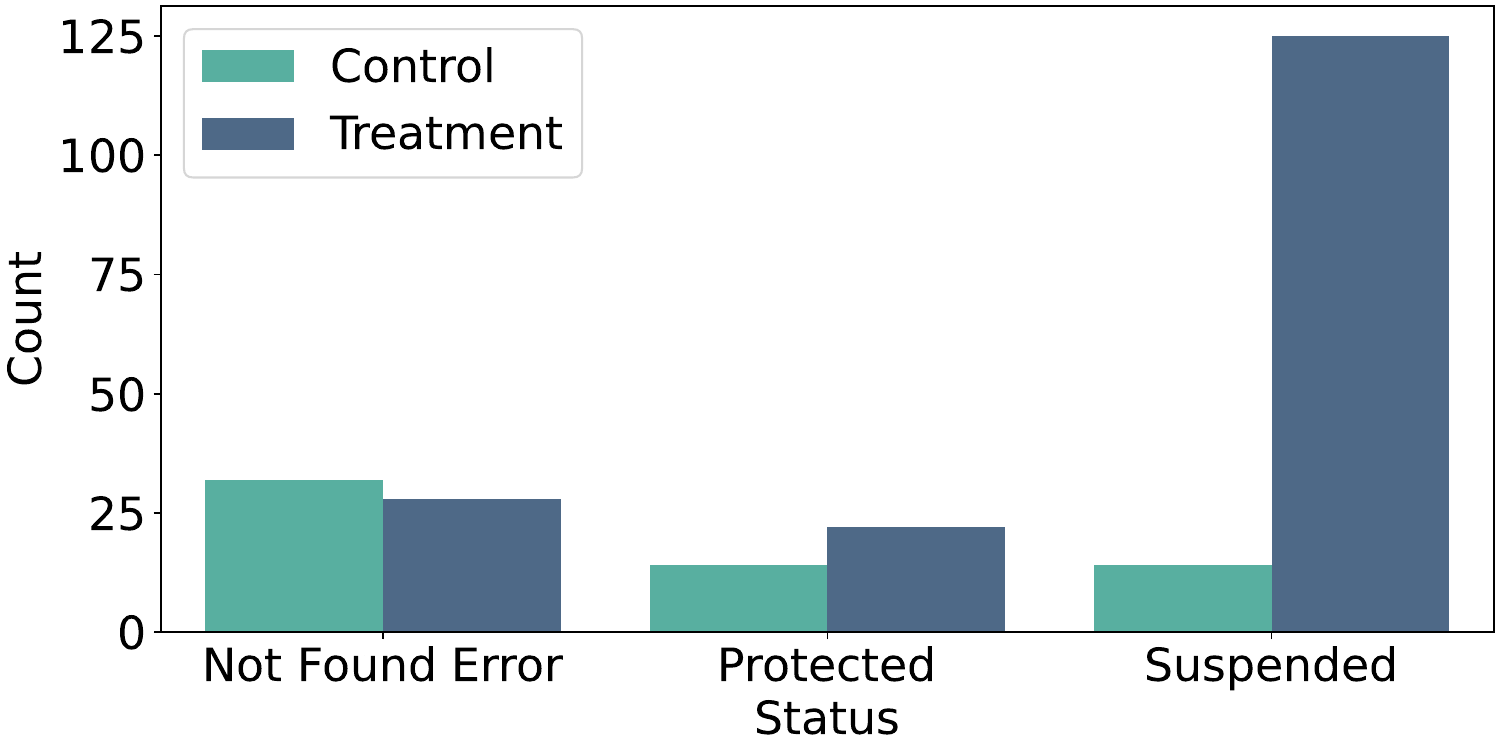}}
    \hfill
    \subfloat[]{\includegraphics[width = .48\textwidth]{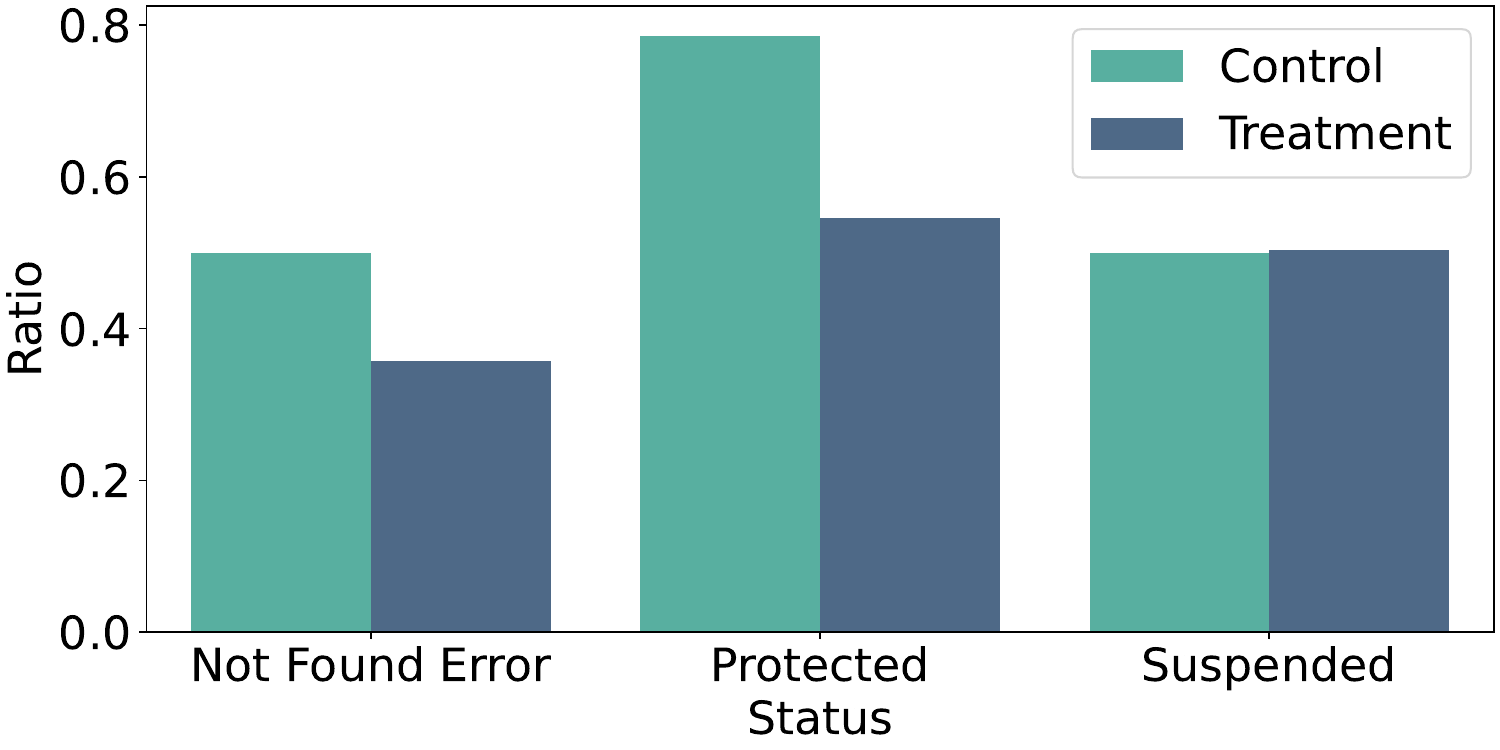}}

    \caption{\textbf{Analysis of account status.} \textbf{(a)}~The count distribution of accounts that were not found (deleted), protected, or suspended in the treatment and control groups based on the API account status query after the post collection. \textbf{(b)}~The ratios of deleted/protected/suspended accounts from which posts were successfully collected during post collection, relative to the total number of accounts with the corresponding status.}
    \label{fig:account_status}
\end{figure}

Out of the 8,200 accounts in the treatment and control groups, only 241 accounts (2.9\%) have no time-series count data returned (\ie, they were protected, suspended, or deleted during count collection). We again use the \X's API to query the statuses of these accounts. It returns three status types: Not Found Error (Deleted), Protected, or Suspended. We find that the accounts in the treatment group have higher numbers regarding suspended and protected accounts, while lower number in deleted accounts (\Cref{fig:account_status}a). Given that \X's API does not return the exact timestamps for the account suspensions, we cannot directly measure the potential effect of location transparency on the suspension by the platform. Alternatively, we further analyse the ratios of suspended/protected/deleted accounts from which the posts were successfully collected during the post collection (\ie, these accounts were deleted/suspended/protected after the location disclosure), relative to the total accounts with corresponding status. \Cref{fig:account_status}b indicates that accounts in the control group are more likely to be deleted or protected (0.5 and 0.786, respectively) after the location transparency compared to the accounts in the treatment group (0.357 and 0.545, respectively), while there is only a tiny difference in suspended accounts (0.5 in the control group vs. 0.504 in the treatment group). However, this does not necessarily mean that the implementation of transparency feature has no effect on account suspension by the platform, as we collected the dataset weeks after the implementation of transparency feature. We advocate for future research to investigate the causal effect of location transparency on account suspension by collecting data with finer temporal resolution around the time of the transparency intervention.

\clearpage
\section{Heterogeneity effects of location transparency}
\label{supp:he_effects}

\subsection{Heterogeneity by revealed non-\US locations}
\label{supp:he_location}
\begin{figure}[ht]
    \centering
    \captionsetup[subfloat]{font={bf, small}, skip=0pt, singlelinecheck=false, labelformat=simple, position=top}
    \subfloat[]{\includegraphics[width = .9\textwidth]{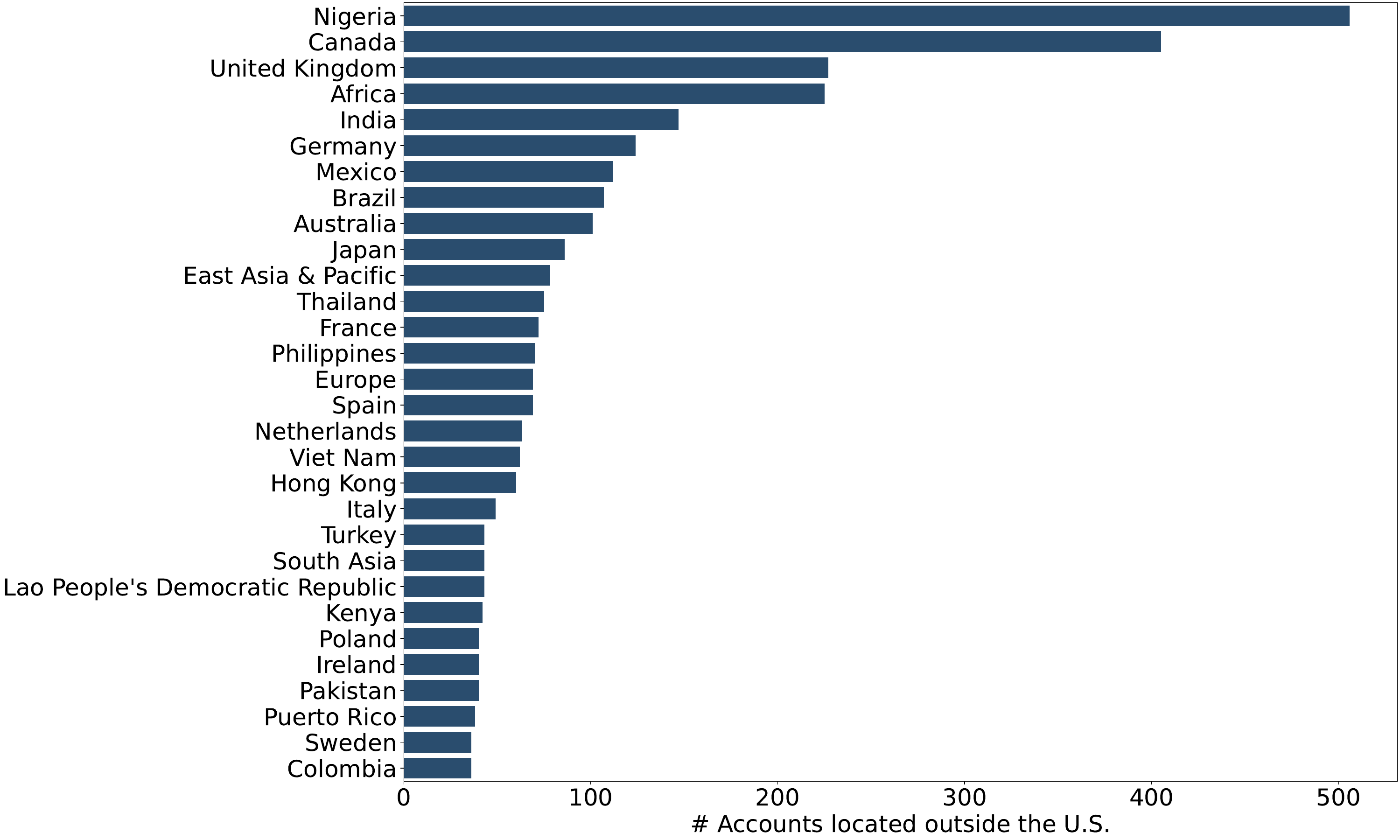}}
    \\
    \subfloat[]{\includegraphics[width = 0.5\textwidth]{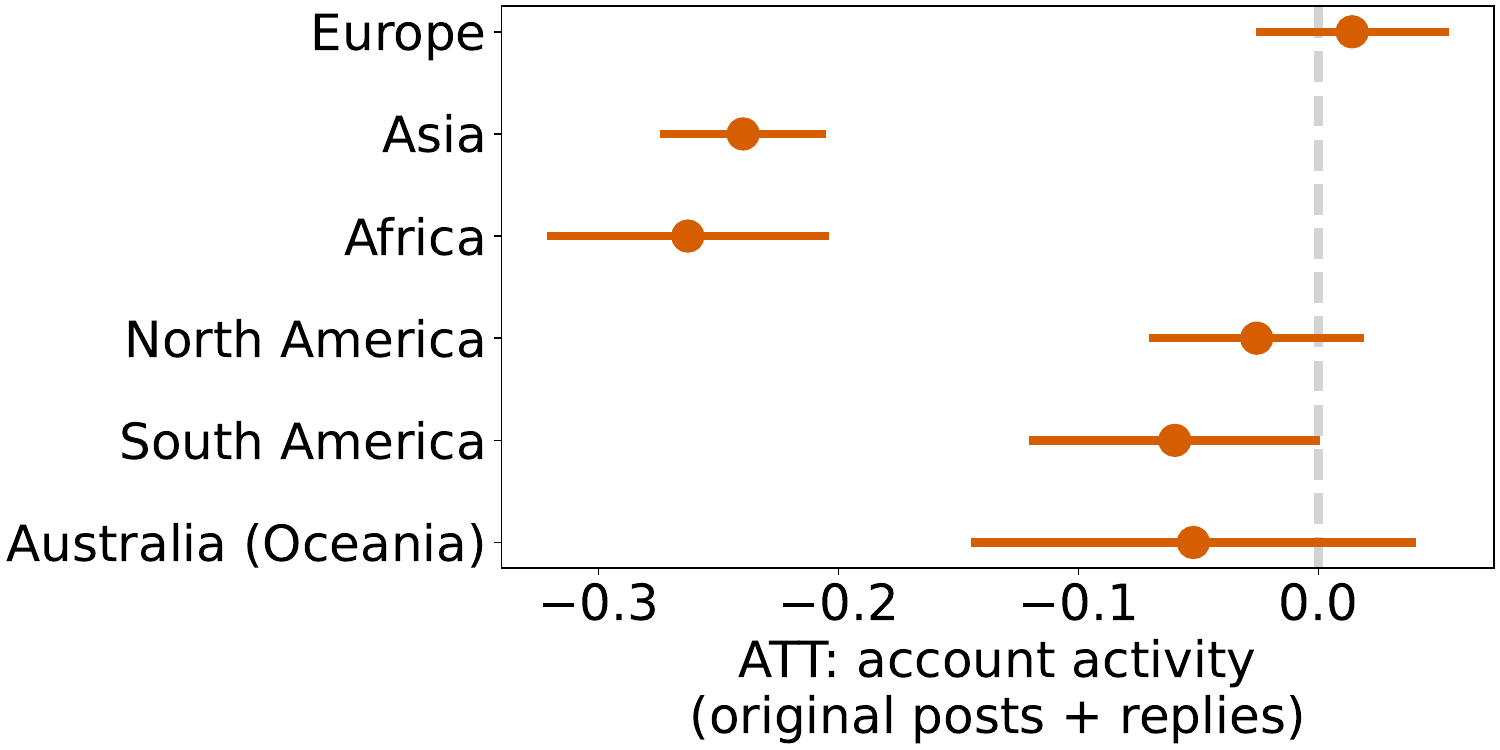}}
    \caption{\textbf{Distribution of account locations and heterogeneity effects of location transparency across areas.} \textbf{(a)}~Distribution of platform-disclosed non-\US locations among location-mismatched accounts. Shown are the 30 most frequently disclosed countries or regions, ranked by the number of accounts. \textbf{(b)}~Aggregated ATT estimates for the number of original posts and replies from location-mismatched accounts in the treatment group and without the use of VPN across areas. The error bars represent 95\% CIs.}
    \label{fig:he_location_supp}
\end{figure}

The distribution of account locations and heterogeneity effects of location transparency across areas are shown in \Cref{fig:he_location_supp}.

\clearpage
\subsection{Heterogeneity by account features and posting behaviours}
\label{supp:he_account}

\begin{figure}[ht]
    \centering
    \includegraphics[width=\linewidth]{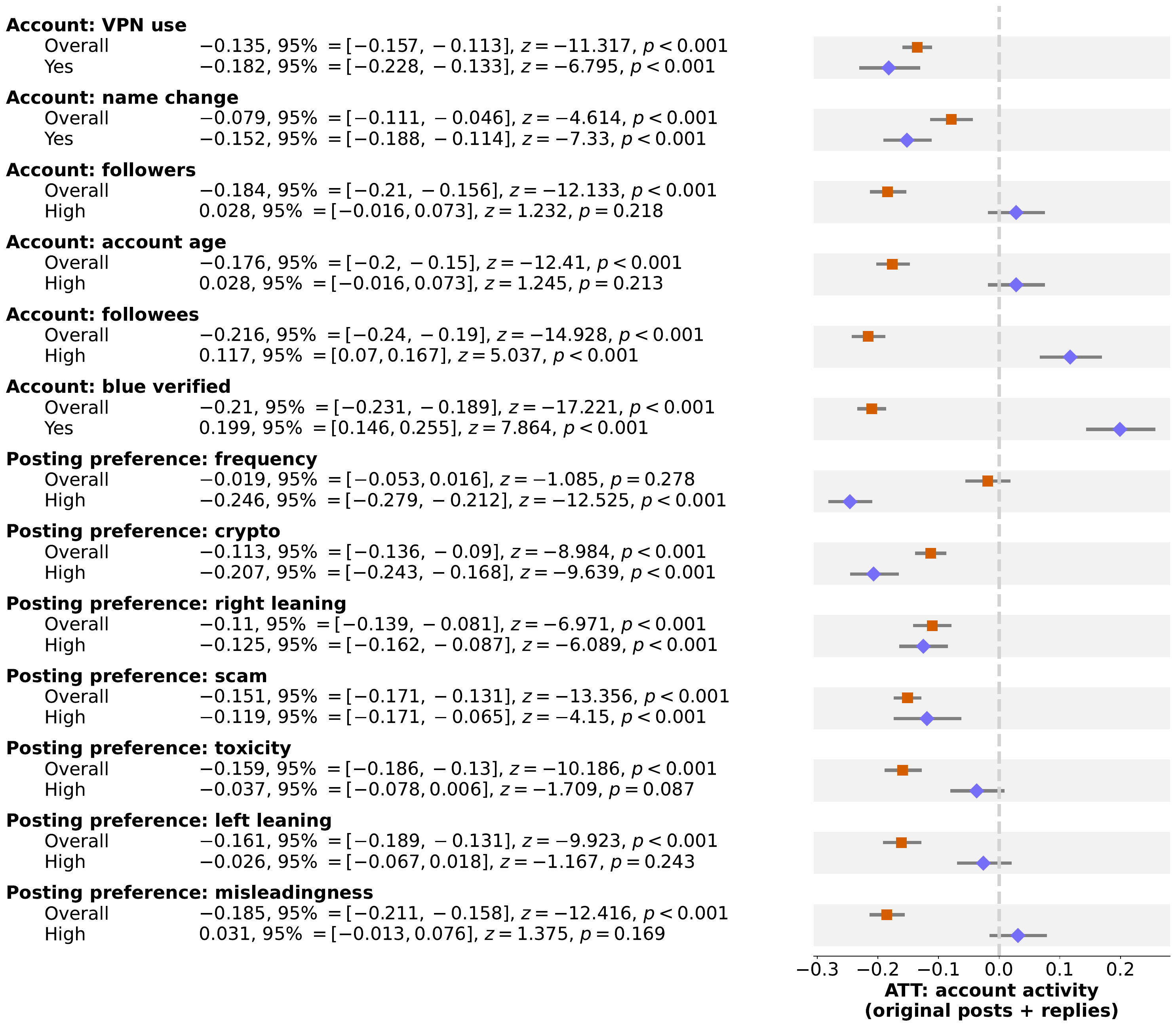}
    \caption{\textbf{Heterogeneity effects of location transparency on account activity by interactions.} Shown are overall aggregated ATT estimates and moderation effects across account characteristics and posting preferences. All estimations are based on 103,677 count observations across 4,937 accounts. The error bars represent 95\% CIs.}
    \label{fig:he_sensitivity_account_interact}
\end{figure}

The summary of post count observations and accounts for each separation by account characteristics and posting preferences are reported in \Cref{tab:he_account}. We further compare the magnitude of treatment effects across author profiles and posting preferences through interaction analysis and moderation effect estimation to support our analysis in the main text (see \Cref{fig:he_sensitivity_account_interact}).

\begin{table}[ht]
    \centering
    \footnotesize
    \caption{\textbf{Summary of count observations and accounts across account characteristics and posting preferences.}}
    \begin{tabular}{lcc}
    \toprule
    {Group} & {\# Count observations} & {\# Accounts}\\
    \midrule
    \multicolumn{3}{l}{\underline{Account: followers}}\\
    \quad High & {$51,828$} & {$2,468$} \\
    \quad Low  & {$51,849$} & {$2,469$} \\
    \multicolumn{3}{l}{\underline{Account: followees}}\\
    \quad High & {$51,828$} & {$2,468$} \\
    \quad Low  & {$51,849$} & {$2,469$} \\
    \multicolumn{3}{l}{\underline{Account: blue verified}}\\
    \quad Yes  & {$28,266$} & {$1,346$} \\
    \quad No   & {$75,411$} & {$3,591$} \\
    \multicolumn{3}{l}{\underline{Account: account age}}\\
    \quad High & {$51,828$} & {$2,468$} \\
    \quad Low  & {$51,849$} & {$2,469$} \\
    \multicolumn{3}{l}{\underline{Account: VPN use}}\\
    \quad Yes  & {$19,677$} & {$937$} \\
    \quad No   & {$84,000$} & {$4,000$} \\
    \multicolumn{3}{l}{\underline{Account: name change}}\\
    \quad Yes  & {$62,874$} & {$2,994$} \\
    \quad No   & {$40,803$} & {$1,943$} \\
    \multicolumn{3}{l}{\underline{Posting preference: frequency}}\\
    \quad High & {$49,476$} & {$2,356$} \\
    \quad Low  & {$54,201$} & {$2,581$} \\
    \multicolumn{3}{l}{\underline{Posting preference: left leaning}}\\
    \quad High & {$51,828$} & {$2,468$} \\
    \quad Low  & {$51,849$} & {$2,469$} \\
    \multicolumn{3}{l}{\underline{Posting preference: right leaning}}\\
    \quad High & {$51,429$} & {$2,449$} \\
    \quad Low  & {$52,248$} & {$2,488$} \\
    \multicolumn{3}{l}{\underline{Posting preference: toxicity}}\\
    \quad High & {$51,828$} & {$2,468$} \\
    \quad Low  & {$51,849$} & {$2,469$} \\
    \multicolumn{3}{l}{\underline{Posting preference: scam}}\\
    \quad High & {$10,836$} & {$516$} \\
    \quad Low  & {$92,841$} & {$4,421$} \\
    \multicolumn{3}{l}{\underline{Posting preference: crypto}}\\
    \quad High & {$21,399$} & {$1,019$} \\
    \quad Low  & {$82,278$} & {$3,918$} \\
    \multicolumn{3}{l}{\underline{Posting preference: misleadingness}}\\
    \quad High & {$51,828$} & {$2,468$} \\
    \quad Low  & {$51,849$} & {$2,469$} \\
    \bottomrule
    \end{tabular}
    \label{tab:he_account}
\end{table} 

\begin{figure}[ht]
    \centering
    \captionsetup[subfloat]{font={bf, small}, skip=0pt, singlelinecheck=false, labelformat=simple, position=top}
    \subfloat[]{\includegraphics[width=\linewidth]{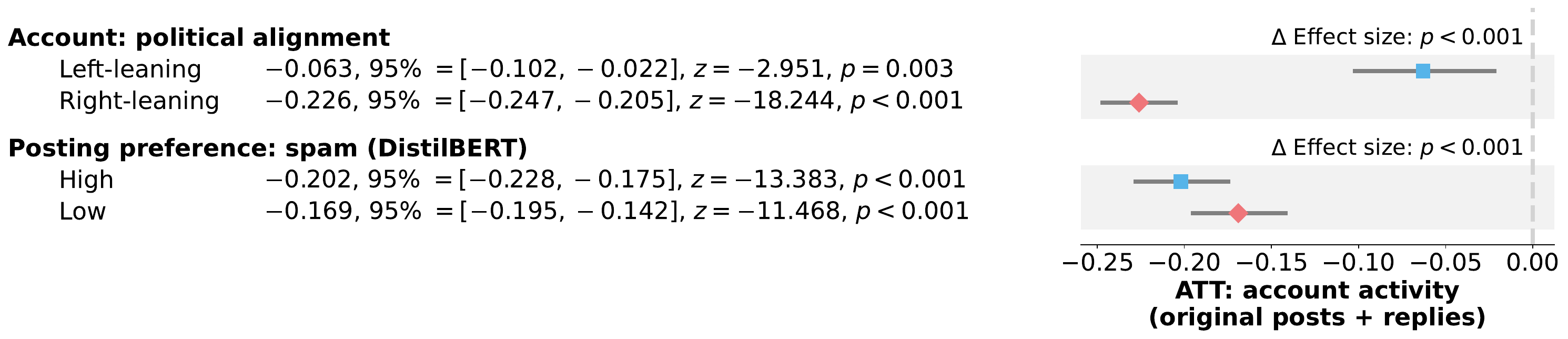}}
    \hfill
    \subfloat[]{\includegraphics[width=\linewidth]{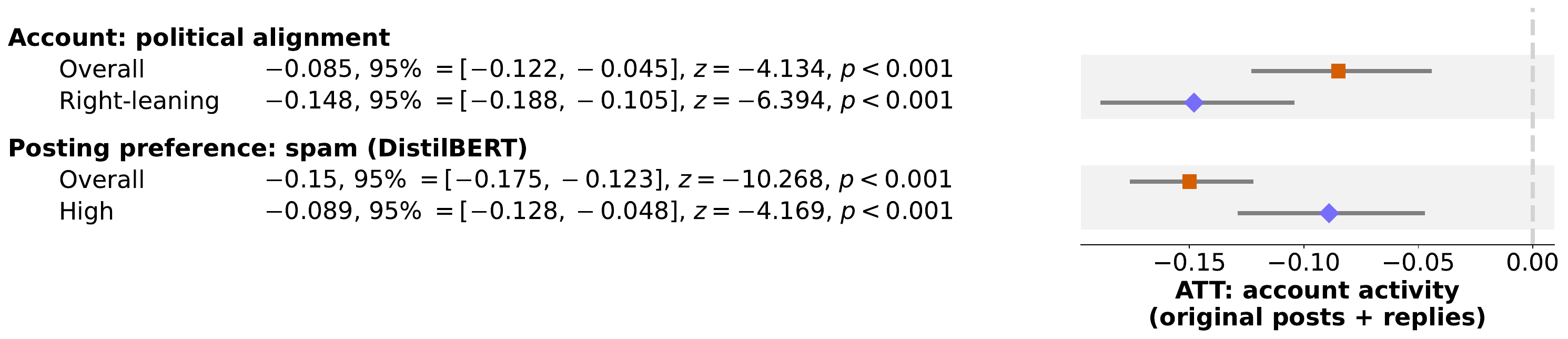}}
    \caption{\textbf{Aggregated ATT estimates for political alignment and spam.} \textbf{(a)}~Subgroup analysis. \textbf{(b)}~Interaction analysis. The estimation for political alignment is based on 100,233 (left: 29,862, right: 70,371) count observations across 4,773 accounts (left: 1,422, right: 3,351). The estimation for spam is based on 103,677 (high: 51,828, low: 51,849) count observations across 4,937 accounts (high: 2,468, low: 2,469). The error bars represent 95\% CIs.}
    \label{fig:he_sensitivity_account_scam}
\end{figure}

\vspace{1em}
\noindent \textbf{Robustness check of account alignment and posting preference on scams.}\\
Based on the LLM-inferred political alignment, we find that right-leaning accounts show a greater reduction (22.6\%), compared to left-leaning accounts (6.3\%, interaction $p<0.001$; \Cref{fig:he_sensitivity_account_scam}). This is consistent with our findings in the main analysis.
Additionally, we adopt a broader DistilBERT-based spam detection model to further validate our findings.\footnote{\url{https://huggingface.co/AventIQ-AI/SMS-Spam-Detection-Model}} The DistilBERT-based model achieves an $F_1$ score of 0.96 on an SMS Spam Collection dataset, and identifies \num{483174} spam posts in our dataset of original posts. While the detected spam encompasses a broader scope than scam posts, we still find a trend of larger reduction in activity among accounts with a high preference for posting spam posts (20.2\% vs. 16.9\%, interaction $p<0.001$; \Cref{fig:he_sensitivity_account_scam}).

\clearpage
\subsection{Heterogeneity by reply targets}
\label{supp:he_replies}

The full estimation results for the separated effects on outbound replies to \US vs. non-\US recipient accounts and across platform-revealed locations are reported in \Cref{tab:did_he_replies} and \Cref{tab:did_he_replies_regions}.

\begin{table*}[ht]
    \centering
    \footnotesize
    \caption{\textbf{DiD estimates for the number of outbound (toxic) replies from accounts in the treatment group to different recipient accounts.} Weekly and account-level fixed effects are included.}
    \begin{tabular}{l*{6}{c}}
    \toprule
    &   DiD estimate   &   $z$  &   $p$   &   95\% CI & \# Obs &   \# Accounts \\
    \midrule
    \multicolumn{7}{l}{\underline{Outbound replies from accounts in the treatment group to different recipient accounts}}\\
    All recipient accounts & $-0.11$ & $z=-4.293$ & $p<0.001$ & $[-0.155, -0.061]$ & $18,864$ & $2,358$\\
    Non-\US & $-0.052$ & $z=-1.016$ & $p=0.31$ & $[-0.145, 0.051]$ & $11,048$ & $1,381$\\
    \US & $-0.103$ & $z=-3.933$ & $p<0.001$ & $[-0.15, -0.053]$ & $18,552$ & $2,319$\\
    \multicolumn{7}{l}{\underline{Outbound toxic replies from accounts in the treatment group to different recipient accounts}}\\
    All recipient accounts & $-0.169$ & $z=-4.598$ & $p<0.001$ & $[-0.233, -0.101]$ & $12,872$ & $1,609$\\
    Non-\US & $0.046$ & $z=0.385$ & $p=0.7$ & $[-0.168, 0.316]$ & $4,488$ & $561$\\
    \US & $-0.179$ & $z=-4.783$ & $p<0.001$ & $[-0.243, -0.11]$ & $12,632$ & $1,579$\\
    \multicolumn{7}{l}{\underline{Outbound toxic replies from accounts in the treatment group to different recipient accounts (ratio)}}\\
    All recipient accounts & $-0.014$ & $t=-2.524$ & $p=0.012$ & $[-0.026, -0.003]$ & $13,080$ & $2,358$\\
    Non-\US & $-0.0$ & $t=-0.015$ & $p=0.988$ & $[-0.031, 0.03]$ & $4,468$ & $1,381$\\
    \US & $-0.017$ & $t=-2.834$ & $p=0.005$ & $[-0.028, -0.005]$ & $12,767$ & $2,319$\\
    \bottomrule
    \end{tabular}
    \label{tab:did_he_replies}
\end{table*}

\begin{table}[ht]
    \centering
    \footnotesize
    \caption{\textbf{Summary of count observations and accounts for DiD estimations across the platform-revealed regions of location-mismatched accounts.}}
    \begin{tabular}{lcc}
    \toprule
    {Group} & {\# Count observations} & {\# Accounts}\\
    \midrule
    \multicolumn{3}{l}{\underline{Europe}}\\
    \quad All   & {$12,856$} & {$1,607$} \\
    \quad Toxic & {$9,144$}  & {$1,143$} \\
    \multicolumn{3}{l}{\underline{Asia}}\\
    \quad All   & {$12,160$} & {$1,520$} \\
    \quad Toxic & {$8,560$}  & {$1,070$} \\
    \multicolumn{3}{l}{\underline{Africa}}\\
    \quad All   & {$10,904$} & {$1,363$} \\
    \quad Toxic & {$7,672$}  & {$959$} \\
    \multicolumn{3}{l}{\underline{North America}}\\
    \quad All   & {$12,024$} & {$1,503$} \\
    \quad Toxic & {$8,768$}  & {$1,096$} \\
    \multicolumn{3}{l}{\underline{South America}}\\
    \quad All   & {$11,040$} & {$1,380$} \\
    \quad Toxic & {$7,912$}  & {$989$} \\
    \multicolumn{3}{l}{\underline{Australia (Oceania)}}\\
    \quad All   & {$10,568$} & {$1,321$} \\
    \quad Toxic & {$7,736$}  & {$967$} \\
    \bottomrule
    \end{tabular}
    \label{tab:did_he_replies_regions}
\end{table}

\vspace{1em}
\noindent \textbf{Outbound replies vs. inbound replies.}\\
Our results indicate that location transparency has no significant effect on audience engagement, as measured by inbound replies directed to accounts in the treatment group (see \Cref{fig:main}b in the main text and \Cref{fig:audience_engagement}a in \Cref{supp:did_estimations}). This is seemingly at odds with the finding that outbound replies from location-mismatched accounts to location-matched accounts are reduced. We consider two factors that could reconcile these two findings: (i) the proportion of location-mismatched accounts among the audience; and (ii) the proportion of call-out replies regarding location mismatch. First, we additionally collect \num{1276172} inbound replies that are sent from \num{476613} accounts, covering a period between four weeks before and four weeks after the transparency intervention. We again use the country extraction approach to identify the location matchingness of these audience accounts and find that \num{16870} location-matched ones and \num{613} location-mismatched ones. We find that location-mismatched users account for only 3.5\% of all identified audience accounts, suggesting that they constitute too small a fraction to drive significant changes in overall engagement. Additionally, the introduction of location transparency could trigger call-out replies regarding location mismatch. By identifying such replies from audiences (detailed below), we find that the proportion of call-out replies directed to accounts in the treatment group increases sharply from 0.001 to 0.007 following the intervention, compared to a mild increase from 0.002 to 0.003 for accounts in the control group. This suggests that the transparency feature could trigger an increase in call-out replies directed to location-mismatched accounts. However, the overall proportion of call-out replies remains very small (0.4\% on average), and is thus insufficient to produce a meaningful change in total inbound replies. Overall, the statistically non-detectable effect of location transparency on audience engagement can be partially explained by both the negligible proportion of location-mismatched accounts among the audience and the very low proportion of call-out replies, neither of which is sufficient to produce a measurable shift in total inbound replies.

\begin{figure}[ht]
    \centering
    \includegraphics[width=.43\linewidth]{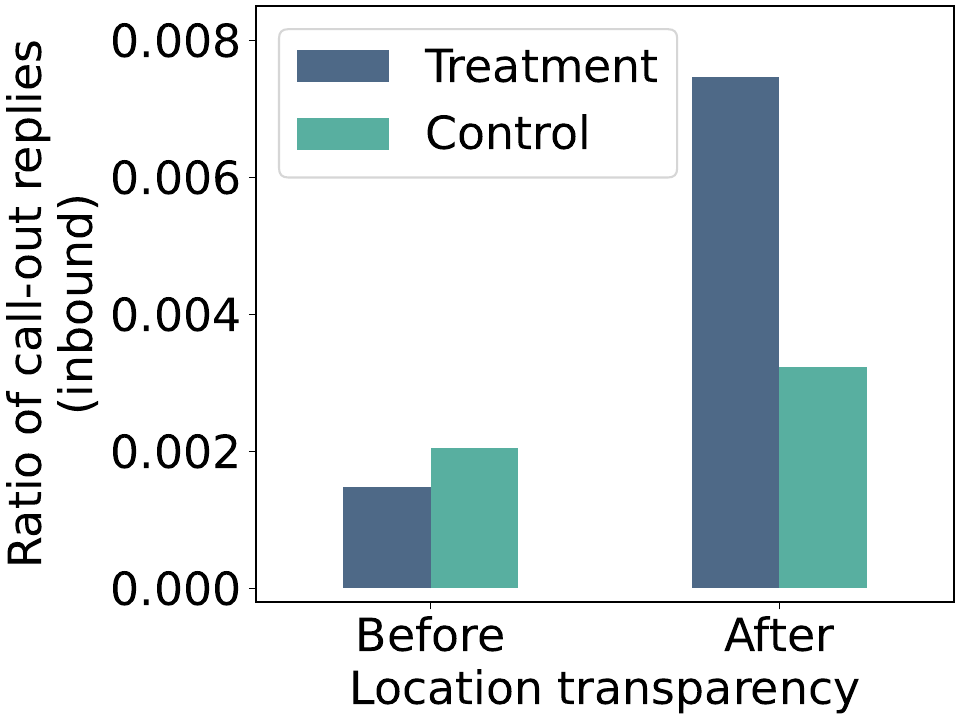}
    \caption{\textbf{The distribution of call-out replies before and after location transparency and across the treatment and control groups.}}
    \label{fig:call_out_replies}
\end{figure}

\vspace{1em}
\noindent \textbf{Identification of call-out replies.}\\
We consider that audiences may directly call out that the accounts' self-claimed locations or countries do not match the disclosed locations. Given the large number of replies from audiences (\num{1276172} in total), we sample half of the collected audience replies for the call-out analysis, employing the GPT-5.4 mini model to identify potential call-out replies directed at the selected accounts (see the full instruction in Prompt~5). Specifically, the model successfully processes \num{606001} replies (47.5\%) and identifies \num{2665} replies that are related to location mismatch. To ensure the reliability of mini model outputs, we select 200 random replies that are identified as location-mismatch replies and use the full GPT-5.4 model to double-check. Based on the outputs from the GPT-5.4 model, the precision score of GPT-5.4 mini model on the identification of call-out replies reaches 0.82. This suggests that the mini model can identify the call-out replies related to location mismatch at scale and with comparable efficacy to its full-sized counterpart. The distribution of call-out replies before and after location transparency and across the treatment and control groups is shown in \Cref{fig:call_out_replies}.

\begin{tcolorbox}[colback=blue!5, colframe=blue!40, title=\textbf{Prompt~5. Instruction for the identification of call-out replies}]
\setlength{\parskip}{0.8em}   
\setlength{\parindent}{0pt}   
\linespread{1}\selectfont     

You are given a X/Twitter reply. Determine whether the reply discusses, questions, or accuses an account of having a misleading, hidden, incorrect, or inconsistent location.

Label the reply as:\\
0 = NotLocationMismatch (no discussion of location mismatch)\\
1 = LocationMismatch (claims, questions, suspects, or suggests that the account's self-claimed location is hidden, misleading, inaccurate, fabricated, or inconsistent with its actual location)\\
2 = Unclear (location is mentioned but the intent is ambiguous)

Output JSON:\\
\{``label'': 0$|$1$|$2\}
\end{tcolorbox}

\end{document}